\documentclass[review]{elsarticle}

\usepackage{lineno,hyperref}
\usepackage{amsmath}

\modulolinenumbers[5]

\usepackage{bm}
\usepackage{graphicx}

\begin{document}

\begin{frontmatter}

\title{General Tensor Structure for Inclusive and Semi-inclusive Electron Scattering\\ from Polarized Spin-1/2 Targets}

\author{T. W. Donnelly}
\address{Center for Theoretical Physics, Department of Physics and Laboratory for Nuclear Science, Massachusetts Institute of Technology, Cambridge, MA 02139}

\author{Sabine Jeschonnek}
\address{The Ohio State University, Physics
Department, Lima, OH 45804}

\author{ J. W. Van Orden }
\address{Department of Physics, Old Dominion University, Norfolk, VA
23529\\ and\\ Jefferson Lab, \footnote{Notice: Authored by Jefferson Science Associates, LLC under U.S. DOE Contract No. DE-AC05-06OR23177. The U.S. Government retains a non-exclusive, paid-up, irrevocable, world-wide license to publish or reproduce this manuscript for U.S. Government purposes.} 
 12000 Jefferson Avenue, Newport
News, VA 23606}

\begin{abstract}

The general structure of semi-inclusive polarized electron scattering from polarized spin-1/2 targets is developed for use at all energy scales, from modest-energy nuclear physics applications to use in very high energy particle physics. The leptonic and hadronic tensors that enter in the formalism are constructed in a general covariant way in terms of kinematic factors that are frame dependent but model independent and invariant response functions which contain all of the model-dependent dynamics. In the process of developing the general problem the relationships to the conventional responses expressed in terms of the helicity components of the exchanged virtual photon are presented. For semi-inclusive electron scattering with polarized electrons and polarized spin-1/2 targets one finds that 18 invariant response functions are required, each depending on four Lorentz scalar invariants. Additionally it is shown how the semi-inclusive cross sections are related via integrations over the momentum of the selected coincidence particle and sums over open channels.

\end{abstract}


\end{frontmatter}

\linenumbers

\section{Introduction\label{sec-intro}}
\begin{figure}
	\centering
	\includegraphics[height=10cm]{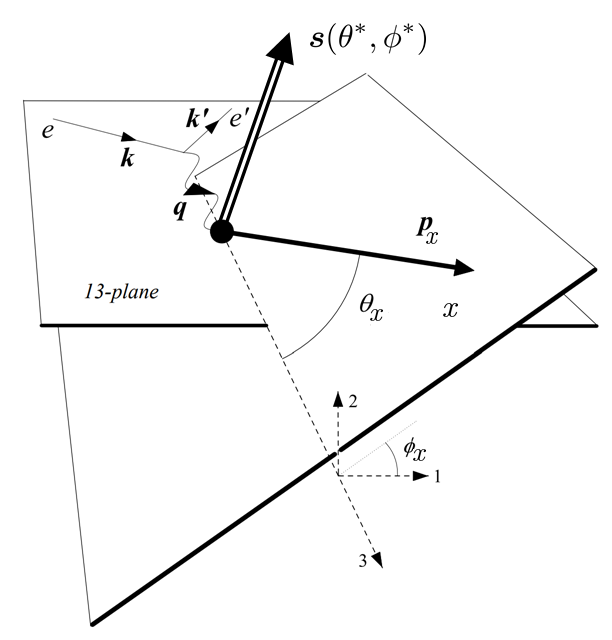} 				
	\caption{Schematic representation of semi-inclusive electron scattering. The coordinate system is chosen such that the electron scattering occurs in the 13-plane and has the 3-momentum transfer along the 3-axis. The particle x detected in coincidence with the scattered electron has 3-momentum ${\mathbf p}_x$ which lies in a plane in general inclined at an azimuthal angle $\phi_x$ with respect to the electron scattering plane and has polar angle $\theta_x$ with respect to ${\mathbf q}$. The polarization of the spin-1/2 target involves the spin 3-vector ${\mathbf s}$ with polar and azimuthal angles $\theta^\ast$ and $\phi^\ast$, respectively, in the chosen coordinate system.
}
	\label{fig:figure1}
\end{figure}

In this study we place our main focus on semi-inclusive polarized electron
scattering from polarized spin-1/2 targets, shown schematically in Fig.~\ref{fig:figure1}. That is, we consider reactions
of the type $\overrightarrow{e}+\overrightarrow{A}(1/2)\rightarrow e^{\prime
}+x+B$ where the incident electron may be polarized, the spin-1/2 target $A$
may be polarized and where we assume that, in addition to the scattered
electron, some (unpolarized) particle $x$ is detected in coincidence. The
sum of all open channels that make up the final state is denoted $B$ and is
assumed not to be detected. Employing notation commonly used in nuclear physics
the reaction may be written $\overrightarrow{A}(1/2)(\overrightarrow{e}%
,e^{\prime }x)B$. We shall discuss how such semi-inclusive reactions are
related to the inclusive cross section, \textit{i.e.,} for reactions of the
type $\overrightarrow{e}+\overrightarrow{A}(1/2)\rightarrow e^{\prime }+X$
or $\overrightarrow{A}(1/2)(\overrightarrow{e},e^{\prime })X$, where $X$
denotes the complete (undetected) final state. We develop the formalism in a
general frame as we wish to be able to relate the response in different
frames of reference, in particular, in the target rest frame and in a frame
where the incident electrons and the spin-1/2 target are both moving and
colliding. Importantly, we show how the cross sections may be written in a
general way in terms of invariant response functions (see below for an
introductory discussion of what motivates this strategy).

\begin{figure}
	\centering
	\includegraphics[height=5cm]{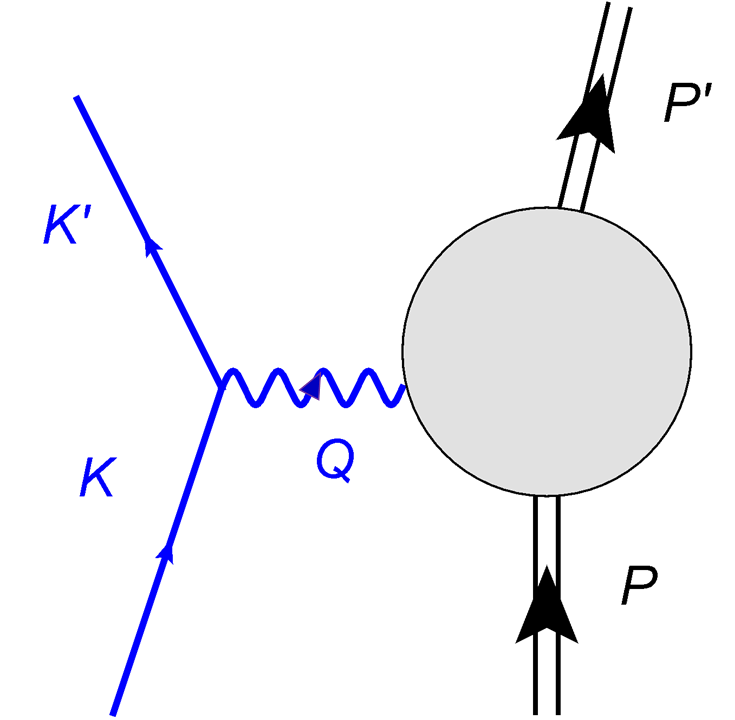} 				
	\caption{Feynman diagram for inclusive electron scattering. The 4-momenta here are discussed in the text.}
	\label{fig:figure2}
\end{figure}

Before entering into the polarized semi-inclusive developments, here we
discuss in general terms a simple, well-known example to help in
understanding the basic motivation for the present study, namely, we
consider the case of unpolarized, inclusive electron scattering from
unpolarized targets. The conventions employed in this work are summarized in
 \ref{sec-conventions}. The electron tensor $\eta _{\mu \nu }$ takes
on its standard form; this will be introduced in Sec. \ref{sec-general} and
here we take it as given. It is symmetric under interchange of $\mu $ and $%
\nu $, \textit{viz.,} $\eta _{\mu \nu }=\eta _{\nu \mu }$. Accordingly, in
forming the contraction of the leptonic and hadronic tensors, $\eta _{\mu
\nu }W^{\mu \nu }$ to obtain the invariant quantity that yields the cross
section for this situation we require only the symmetric part of the
hadronic tensor $W^{\mu \nu }$. The hadronic piece of the problem is
indicated in Fig.~\ref{fig:figure2}: here the virtual photon having 4-momentum $Q^{\mu }$
interacts with the target having 4-momentum $P^{\mu }$, leading to a final
state with 4-momentum $P^{\prime \mu }$. Since we are assuming that the
process involves inclusive scattering, nothing in the hadronic final state is
detected. Momentum conservation allows us to eliminate the total final-state
momentum, $P^{\prime \mu }=Q^{\mu }+P^{\mu }$, and hence we have two
independent 4-momenta with which the hadronic tensor is to be constructed,
namely, $Q^{\mu }$ and $P^{\mu }$. The Lorentz scalars that can be built
from these two are $Q^{2}$, $Q\cdot P$ and $P^{2}$; since $P^{2}=M^{2}$,
with $M$ the mass of the target, is presumed to be known we have only the
two remaining dynamical Lorentz scalars upon which the hadronic tensor can
depend. Typically one uses other (perhaps not invariant) quantities such as $%
(Q^{2},x)$ or $(q,\omega )$ for the dynamical variables --- these variables
will be introduced in due course.

The final step in building the hadronic tensor is then to determine the most
general form it can take, given the type of reaction being assumed.
Certainly one uses the two dynamical 4-vectors to do this. It proves
convenient to use $Q^{\mu }$ with, instead of $P^{\mu }$, a linear
combination of $P^{\mu }$ with $Q^{\mu }$%
\begin{equation}
U^{\mu }\equiv \frac{1}{M}\left( P^{\mu }-\left( \frac{Q\cdot P}{Q^{2}}%
\right) Q^{\mu }\right) ,  \label{eq-intro-1}
\end{equation}%
since it has the property that $Q\cdot U=0$, by construction. Any choice of
two independent 4-vectors will work, although experience shows that such
projected quantities help in simplifying the arguments. The symmetric
hadronic tensor for unpolarized, inclusive scattering may then be written in
terms of $Q^{\mu }Q^{\nu }$, $U^{\mu }U^{\nu }$ and $Q^{\mu }U^{\nu }+Q^{\nu
}U^{\mu }$ together with $g^{\mu \nu }$:%
\begin{equation}
W^{\mu \nu }=X_{1}g^{\mu \nu }+X_{2}Q^{\mu }Q^{\nu }+X_{3}U^{\mu }U^{\nu
}+X_{4}\left( Q^{\mu }U^{\nu }+Q^{\nu }U^{\mu }\right)   \label{eq-intro-2}
\end{equation}%
which contains four contributions involving these basis tensors each
multiplied by an invariant response function that depends on the two
dynamical Lorentz scalars, \textit{i.e.,} $X_{i}=X_{i}\left( Q^{2},Q\cdot
P\right) $ where $i=1,\ldots ,4$. Since the electromagnetic current is
conserved the constraint%
\begin{equation}
Q_{\mu }W^{\mu \nu }=0  \label{eq-intro-3}
\end{equation}%
must be satisfied, which leads to%
\begin{equation}
\left( X_{1}+X_{2}Q^{2}\right) Q^{\nu }+\left( X_{4}Q^{2}\right) U^{\nu }=0.
\label{eq-intro-4}
\end{equation}%
The theorem of linear algebra and the fact that $Q^{\nu }$ and $U^{\nu }$
are linearly independent 4-vectors then immediately yields the following:%
\begin{equation}
X_{1}+X_{2}Q^{2}=X_{4}=0.  \label{eq-intro-5}
\end{equation}%
Changing notation to the more conventional one, $X_{1}\equiv -W_{1}$ and $%
X_{3}\equiv W_{2}$, one has then proven the well-known result%
\begin{equation}
W^{\mu \nu }=-W_{1}\left( g^{\mu \nu }-\frac{Q^{\mu }Q^{\nu }}{Q^{2}}\right)
+W_{2}U^{\mu }U^{\nu },  \label{eq-intro-6}
\end{equation}%
namely, the hadronic response for this particular situation has two terms
involving two invariant response functions, each a function of two Lorentz
invariants. Upon contracting with the leptonic tensor one then recovers the
standard result%
\begin{equation}
\sigma \sim W_{2}+2W_{1}\tan ^{2}\theta _{e}/2  \label{eq-intro-7}
\end{equation}%
where $\theta _{e}$ is the electron scattering angle. Note that $W_{1,2}$
are invariant response functions, but that the factor $\tan ^{2}\theta _{e}/2
$ depends on the particular frame of reference. Moreover, this result is
often recast in a form where the helicity projections of the virtual photon
are made manifest (see later examples of why this can be important). For
this situation one finds that the cross section may be written%
\begin{equation}
\sigma \sim v_{L}W^{L}+v_{T}W^{T}  \label{eq-intro-8}
\end{equation}%
where the quantities $v_{L}$ and $v_{T}$ are the well-known leptonic
(Rosenbluth) factors and the $W^{L}$ and $W^{T}$ the corresponding hadronic
responses. All of these quantities, however, are not Lorentz invariants and
accordingly are different in different frames. A strong motivation in the
present study is to establish the relationships between the two ways of
representing the hadronic response and thereby to provide a way to relate
the hadronic physics between any two frames.

This simplest example is a textbook case (see, for example, \cite{thebook}). While constituting relatively straightforward extensions
to what has been summarized here, for the general problem that involves
polarized electrons, polarized spin-1/2 targets and semi-inclusive reactions
the developments are much more complicated.

To summarize, some of the basic motivations for this study are the following:

\begin{itemize}
\item As in the simple example discussed above, in this study we will
develop the general formalism for semi-inclusive electron scattering of
polarized electrons on polarized spin-1/2 targets. We anticipate
applications to both particle and nuclear physics and to both relatively low
energies and to the high-energy regime (HER).

\item We shall see that there are four sectors which may be separated by
employing the polarizations. When unpolarized electrons are involved only
symmetric tensors enter, whereas when the incident electrons are
longitudinally polarized only anti-symmetric tensors occur.

\item The four types of polarization (electrons polarized or not with target
polarized or not) may be separated using those polarizations. We shall see
that there are four symmetric invariant responses for the fully unpolarized
case, one anti-symmetric invariant response when the electron is polarized
but the target is not,  eight symmetric invariant responses when the
electron is unpolarized but the target is polarized, and five anti-symmetric
invariant responses when both electrons and target are polarized.

\item These 18 invariant response functions will be shown to be functions of
four Lorentz scalar invariants. The 18 responses may be sub-divided into two
sets of nine according to their properties under parity and time-reversal;
these two sets typically behave quite differently.

\item We also detail how the hadronic response may be characterized using
the helicity decomposition of the virtual photon to label the various
contributions. We shall detail how this representation relates to the
decomposition in terms of invariant response functions.

\item A prime motivation for this study is to have the semi-inclusive cross
section written in a completely general frame of reference. This then allows
one to relate the results in (say) the collider frame to the target rest
frame, or to relate the results in the rest frame to those in the
photon-target center-of-momentum frame. This can prove to be essential when
models are being developed for the hadronic physics that are
non-relativistic and hence cannot be boosted --- polarized $^{3}$He would be
one such example --- since only in the target rest frame will such models
make sense.

\item Finally, we provide some discussion of how inclusive (polarized)
scattering emerges via specific integrals over semi-inclusive cross sections
with appropriate sums over all open channels.
\end{itemize}

With the above basic motivations for the present study we briefly discuss two examples where the ideas are relevant, one from particle physics and one from nuclear physics. For the former consider charged pion production from a polarized proton target. For single-pion production one then has the (exclusive) reaction $\overrightarrow{e}+\overrightarrow{p}\rightarrow e^{\prime}+n+\pi^+$ with a neutron and a positive pion in the final state. As a semi-inclusive reaction one then has either $\overrightarrow{p}(\overrightarrow{e},e^{\prime}n)\pi^+$ where particle x is a neutron and the pion is undetected or $\overrightarrow{p}(\overrightarrow{e},e^{\prime}\pi^+)n$ where particle x is a $\pi^+$ and the neutron is undetected. In fact these are the same reaction and accordingly they constitute a single channel. Clearly there are experimental considerations involved in which particle is the one detected in coincidence; however, theoretically they are not distinguishable. For higher-energy kinematics one reaches a threshold where additional channels open. For instance, once the relevant threshold is reached, two-pion production becomes possible, $\overrightarrow{e}+\overrightarrow{p}\rightarrow e^{\prime}+n+\pi^+ +\pi^0$ and then $\overrightarrow{e}+\overrightarrow{p}\rightarrow e^{\prime}+p+\pi^- +\pi^+$, and so one, with more and more particles in the final state. Of those a given semi-inclusive reaction is to be taken as having some given particle detected in coincidence with the scattered electron and all other particles undetected. 

A second example, taken from nuclear physics, is where the polarized electron is scattered from a polarized $^3$He target. Let us focus on the reaction $^3\overrightarrow{{\rm He}}(\overrightarrow{e},e^{\prime}p)$ where a proton is assumed to be detected in coincidence with the scattered electron. The unobserved part of the final state depends on the specific kinematics of the reaction. At threshold one has the (exclusive) two-body reaction $\overrightarrow{e}+^3\overrightarrow{{\rm He}}\rightarrow e^{\prime}+p+d$ and then for slightly higher missing energies the three-body breakup reaction $\overrightarrow{e}+^3\overrightarrow{{\rm He}}\rightarrow e^{\prime}+p+p+n$. Alternatively one could have a neutron as the particle detected in coincidence with the scattered electron, $^3\overrightarrow{{\rm He}}(\overrightarrow{e},e^{\prime}n)$. In this case the two-body channel does not occur, although the three-body breakup channel does. In fact, for the latter the final state is the same and this will have consequences later when we discuss the issue of avoiding double counting. As in the particle physics example above, as the energy increases a threshold is reached where pion production can occur and the final state becomes even more complicated. Nevertheless, the semi-inclusive reaction is well defined: the point is that a specific particle is assumed to be the one called x, namely, the one that is detected, whereas all other particles in the final state must be summed while avoiding double counting.

The paper is organized in the following way: in Sec.~\ref{sec-general} some general developments are summarized which involve the contraction of the leptonic and hadronic tensors and include the specific forms for the electron scattering tensors in the Extreme Relativistic Limit (ERL$_e$). This is followed in Sec.~\ref{sec-had} with the detailed construction of the general hadronic tensors for the semi-inclusive reaction. In Sec.~\ref{subsec-basic4} the basic 4-vectors used in building the hadronic tensors are introduced, followed in Sec.~\ref{subsec-GenTen} with the 18 types of tensors that constitute the problem, and in Sec.~\ref{subsec-comp} with specific components of responses categorized by the projections of the exchanged virtual photon’s helicity. In Sec.~\ref{sec-Semi} the semi-inclusive cross section is given for a general situation where the polarized spin-1/2 target is moving in some arbitrary direction --- this for use in collider physics. For completeness the simpler situation of polarized inclusive electron scattering from a (moving) polarized spin-1/2 target is presented in Sec.~\ref{sec-CrossSec}. These general developments are then specialized to the target rest frame in Sec.~\ref{sec-restSystem}. To conclude the body of the paper a summary is given in Sec.~\ref{sec-summary} and to extend some aspects of the problem six appendices are included detailing the conventions used (A), expressing the contraction of the tensors entirely in terms of invariants (B), inverting the invariant response representations in terms of photon helicity projections (C), detailing the nature of the cross section as the available phase-space increases and more channels become open (D), including some connections with conventional kinematic variables (E) and discussing inclusive scattering in more detail to make connections with (more) familiar material (F).

\section{General Developments}\label{sec-general}

We begin with some general developments that are common to all electron
scattering formalism at the level of the plane-wave Born approximation. The
general cross section is proportional to the contraction of the leptonic and
hadronic tensors $\eta _{\mu \nu }$ and $W^{\mu \nu }$, respectively
\begin{equation}
\eta _{\mu \nu }W^{\mu \nu }.  \label{eq-gen1}
\end{equation}%
Being composed of
bilinear products of the corresponding leptonic and hadronic current matrix
elements $\left( j_{fi}\right) _{\mu }$ and $\left( J_{fi}\right) ^{\mu }$,
respectively, in the forms  
\begin{eqnarray}
\eta _{\mu \nu } &\sim &\overline{\underset{if}{\sum }}\left( j_{fi}\right)
_{\mu }^{\ast }\left( j_{fi}\right) _{\nu }  \label{eq-gen2} \\
W^{\mu \nu } &\sim &\overline{\underset{if}{\sum }}\left( J_{fi}\right)
^{\mu \ast }\left( J_{fi}\right) ^{\nu },  \label{eq-gen3}
\end{eqnarray}%
with appropriate averages over initial and sums over final states, one has
immediately that%
\begin{eqnarray}
\eta _{\nu \mu } &=&\left( \eta _{\mu \nu }\right) ^{\ast }  \label{eq-gen4}
\\
W^{\nu \mu } &=&\left( W^{\mu \nu }\right) ^{\ast }.  \label{eq-gen5}
\end{eqnarray}%
Instead of $\eta _{\mu \nu }$ we employ the following convention for the
leptonic tensor (see \cite{Donnelly:1985ry}) 
\begin{eqnarray}
\chi ^{\mu \nu } &\equiv &4m_{e}^{2}\eta ^{\mu \nu }  \label{eq-gen13a} \\
&=&\chi _{unpol}^{\mu \nu }+\chi _{pol}^{\mu \nu }.  \label{eq-gen13b}
\end{eqnarray}%
Also, since the electromagnetic current is conserved,%
\begin{equation}
Q^{\mu }\left( j_{fi}\right) _{\mu }=Q_{\mu }\left( J_{fi}\right) ^{\mu }=0,
\label{eq-gen6}
\end{equation}%
one has that%
\begin{equation}
Q^{\mu }\chi _{\mu \nu }=\chi _{\mu \nu }Q^{\nu }=Q_{\mu }W^{\mu \nu
}=W^{\mu \nu }Q_{\nu }=0.  \label{eq-gen7}
\end{equation}

Since one can decompose the tensors into symmetric and anti-symmetric
contributions (\textit{i.e.,} under exchange of $\mu $ and $\nu $), namely,%
\begin{eqnarray}
\chi _{\mu \nu }^{s} &\equiv &\frac{1}{2}\left( \chi _{\mu \nu }+\chi _{\nu
\mu }\right)  \label{eq-hadd1a} \\
\chi _{\mu \nu }^{a} &\equiv &\frac{1}{2}\left( \chi _{\mu \nu }-\chi _{\nu
\mu }\right)  \label{eq-hadd1b} \\
W_{s}^{\mu \nu } &\equiv &\frac{1}{2}\left( W^{\mu \nu }+W^{\nu \mu }\right)
\label{eq-hadd2} \\
W_{a}^{\mu \nu } &\equiv &\frac{1}{2}\left( W^{\mu \nu }-W^{\nu \mu }\right)
\label{eq-hadd3}
\end{eqnarray}%
with%
\begin{eqnarray}
\chi ^{\mu \nu } &=&\chi _{s}^{\mu \nu }+\chi _{a}^{\mu \nu }
\label{eq-hadd3a} \\
W^{\mu \nu } &=&W_{s}^{\mu \nu }+W_{a}^{\mu \nu }.  \label{eq-hadd4}
\end{eqnarray}%
Clearly one has the individual continuity equation relationships%
\begin{equation}
Q_{\mu }W_{s}^{\mu \nu }=Q_{\mu }W_{a}^{\mu \nu }=0,  \label{eq-hadd5}
\end{equation}%
and also only symmetric (anti-symmetric) leptonic tensors will contract with
symmetric (anti-symmetric) hadronic tensors when forming the cross section,
the last going as%
\begin{equation}
\chi _{\mu \nu }W^{\mu \nu }=\chi _{\mu \nu }^{s}W_{s}^{\mu \nu }+\chi _{\mu
\nu }^{a}W_{a}^{\mu \nu }.  \label{eq-hadd6}
\end{equation}%
We also have from Eqs. (\ref{eq-gen4}) and (\ref{eq-gen5}) that 
\begin{eqnarray}
\chi _{s}^{\mu \nu } &=&\mathrm{Re}\chi ^{\mu \nu }  \label{eq-hadd6x1} \\
\chi _{a}^{\mu \nu } &=&i\mathrm{Im}\chi ^{\mu \nu }  \label{eq-hadd6x2} \\
W_{s}^{\mu \nu } &=&\mathrm{Re}W^{\mu \nu }  \label{eq-hadd6a} \\
W_{a}^{\mu \nu } &=&i\mathrm{Im}W^{\mu \nu };  \label{eq-hadd6c}
\end{eqnarray}%
we shall make use of this when constructing explicit forms for the tensors
by including the factor $i$ in the appropriate places.

Furthermore, one can isolate contributions that contain the target spin from
those that do not by forming the unpolarized (spin sum) terms and polarized
(spin difference) terms, so that the total becomes%
\begin{eqnarray}
\chi _{s,a}^{\mu \nu } &=&\left( \chi _{s,a}^{\mu \nu }\right)
_{unpol}+\left( \chi _{s,a}^{\mu \nu }\right) _{pol}  \label{eq-hadd7a} \\
W_{s,a}^{\mu \nu } &=&\left( W_{s,a}^{\mu \nu }\right) _{unpol}+\left(
W_{s,a}^{\mu \nu }\right) _{pol}  \label{eq-hadd7}
\end{eqnarray}%
with all four contributions individually satisfying the continuity equation
constraint:%
\begin{eqnarray}
Q_{\mu }\left( \chi _{s,a}^{\mu \nu }\right) _{unpol} &=&Q_{\mu }\left( \chi
_{s,a}^{\mu \nu }\right) _{pol}=0  \label{eq-hadd8a} \\
Q_{\mu }\left( W_{s,a}^{\mu \nu }\right) _{unpol} &=&Q_{\mu }\left(
W_{s,a}^{\mu \nu }\right) _{pol}=0.  \label{eq-hadd8}
\end{eqnarray}%
When only the incident electrons may be polarized but the scattered
electron's polarization is assumed not to be measured one can show that the
leptonic tensor contributions that do not involve the electron polarization
are only symmetric, while those that do involve the electron polarization
are only anti-symmetric (see \cite{Donnelly:1985ry}).

We shall adopt the convention where $\mathbf{q}$ points along the
3-direction so that the 4-vector momentum transfer is%
\begin{equation}
Q^{\mu }=\left( \omega ,0,0,q\right)   \label{eq-gen8}
\end{equation}%
with energy transfer $\omega = \nu$ (the former is commonly employed in nuclear physics while the latter is almost always chosen for use in particle physics; we use the two interchangeably) and 3-momentum transfer $q=|\mathbf{q}|$.
One can show that for electron scattering the 4-momentum transfer must be
spacelike: %
\begin{equation}
Q^{2}=\omega ^{2}-q^{2}\leq 0;  \label{eq-gen9}
\end{equation}%
(see the comment in  \ref{sec-conventions}). 
We shall define the following dimensionless quantities that prove to be useful later%
\begin{eqnarray}
\nu^{\prime}  &\equiv &\frac{\omega }{q} = \frac{\nu }{q}  \label{eq-gen11} \\
\rho  &\equiv &\frac{-Q^{2}}{q^{2}}=1-{\nu^{\prime}} ^{2}  \label{eq-gen11a} \\
\rho^{\prime}  &\equiv &\frac{q}{\epsilon + \epsilon^{\prime}}  \label{eq-gen11aa}
\end{eqnarray}%
and have from Eq. (\ref{eq-gen9}) that%
\begin{eqnarray}
0 &\leq &\nu^{\prime} \leq 1  \label{eq-gen12} \\
0 &\leq &\rho \leq 1.  \label{eq-gen12a} \\
0 &\leq &\rho^{\prime} \leq 1.  \label{eq-gen12aa}
\end{eqnarray}
The continuity equation constraints above then imply that%
\begin{eqnarray}
\left( \chi _{s,a}^{3\nu }\right) _{unpol} &=&\nu^{\prime}%
 \left( \chi _{s,a}^{0\nu }\right) _{unpol}  \label{eq-gen10a} \\
\left( \chi _{s,a}^{3\nu }\right) _{pol} &=&\nu^{\prime}
\left( \chi _{s,a}^{0\nu }\right) _{pol}  \label{eq-gen10b} \\
\left( W_{s,a}^{3\nu }\right) _{unpol} &=&\nu^{\prime}
\left( W_{s,a}^{0\nu }\right) _{unpol}  \label{eq-gen10c} \\
\left( W_{s,a}^{3\nu }\right) _{pol} &=&\nu^{\prime}
\left( W_{s,a}^{0\nu }\right) _{pol}.  \label{eq-gen10d}
\end{eqnarray}%

\subsection{Contraction of Tensors\label{sec-contract}}

We now proceed to contract the leptonic and hadronic tensors involving the
separated symmetric (10) and anti-symmetric (6) contractions%
\begin{eqnarray}
\left( \chi _{s}\right) _{\mu \nu }W_{s}^{\mu \nu } &=&\left( \chi
_{s}\right) _{00}W_{s}^{00}+2\left( \chi _{s}\right) _{03}W_{s}^{03}+\left(
\chi _{s}\right) _{33}W_{s}^{33}  \notag \\
&&+\left( \chi _{s}\right) _{11}W_{s}^{11}+\left( \chi _{s}\right)
_{22}W_{s}^{22}+2\left( \chi _{s}\right) _{12}W_{s}^{12}  \notag \\
&&+2\left\{ \left( \chi _{s}\right) _{01}W_{s}^{01}+\left( \chi _{s}\right)
_{31}W_{s}^{31}\right.  \notag \\
&&+\left. \left( \chi _{s}\right) _{02}W_{s}^{02}+\left( \chi _{s}\right)
_{32}W_{s}^{32}\right\}  \label{q-cont1} \\
\left( \chi _{a}\right) _{\mu \nu }W_{a}^{\mu \nu } &=&2\left\{ \left( \chi
_{a}\right) _{03}W_{a}^{03}+\left( \chi _{a}\right) _{12}W_{a}^{12}\right. 
\notag \\
&&+\left( \chi _{a}\right) _{01}W_{a}^{01}+\left( \chi _{a}\right)
_{31}W_{a}^{31}  \notag \\
&&\left. +\left( \chi _{a}\right) _{02}W_{a}^{02}+\left( \chi _{a}\right)
_{32}W_{a}^{32}\right\}  \label{eq-cont2}
\end{eqnarray}%
where we have employed the symmetries under $\mu \leftrightarrow \nu $.
Also, using the continuity equation constraints we find that%
\begin{eqnarray}
\left( \chi _{s}\right) _{\mu \nu }W_{s}^{\mu \nu } &=&\rho ^{2}\left( \chi
_{s}\right) _{00}W_{s}^{00}+\left( \chi _{s}\right) _{11}W_{s}^{11}+\left(
\chi _{s}\right) _{22}W_{s}^{22}  \notag \\
&&+2\left\{ \left( \chi _{s}\right) _{12}W_{s}^{12}+\rho \left( \chi
_{s}\right) _{01}W_{s}^{01}+\rho \left( \chi _{s}\right)
_{02}W_{s}^{02}\right\}  \label{eq-cont3} \\
\left( \chi _{a}\right) _{\mu \nu }W_{a}^{\mu \nu } &=&2\left\{ \left( \chi
_{a}\right) _{12}W_{a}^{12}+\rho \left( \chi _{a}\right)
_{01}W_{a}^{01}+\rho \left( \chi _{a}\right) _{02}W_{a}^{02}\right\} ,
\label{eq-cont4}
\end{eqnarray}%
namely, 6 symmetric and 3 anti-symmetric contributions for a total of 9, as expected. Note
that $\left( \chi _{a}\right) _{03}W_{a}^{03}=0$, since $\left( \chi
_{a}\right) _{03}=-\nu^{\prime} \left( \chi _{a}\right) _{00}=0$. We have from past
work \cite{Donnelly:1985ry} that equivalently the contractions may be re-written in terms
of the following real leptonic 
kinematic factors and responses:%
\begin{equation}
\chi _{s}^{00}=\chi _{unpol}^{00}\equiv \frac{1}{2}v_{0}  \label{eq-gen14}
\end{equation}%
we have%
\begin{eqnarray}
V_{L} &\equiv &\rho ^{2}  \label{eq-gg1} \\
V_{T} &\equiv &\frac{2}{v_{0}}\cdot \frac{1}{2}\left( \chi _{s}^{22}+\chi
_{s}^{11}\right)  \label{eq-gg2} \\
V_{TT} &\equiv &\frac{2}{v_{0}}\cdot \frac{1}{2}\left( \chi _{s}^{22}-\chi
_{s}^{11}\right)  \label{eq-gg3} \\
V_{TL} &\equiv &\frac{2}{v_{0}}\cdot \frac{1}{\sqrt{2}}\rho \left( -\chi
_{s}^{01}\right)  \label{eq-gg4} \\
V_{T^{\prime }} &\equiv &\frac{2}{v_{0}}\cdot \left( -i\chi _{a}^{12}\right)
\label{eq-gg5} \\
V_{TL^{\prime }} &\equiv &\frac{2}{v_{0}}\cdot \frac{1}{\sqrt{2}}\rho \left(
i\chi _{a}^{02}\right)  \label{eq-gg6} \\
V\underline{_{TT}} &\equiv &\frac{2}{v_{0}}\cdot \left( \chi _{s}^{12}\right)
\label{eq-gg7} \\
V_{\underline{TL}} &\equiv &\frac{2}{v_{0}}\cdot \frac{1}{\sqrt{2}}\rho
\left( -\chi _{s}^{02}\right)  \label{eq-gg8} \\
V_{\underline{TL}^{\prime }} &\equiv &\frac{2}{v_{0}}\cdot \frac{1}{\sqrt{2}}%
\rho \left( -i\chi _{a}^{01}\right)  \label{eq-gg9}
\end{eqnarray}%
for the leptonic factors, and%
\begin{eqnarray}
W^{L} &\equiv &\left( W_{fi}^{00}\right) _{s}  \label{eq-g1} \\
W^{T} &\equiv &\left( W_{fi}^{22}\right) _{s}+\left( W_{fi}^{11}\right) _{s}
\label{eq-g2} \\
W^{TT} &\equiv &\left( W_{fi}^{22}\right) _{s}-\left( W_{fi}^{11}\right) _{s}
\label{eq-g3} \\
W^{TL} &\equiv &2\sqrt{2}\left( W_{fi}^{01}\right) _{s}=2\sqrt{2}\mathrm{Re}%
W_{fi}^{01}  \label{eq-g4} \\
W^{T^{\prime }} &\equiv &2i\left( W_{fi}^{12}\right) _{a}=-2\mathrm{Im}%
W_{fi}^{12}  \label{eq-g5} \\
W^{TL^{\prime }} &\equiv &2\sqrt{2}i\left( W_{fi}^{02}\right) _{a}=-2\sqrt{2}%
\mathrm{Im}W_{fi}^{02}  \label{eq-g6} \\
W^{\underline{TT}} &\equiv &2\left( W_{fi}^{12}\right) _{s}=2\mathrm{Re}%
W_{fi}^{12}  \label{eq-g7} \\
W^{\underline{TL}} &\equiv &2\sqrt{2}\left( W_{fi}^{02}\right) _{s}=2\sqrt{2}%
\mathrm{Re}W_{fi}^{02}  \label{eq-g8} \\
W^{\underline{TL}^{\prime }} &\equiv &-2\sqrt{2}i\left( W_{fi}^{01}\right)
_{a}=2\sqrt{2}\mathrm{Im}W_{fi}^{01}  \label{eq-g9}
\end{eqnarray}%
for the hadronic parts of the response. As in the cited work the notation
here is the following: the quantities labelled $L$ refer to contributions
involving the $\mu \nu =00$ parts of the tensors; those labelled $T$, $TT$, $%
T^{\prime }$ and \underline{$TT$} involve only transverse components of the
tensors; and those labelled $TL$, $TL^{\prime }$, \underline{$TL$} and 
\underline{$TL$}$^{\prime }$ involve interferences having real or imaginary
parts of the $\mu \nu =01$ and $02$ components of the tensors. Unprimed
quantities arise from symmetric tensors, \textit{viz.,} those that do not
involve polarized electrons, whereas those with primes only occur when
electron polarizations enter. The underlined quantities labelled \underline{$%
TT$} and \underline{$TL$} occur only when the electron beam is polarized 
\emph{and} the polarization of the scattered electron is measured (see \cite{Donnelly:1985ry});
since we will not consider this situation in the present study, these
contributions are henceforth dropped. Finally, the sector labelled 
\underline{$TL$}$^{\prime }$ does occur when only the electron beam is
polarized, although at high energies these can also safely be ignored since
they go as $1/\gamma $ where $\gamma $ is the usual ratio of energy to mass
for the electron and thus are also neglected in the present work leaving 6
classes of response. Accordingly, for the situation of interest in the
present study the full contraction of the leptonic and hadronic tensors may
then be written in terms of these real quantities, 4 involving symmetric
contributions and 2 involving anti-symmetric contributions:%
\begin{eqnarray}
\mathcal{C} &=&v_{0}\left[ V_{L}W^{L}+V_{T}W^{T}+V_{TL}W^{TL}+V_{TT}W^{TT}%
\right.  \notag \\
&&\left. + h \left( V_{T^{\prime }}W^{T^{\prime }}+V_{TL^{\prime }}W^{TL^{\prime }} \right )
\right] ,  \label{eq-g4aa}
\end{eqnarray}%
where $\mathcal{C}$ is a Lorentz invariant. The electron helicity is denoted by $h$. We again note that, while the entire right-hand side of the equation forms a Lorentz invariant, the individual factors are all frame-dependent.

We next proceed to develop the various tensors and related parts of the
response.

\subsection{Leptonic Tensors\label{sec-lept}}

The leptonic tensor may be built from the 4-momenta of the incident electron
beam and of the scattered electron, $K^{\mu }$ and $K^{\prime \mu }$,
respectively. The incident electron has 3-momentum $k$ and on-shell energy $%
\epsilon =\sqrt{m_{e}^{2}+k^{2}}$, the scattered electron has 3-momentum $%
k^{\prime }$ and energy $\epsilon ^{\prime }=\sqrt{m_{e}^{2}+k^{\prime 2}}$,
and $\theta _{e}$ denotes the electron scattering angle. Alternatively, it
is often convenient to re-express the tensor in terms of two other
4-vectors, the 4-momentum transfer%
\begin{equation}
Q^{\mu }=K^{\mu }-K^{\prime \mu }  \label{eq-lept1}
\end{equation}%
and%
\begin{equation}
R^{\mu }\equiv \frac{1}{2}\left( K^{\mu }+K^{\prime \mu }\right) .
\label{eq-lept2}
\end{equation}%
The leptonic tensor for electron scattering in the plane-wave Born
approximation is well-known from previous work. Here we draw on the
developments in \cite{Donnelly:1985ry} where a general form was presented
that allows for the electron mass to be retained and where both the incident
and scattered electrons can be polarized. Hereafter we will restrict our
attention to the situation where the electrons have energies that are much
greater than their mass, namely, the so-called Extreme Relativistic Limit
(ERL$_{e}$). Accordingly, we take $\epsilon \approx k$ and $\epsilon
^{\prime }\approx k^{\prime }$. Additionally, we shall assume that only the
incident beam is polarized (in fact longitudinally; see \cite{Donnelly:1985ry}) and then some
simplifications for the leptonic tensor are seen to occur. In particular,
the cases $V\underline{_{TT}}$, $V\underline{_{TL}}$ and $V\underline{_{TL^{\prime}}}$
in Eqs. (\ref{eq-gg7}-\ref{eq-gg9}) are absent and for the six cases that
remain one has (see \cite{Donnelly:1985ry})  
\begin{eqnarray}
&&V_{L}\underset{ERL_{e}}{\longrightarrow }v_{L}  \label{eq-i1} \\
&&V_{T}\underset{ERL_{e}}{\longrightarrow }v_{T}  \label{eq-i2} \\
&&V_{TT}\underset{ERL_{e}}{\longrightarrow }v_{TT}  \label{eq-i3} \\
&&V_{TL}\underset{ERL_{e}}{\longrightarrow }v_{TL}  \label{eq-i4} \\
&&V_{T^{\prime }}\underset{ERL_{e}}{\longrightarrow }hv_{T^{\prime }}
\label{eq-i5} \\
&&V_{TL^{\prime }}\underset{ERL_{e}}{\longrightarrow }hv_{TL^{\prime }}
\label{eq-i7}
\end{eqnarray}%
where $h\equiv \pm 1$ is the incident electron's helicity. In detail we have%
\begin{equation}
\chi ^{\mu \nu }\equiv \chi _{unpol}^{\mu \nu }+\chi _{pol}^{\mu \nu },
\label{eq-lept8b}
\end{equation}%
where%
\begin{eqnarray}
\chi _{unpol}^{\mu \nu } &=&\chi _{s}^{\mu \nu }=K^{\mu }K^{\prime \nu
}+K^{\prime \mu }K^{\nu }+\frac{1}{2}Q^{2}g^{\mu \nu }  \label{eq-lept9} \\
&\equiv &\chi _{1,s}^{\mu \nu }+\chi _{2,s}^{\mu \nu }  \label{eq-lept10} \\
\chi _{1,s}^{\mu \nu } &=&\frac{1}{2}Q^{2}\left( g^{\mu \nu }-\frac{Q^{\mu
}Q^{\nu }}{Q^{2}}\right)   \label{eq-lept11} \\
\chi _{2,s}^{\mu \nu } &=&2R^{\mu }R^{\nu }  \label{eq-lept12} \\
\chi _{pol}^{\mu \nu } &=&\chi _{a}^{\mu \nu }=-ih\epsilon ^{\mu \nu
\alpha \beta }K_{\alpha }K_{\beta }^{\prime }  \label{eq-lept13} \\
&=&-ih\epsilon ^{\mu \nu \alpha \beta }Q_{\alpha }R_{\beta }.
\label{eq-lept14}
\end{eqnarray}

As noted above, typically we work in a coordinate system where the
3-momentum transfer lies in the 3-direction, and hence%
\begin{equation}
Q^{\mu }=\left( \omega ,0,0,q\right) =q\left( \nu^{\prime} ,0,0,1\right) 
\label{eq-lept16}
\end{equation}%
with $\nu^{\prime} \equiv \omega /q$, as usual. This implies that%
\begin{equation}
\chi _{m}^{3\alpha }=\nu^{\prime} \chi _{m}^{0\alpha }  \label{eq-lept17}
\end{equation}%
for $m=\left( 1,s\right) ,$ $\left( 2,s\right) $ or $\left( a\right) $. We
have%
\begin{eqnarray}
-Q^{2} &=&\left\vert Q^{2}\right\vert = q^2 \rho =4kk^{\prime }\sin ^{2}\theta _{e}/2
\label{eq-lept31} \\
v_{0} &\equiv &(\epsilon +\epsilon ^{\prime })^{2}-q^{2}= q^2 \left( \frac{1}{(\rho^{\prime})^2} - 1 \right) =4kk^{\prime }\cos
{}^{2}\theta _{e}/2.  \label{eq-lept32}
\end{eqnarray}%
Using these identities one finds for the required components of the
symmetric (electron unpolarized) and anti-symmetric (electron longitudinally
polarized) tensors%
\begin{eqnarray}
\chi _{s}^{00}-2\nu \chi _{s}^{03}+\nu ^{2}\chi _{s}^{33} &\equiv &\frac{1}{2%
}v_{0}\times v_{L}  \label{eq-lept38} \\
\frac{1}{2}\left( \chi _{s}^{22}+\chi _{s}^{11}\right)  &\equiv &\frac{1}{2}%
v_{0}\times v_{T}  \label{eq-lept39} \\
\frac{1}{2}\left( \chi _{s}^{22}-\chi _{s}^{11}\right)  &\equiv &\frac{1}{2}%
v_{0}\times v_{TT}  \label{eq-lept40} \\
\frac{1}{\sqrt{2}}\left( \chi _{s}^{01}-\nu \chi _{s}^{31}\right)  &\equiv &-%
\frac{1}{2}v_{0}\times v_{TL}  \label{eq-lept41} \\
\chi _{a}^{12} &\equiv &ih\frac{v_{0}}{2}\times v_{T^{\prime }}
\label{eq-lept48} \\
\frac{1}{\sqrt{2}}\left( \chi _{a}^{02}-\nu \chi _{a}^{32}\right)  &\equiv
&-ih\frac{v_{0}}{2}\times v_{TL^{\prime }},  \label{eq-lept49}
\end{eqnarray}%
which yields the standard results:%
\begin{eqnarray}
v_{L} &=&\rho ^{2}  \label{eq-lept42} \\
v_{T} &=&\frac{1}{2}\rho +\tan ^{2}\theta _{e}/2  \label{eq-lept43} \\
v_{TT} &=&-\frac{1}{2}\rho   \label{eq-lept44} \\
v_{TL} &=&-\frac{1}{\sqrt{2}}\rho \sqrt{\rho +\tan ^{2}\theta _{e}/2}
\label{eq-lept45} \\
v_{T^{\prime }} &=&\tan \theta _{e}/2\sqrt{\rho +\tan ^{2}\theta _{e}/2}
\label{eq-lept50} \\
v_{TL^{\prime }} &=&-\frac{1}{\sqrt{2}}\rho \tan \theta _{e}/2
\label{eq-lept51}
\end{eqnarray}

One may also re-write the leptonic factors in a way that involves the
so-called photon longitudinal polarization. One begins with the transverse
term in Eq. (\ref{eq-lept43})%
\begin{eqnarray}
v_{T} &=&\frac{1}{2}\rho +\tan ^{2}\theta _{e}/2  \label{eq-lept52} \\
&=&\frac{1}{2}\rho \left[ 1+\frac{2}{\rho }\tan ^{2}\theta _{e}/2\right] ,
\label{eq-lept53}
\end{eqnarray}%
thereby defining the photon longitudinal polarization%
\begin{equation}
\mathcal{E}\equiv \left[ 1+\frac{2}{\rho }\tan ^{2}\theta _{e}/2\right]
^{-1},  \label{eq-lept54}
\end{equation}%
which implies that%
\begin{equation}
\tan ^{2}\theta _{e}/2=\frac{\rho }{2}\left( \mathcal{E}^{-1}-1\right) .
\label{eq-lept55}
\end{equation}%
If one defines the ratios%
\begin{equation}
u_{X}\equiv \frac{v_{X}}{v_{T}}  \label{eq-lept56}
\end{equation}%
with $X=L,T,TT,TL,T^{\prime }$ and $TL^{\prime }$ and substitutes in the
above equations for $v_{X}$ for the factor $\tan \theta _{e}/2$ one finds
that%
\begin{eqnarray}
u_{L} &=&2\rho \mathcal{E}  \notag \\
u_{T} &=&1  \notag \\
u_{TT} &=&-\mathcal{E}  \notag \\
u_{TL} &=&-\sqrt{\rho }\sqrt{\mathcal{E}\left( 1+\mathcal{E}\right) }
\label{eq-lept57} \\
u_{T^{\prime }} &=&\sqrt{1-\mathcal{E}^{2}}  \notag \\
u_{TL^{\prime }} &=&-\sqrt{\rho }\sqrt{\mathcal{E}\left( 1-\mathcal{E}%
\right) }.  \notag
\end{eqnarray}%
The invariant in Eq. (\ref{eq-g4aa}) in this notation in the ERL$_{e}$ then
becomes%
\begin{eqnarray}
\mathcal{C} &=&v_{0}v_{T}\left[ 2\rho \mathcal{E}W^{L}+W^{T}-\mathcal{E}%
W^{TT}-\sqrt{\rho }\sqrt{\mathcal{E}\left( 1+\mathcal{E}\right) }%
W^{TL}\right.  \notag \\
&&\left. + h \left(\sqrt{1-\mathcal{E}^{2}}W^{T^{\prime }}-\sqrt{\rho }\sqrt{\mathcal{%
E}\left( 1-\mathcal{E}\right) }W^{TL^{\prime }} \right) \right] .  \label{eq-lept58}
\end{eqnarray}

\section{Hadronic Tensors\label{sec-had}}

In this section we proceed to build the most general tensors for
semi-inclusive electron scattering from polarized spin-1/2 targets. A
general frame will be assumed to begin with and later the special choice of
the rest frame will be discussed. The strategy is to write the tensors in
terms of invariant response functions. Accordingly, if one (say) has a model
for the cross section in the rest frame, then the set of invariant response
functions can be deduced and one immediately has the corresponding cross
section in a general frame, for instance, in the collider frame that will
provide a focus for some of our discussions. This way there is no need to
deal with the difficulties of requiring modeling that is covariant,
something that is rarely possible.

We begin by introducing the basic 4-vectors upon which the general hadronic
tensor is built.

\subsection{Basic Hadronic 4-Vectors\label{subsec-basic4}}

We build the hadronic tensors in an arbitrary frame using the basic
4-vectors that characterize semi-inclusive electron scattering from
(possibly) polarized spin-1/2 targets. The strategy in the following is to
expand the general second-rank tensors that are needed in the four sectors
symmetric/unpolarized, anti-symmetric/unpolarized, symmetric/polarized and
anti-symmetric/polarized, and impose the continuity equation in each sector
-- note that, since each sector may be isolated by controlling the electron
and target spins, they may be considered independently. Upon contracting
with $Q^{\mu }$ in each case one may expand in a basis set of four
independent 4-vectors such as those below to determine which contributions
enter and which do not.

\begin{figure}
	\centering
	\includegraphics[height=5cm]{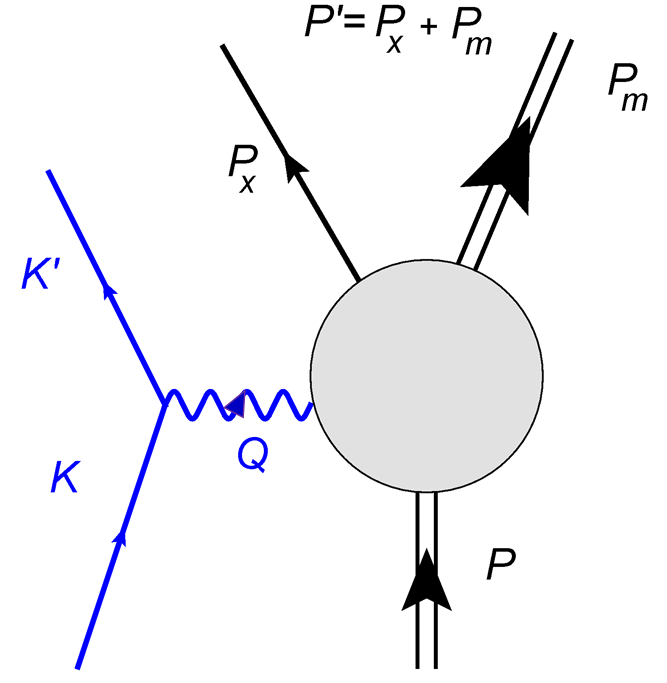} 				
	\caption{Feynman diagram for semi-inclusive electron scattering. The 4-momenta here are discussed in the text. In particular, particle x is assumed to be detected in coincidence with the scattered electron and thus $P_x^\mu$ is assumed to be known. Since the total final-state momentum ${P'}^\mu$ is known (see Fig.~\ref{fig:figure2} for inclusive scattering) this implies that the missing 4-momentum is also known via the relationship $P_m^\mu = {P^{\prime}}^\mu - P_x^\mu$.
	}
	\label{fig:figure3}
\end{figure}

We begin this discussion of the 4-vectors with a specific choice of
coordinate system; see Fig.~\ref{fig:figure1}. Since we want to
retain the usual meaning for the leptonic and hadronic factors discussed in
Sec. \ref{sec-contract}, it is important to employ this system for the
developments to follow. In this system we have the following 4-vectors:%
\begin{eqnarray}
Q^{\mu } &=&\left( \omega ,\mathbf{q}\right)  \label{eq-f1} \\
P^{\mu } &=&\left( E_{p},\mathbf{p}\right)  \label{eq-f2} \\
P_{x}^{\mu } &=&\left( E_{x},\mathbf{p}_{x}\right)  \label{eq-f3} \\
S^{\mu } &=&\left( S^{0},\mathbf{s}\right)  \label{eq-f4}
\end{eqnarray}%
with 3-vectors%
\begin{eqnarray}
\mathbf{q} &=&q\mathbf{u}_{3}  \label{eq-f5} \\
\mathbf{p} &=&p\left( \sin \theta \cos \phi \mathbf{u}_{1}+\sin
\theta \sin \phi \mathbf{u}_{2}+\cos \theta \mathbf{u}_{3}\right)
\label{eq-f6} \\
\mathbf{p}_{x} &=&p_{x}\left( \sin \theta _{x}\cos \phi _{x}\mathbf{u%
}_{1}+\sin \theta _{x}\sin \phi _{x}\mathbf{u}_{2}+\cos \theta _{x}%
\mathbf{u}_{3}\right)  \label{eq-f7} \\
\mathbf{s} &=&s\left( \sin \theta ^{\ast }\cos \phi ^{\ast }\mathbf{u%
}_{1}+\sin \theta ^{\ast }\sin \phi ^{\ast }\mathbf{u}_{2}+\cos \theta
^{\ast }\mathbf{u}_{3}\right) ,  \label{eq-f8}
\end{eqnarray}%
where $\mathbf{u}_{1}$, $\mathbf{u}_{2}$ and $\mathbf{u}_{3}$ are unit vectors (see Fig.~ \ref{fig:figure1}) defined such that $\mathbf{u}_{3}$ is along the direction of the 3-momentum transfer, the lepton scattering plane is the 13-plane and $\mathbf{u}_{2}$ is normal to that plane.
The target (mass $M$) and particle detected in coincidence with the
scattered electron (mass $M_{x}$) are both on-shell and thus%
\begin{eqnarray}
E_{p} &=&\sqrt{p^{2}+M^{2}}  \label{eq-f8a} \\
E_{x} &=&\sqrt{p_{x}^{2}+M_{x}^{2}}.  \label{eq-f8b}
\end{eqnarray}

The target spin 4-vector may be developed further by exploiting the two
conditions it must satisfy, namely%
\begin{equation}
P\cdot S=0  \label{eq-ff1}
\end{equation}%
and%
\begin{equation}
S^{2}=\left( S^{0}\right) ^{2}-s^{2}=-1,  \label{eq-ff2}
\end{equation}%
which may be verified by going to the target rest frame. If we define%
\begin{equation}
\bm{\beta }_{p}\equiv \mathbf{p}/E_{p}  \label{eq-ff3}
\end{equation}%
so that%
\begin{equation}
\gamma _{p}=\frac{1}{\sqrt{1-\beta _{p}^{2}}}=E_{p}/M  \label{eq-ff4}
\end{equation}%
and let $\chi $ be the angle between $\mathbf{p}$ and $\mathbf{s}$,
then Eq. (\ref{eq-ff1}) implies that%
\begin{equation}
S^{0}=\bm{\beta }_{p}\cdot \mathbf{s=}\beta _{p}s\cos \chi ,
\label{eq-ff5}
\end{equation}%
where%
\begin{equation}
\cos \chi =\cos \theta \cos \theta ^{\ast }+\sin \theta \sin \theta ^{\ast
}\cos \left( \phi -\phi ^{\ast }\right) .  \label{eq-ff6}
\end{equation}%
Equation (\ref{eq-ff2}) implies that%
\begin{equation}
s^{2}\left( 1-\beta _{p}^{2}\cos ^{2}\chi \right) =1  \label{eq-ff7}
\end{equation}%
which yields%
\begin{eqnarray}
s &\equiv &\left\vert \mathbf{s}\right\vert =\frac{1}{\sqrt{1-\beta
_{p}^{2}\cos ^{2}\chi }}  \label{eq-ff7a} \\
\mathbf{s} &\equiv &h^{\ast }s\left( \sin \theta ^{\ast }\cos \phi
^{\ast }\mathbf{u}_{1}+\sin \theta ^{\ast }\sin \phi ^{\ast }\mathbf{%
u}_{2}+\cos \theta ^{\ast }\mathbf{u}_{3}\right)  \label{eq-ff7a1} \\
S^{0} &=&h^{\ast }\beta _{p}s\cos \chi  \label{eq-ff7a2}
\end{eqnarray}%
where we have now introduced $h^{\ast }=\pm $, namely, a convenient factor
that allows the target spin to be flipped while keeping the axis of
quantization for the spin fixed. Accordingly, the target spin 4-vector may be
written%
\begin{equation}
S^{\mu }=\frac{h^{\ast }}{\sqrt{1-\beta _{p}^{2}\cos ^{2}\chi }}\left( \beta
_{p}\cos \chi ,\sin \theta ^{\ast }\cos \phi ^{\ast },\sin \theta ^{\ast
}\sin \phi ^{\ast },\cos \theta ^{\ast }\right) ,  \label{eq-ff8}
\end{equation}

Thus we see that as building blocks we may employ the 4-momentum transfer $%
Q^{\mu }$, the target 4-momentum $P^{\mu }$, the 4-momentum of some particle
detected in the final state $P_{x}^{\mu }$, and the 4-vector that
characterizes the target spin, $S^{\mu }$. As usual, it is convenient to
replace the last three with projected 4-vectors, \textit{i.e.,} vectors that
are by construction orthogonal to $Q^{\mu }$. When the spin is not involved,
namely the target is unpolarized (see below for the polarized case), we use
the following as basic polar vectors
\begin{align}
& Q^{\mu }  \label{eq-had1} \\
& U^{\mu }\equiv \frac{1}{M}\left( P^{\mu }-\left( \frac{Q\cdot P}{Q^{2}}%
\right) Q^{\mu }\right)  \label{eq-had2} \\
& V^{\mu }\equiv \frac{1}{M}\left( P_{x}^{\mu }-\left( \frac{Q\cdot P_{x}}{%
Q^{2}}\right) Q^{\mu }\right) ,  \label{eq-had4}
\end{align}%
where the factors $M$ in Eqs. (\ref{eq-had2}) and (\ref{eq-had4}) are
included for convenience to make the 4-vectors dimensionless (see below) and where, by construction, one has%
\begin{equation}
Q\cdot U=Q\cdot V=0  \label{eq-had5}
\end{equation}%
and%
\begin{eqnarray}
U^{2} &=&1-\frac{\left( Q\cdot P\right) ^{2}}{M^{2}Q^{2}}  \label{eq-had5x1}
\\
V^{2} &=&\frac{1}{M^{2}}\left( M_{x}^{2}-\frac{\left( Q\cdot P_{x}\right)
^{2}}{Q^{2}}\right)  \label{eq-had5x2} \\
U\cdot V &=&\frac{1}{M^{2}}\left( P\cdot P_{x}-\frac{\left( Q\cdot P\right)
\left( Q\cdot P_{x}\right) }{Q^{2}}\right) .  \label{eq-had5x3}
\end{eqnarray}%
Note that we have chosen to use the target mass $M$ above and not the mass
of particle $x$, namely $M_{x}$, since we want to allow the latter to be
general enough to include the photon. Furthermore, we can replace $V^{\mu }$
with a 4-vector that is orthogonal not only to $Q^{\mu }$ but to $U^{\mu }$
as well:%
\begin{equation}
X^{\mu }\equiv V^{\mu }-\left( \frac{U\cdot V}{U^{2}}\right) U^{\mu },
\label{eq-had5x4}
\end{equation}%
where then%
\begin{equation}
Q\cdot U=Q\cdot X=U\cdot X=0  \label{eq-had5x5}
\end{equation}%
and%
\begin{equation}
X^{2}=V^{2}-\frac{\left( U\cdot V\right) ^{2}}{U^{2}}.  \label{eq-had5x6}
\end{equation}%
We can also define a fourth 4-vector via%
\begin{equation}
D^{\mu }\equiv \frac{1}{M}\epsilon ^{\mu \alpha \beta \gamma }Q_{\alpha
}U_{\beta }X_{\gamma }=\frac{1}{M^{3\frac{{}}{{}}}}\epsilon ^{\mu \alpha
\beta \gamma }Q_{\alpha }P_{\beta }P_{x\gamma }  \label{eq-had5a}
\end{equation}%
which is dual to the above set, behaves as an axial-vector and satisfies%
\begin{equation}
Q\cdot D=U\cdot D=X\cdot D=0.  \label{eq-had5b}
\end{equation}%
This yields a set of four 4-vectors that can be used to span 4-dimensional
space.

We can define the following invariants%
\begin{eqnarray}
I_{1} &\equiv &I_p \equiv \frac{Q\cdot P}{Q^{2}} =\left( \omega E_p - qp\cos{\theta} \right)/Q^2 \label{eq-f9} \\
I_{2} &\equiv &I_x \equiv \frac{Q\cdot P_{x}}{Q^{2}} =\left( \omega E_x - qp_x\cos{\theta_x} \right)/Q^2 \label{eq-f10} \\
I_{3} &\equiv &I_{pp_x} \equiv \frac{P\cdot P_{x}}{Q^{2}} \notag\\
& & =\left( E_p E_x - pp_x (\sin \theta \sin \theta_x \cos(\phi - \phi_x) +\cos{\theta}\cos{\theta_x} \right)/Q^2, \label{eq-f14a}
\end{eqnarray}%
where as usual $-Q^2=4kk'\sin^2 \theta_e/2$ in the ERL$_e$ (see Sec. \ref{sec-lept}). We note here that for
the Lorentz scalars upon which the invariant response functions discussed
above depend we have the following: two of them are fixed, namely, $%
P^{2}=M^{2}$and $P_{x}^{2}=M_{x}^{2}$, one can be chosen to be $Q^{2}$, and three
can be chosen to be those given in Eqs. (\ref{eq-f9}--\ref{eq-f14a}),
for a total of four dynamical scalars, namely, ($Q^{2}$, $I_{1}$, $I_{2}$
and $I_{3}$). 
Eqs. (\ref{eq-had2}--\ref{eq-had4}) then yield the projected 4-vectors employed
above:%
\begin{eqnarray}
U^{\mu } &\equiv &\frac{1}{M}\left( P^{\mu }-I_{1}Q^{\mu }\right)
\label{eq-f12} \\
V^{\mu } &\equiv &\frac{1}{M}\left( P_{x}^{\mu }-I_{2}Q^{\mu }\right).
\label{eq-f13} 
\end{eqnarray}%
We also have that
\begin{eqnarray}
U^2 &= &1-\frac{Q^2 I_1^2}{M^2}  \label{eq-fff1} \\
U\cdot V &= &\frac{Q^2}{M^2} \left( I_3 -I_1 I_2 \right) \label{eq-fff2}
\end{eqnarray}%
and therefore that
\begin{equation}
I_4 \equiv \frac{U\cdot V}{U^2} = \frac{Q^2 \left( I_3 -I_1 I_2 \right)}{M^2 - Q^2 I_!^2}, \label{eq-fff3}
\end{equation}
which yields explicit expressions for the basis 4-vectors:
\begin{eqnarray}
U^0 &= &\frac{1}{M} \left( E_p - I_1 \omega \right) \label{eq-fff4} \\
U^i &= &\frac{1}{M} \left( \mathbf{p} -I_1 \mathbf{q} \right)^i \label{eq-fff5} \\
X^0 &= &\frac{1}{M} \left(  E_x - I_4 E_p - \left[ I_2 - I_1 I_4 \right] \omega  \right) \label{eq-fff6} \\
X^i &= &\frac{1}{M} \left( \mathbf{p}_x - I_4 \mathbf{p} -\left[ I_2 - I_1 I_4 \right]  \mathbf{q}   \right)^i. \label{eq-fff7} 
\end{eqnarray}

The four Lorentz scalars above can be replaced by (frame dependent) variables
that are traditionally employed in specific sub-fields. For instance, in
high-energy physics one often uses $\left( Q^{2},\nu \right) $ where $\nu
\equiv \omega $ or $\left( Q^{2},x\right) $ for the first two scalars, $x$
being defined by $x\equiv 1/\left( 2I_{1}\right) $. In nuclear physics where
the energy and 3-momentum transfer are more natural one typically uses $%
\left( q,\omega \right) $ instead. Furthermore, in nuclear physics it is
generally better for the remaining two dynamical variables to use the
missing energy $E_{m}$ and missing momentum $p_{m}$ when studying
semi-inclusive electroweak reactions (see later) \cite{VanOrden:2019krz}.

When the target spin is involved we can also define another Lorentz invariant
\begin{equation}
I_{s} \equiv \frac{Q\cdot S}{Q^{2}}  \label{eq-f11}
\end{equation}%
and the corresponding projected 4-vector
\begin{equation}
\Sigma ^{\mu } \equiv S^{\mu }-I_{s}Q^{\mu }  \label{eq-f14}
\end {equation}
where $Q\cdot \Sigma = 0$ and for completeness we note that
\begin{equation}
I_{s}  =\left( \omega S^0 - qs\cos{\theta^{\ast}} \right)/Q^2. \label{eq-fs}
\end{equation}
Note that the spin 4-vector does not enter as a dynamical
Lorentz scalar since it occurs as part of the projection operator%
\begin{equation}
\mathcal{P}_{spin}\equiv \frac{1}{2}\left( 1+\gamma _{5}\gamma _{\mu }S^{\mu
}\right)  \label{eq-f14x}
\end{equation}%
and either does not enter (unpolarized) or occurs explicitly (polarized)
where, being part of the projection operator, it only enters linearly. 

We can also define two 4-vectors that contain the spin 4-vector linearly and
are dual to specific combinations of the others, namely,
\begin{align}
\overline{X}^{\mu }& \equiv \frac{1}{M}\epsilon ^{\mu \alpha \beta \gamma
}S_{\alpha }Q_{\beta }U_{\gamma }=\frac{1}{M^{2}}\epsilon ^{\mu \alpha
\beta \gamma }S_{\alpha }Q_{\beta }P_{\gamma }  \label{eq-had9} \\
\overline{U}^{\mu }& \equiv \frac{1}{M}\epsilon ^{\mu \alpha \beta \gamma
}S_{\alpha }Q_{\beta }X_{\gamma }=\frac{1}{M}\epsilon ^{\mu \alpha \beta
\gamma }S_{\alpha }Q_{\beta }V_{\gamma }-\left( \frac{U\cdot V}{U^{2}}%
\right) \overline{X}^{\mu }.  \label{eq-had7}
\end{align}%
One has that%
\begin{eqnarray}
Q\cdot \overline{U} &=&X\cdot \overline{U}=\Sigma \cdot \overline{U}=0
\label{eq-had10} \\
Q\cdot \overline{X} &=&U\cdot \overline{X}=\Sigma \cdot \overline{X}=0
\label{eq-had10b}
\end{eqnarray}%
and additionally that%
\begin{eqnarray}
 I_{0} \equiv U\cdot \overline{U} &=&-X\cdot \overline{X}  \label{eq-had11a} \\
&=&\frac{1}{M}\epsilon ^{\alpha \beta \gamma \delta }\Sigma _{\alpha
}Q_{\beta }U_{\gamma }X_{\delta }  \label{eq-had11b} \\
&=&\frac{1}{M^{3}}\epsilon ^{\alpha \beta \gamma \delta }S_{\alpha
}Q_{\beta }P_{\gamma }P_{x\delta },  \label{eq-had11c}
\end{eqnarray}%
namely, a dimensionless invariant. Note that a tensor of the form%
\begin{equation}
\overline{Q}^{\mu }\equiv \epsilon ^{\mu \alpha \beta \gamma }\Sigma
_{\alpha }U_{\beta }X_{\gamma }  \label{eq-had12b}
\end{equation}%
is redundant, since it can be shown that%
\begin{equation}
\overline{Q}^{\mu }=-\frac{MI_{0}}{Q^{2}}Q^{\mu }  \label{eq-had12c}
\end{equation}%
where $Q^{\mu }$ will be used instead as a building block.

Next, let us consider the 4-vector $\overline{X}^{\mu }$ defined in Eq. (\ref%
{eq-had9}). In contracting with $\epsilon ^{\mu \alpha \beta \gamma }$ the
contributions in $\Sigma _{\alpha }$ and $U_{\gamma }$ containing $Q_{\alpha
}$ and $Q_{\gamma }$, respectively, may be ignored due to the explicit
factor $Q_{\beta }$, and hence we can write%
\begin{eqnarray}
\overline{X}^{\mu } &=&\frac{1}{M^{2}}\epsilon ^{\mu \alpha \beta \gamma
}S_{\alpha }Q_{\beta }P_{\gamma }  \label{eq-f42a} \\
&=&-\frac{1}{M^{2}}\left[ \omega \epsilon ^{\mu 0\alpha \gamma }-q\epsilon
^{\mu 3\alpha \gamma }\right] S_{\alpha }P_{\gamma },  \label{eq-f42b}
\end{eqnarray}%
the latter expression in the 123-system. Upon developing this expression one can show that
\begin{eqnarray}
\overline{X}^{i} &=&\frac{1}{M^{2}}\left( \left[ \left( \omega \mathbf{p}%
-E_p \mathbf{q}\right) \times \mathbf{s}\right] +S^{0}\left( 
\mathbf{q}\times \mathbf{p}\right) \right) ^{i}\;,i=1,2,3
\label{eq-ff29} \\
\overline{X}^{0} &=&\frac{1}{\nu^{\prime} }\overline{X}^{3}.  \label{eq-ff30}
\end{eqnarray}
And finally we have the 4-vector $\overline{U}^{\mu }$ defined in Eq. (\ref%
{eq-had7})%
\begin{eqnarray}
\overline{U}^{\mu } &=&\frac{1}{M}\epsilon ^{\mu \alpha \beta \gamma
}S_{\alpha }Q_{\beta }X_{\gamma }  \label{eq-fg1} \\
&=&\frac{1}{M}\epsilon ^{\mu \alpha \beta \gamma }S_{\alpha }Q_{\beta
}V_{\gamma }-\left( \frac{U\cdot V}{U^{2}}\right) \overline{X}^{\mu }
\label{eq=-fg2} \\
&=&\overline{T}^{\mu }-\left( \frac{U\cdot V}{U^{2}}\right) \overline{X}%
^{\mu },  \label{eq-fg3}
\end{eqnarray}%
where%
\begin{eqnarray}
\overline{T}^{\mu } &\equiv &\frac{1}{M^{2}}\epsilon ^{\mu \alpha \beta
\gamma }S_{\alpha }Q_{\beta }P_{x\gamma }  \label{eq-fg4} \\
&=&-\frac{1}{M^{2}}\left[ \omega \epsilon ^{\mu 0\alpha \gamma }-q\epsilon
^{\mu 3\alpha \gamma }\right] S_{\alpha }P_{x\gamma }.  \label{eq-fg5}
\end{eqnarray}%
As above we can develop this expression to find that
\begin{eqnarray}
\overline{T}^{i} &=&\frac{1}{M^{2}}\left( \left[ \left( \omega \mathbf{p}%
_{x}-E_{x}\mathbf{q}\right) \times \mathbf{s}\right] +S^{0}\left( 
\mathbf{q}\times \mathbf{p}_{x}\right) \right) ^{i},\;i=1,2,3
\label{eq-fg12} \\
\overline{T}^{3} &=&\nu^{\prime} \overline{T}^{0}.  \label{eq-fg15}
\end{eqnarray}

This completes the specification of the basis 4-vectors that will be used in the next section to obtain the most general form for the hadronic tensor.

\subsection{Hadronic Tensors in a General Reference Frame\label%
{subsec-GenTen}}

\subsubsection{Second-Rank Tensors: Symmetric, Unpolarized\label%
{subsubsec-symunpol}}

Given the above 4-vector building blocks, we now proceed to construct
second-rank hadronic tensors with the appropriate symmetries. We begin with
the symmetric cases where no target polarization is involved.%
\begin{align}
W_{1,s}^{\mu \nu }& \equiv g^{\mu \nu }-\frac{Q^{\mu }Q^{\nu }}{Q^{2}}
\label{eq-had14a} \\
W_{2,s}^{\mu \nu }& \equiv U^{\mu }U^{\nu }  \label{eq-had16} \\
W_{3,s}^{\mu \nu }& \equiv X^{\mu }X^{\nu }  \label{eq-had18} \\
W_{4,s}^{\mu \nu }& \equiv U^{\mu }X^{\nu }+X^{\mu }U^{\nu }.
\label{eq-had19}
\end{align}%
Here the motivation for including the factors $M$ becomes clear: all four of
the tensors above are dimensionless. Note that no contributions of the form $%
Q^{\mu }U^{\nu }+U^{\mu }Q^{\nu }$ or $Q^{\mu }X^{\nu }+X^{\mu }Q^{\nu }$
are used, since upon contraction with the lepton tensor these would yield
zero, and that $D^{\mu }$ does not enter in similar forms since the results
would correspond to second-rank tensors that are of vector/axial-vector or
axial/axial character rather than vector/vector as required and we have no
pseudoscalars to use as multiplicative factors to produce this behavior in
the unpolarized situation where the spin does not enter. We have the
following upon contracting with $Q_{\mu }$:%
\begin{equation}
Q_{\mu }W_{m,s}^{\mu \nu }=0  \label{eq-had19a}
\end{equation}%
for $m=1,2,3,4$. The general tensor of this type is obtained by summing over
the 4 contributions, where each is multiplied by a Lorentz scalar, invariant
response function, $A_{m}$, that in turn depends only on the Lorentz
invariants in the problem, namely%
\begin{equation}
\left( W_{s}^{\mu \nu }\right) _{unpol}=\sum_{m=1}^{4}A_{m}W_{m,s}^{\mu \nu }
\label{eq-had47}
\end{equation}%
and from Eq. (\ref{eq-had19a}) one has%
\begin{equation}
Q_{\mu }\left( W_{s}^{\mu \nu }\right) _{unpol}=0  \label{eq-had48}
\end{equation}%
as required for the overall symmetric, unpolarized tensor by the continuity
equation. Thus the symmetric, unpolarized second-rank hadronic tensor may
then be written%
\begin{eqnarray}
\left( W_{s}^{\mu \nu }\right) _{unpol} &=&-W_{1}\left( g^{\mu \nu }-\frac{%
Q^{\mu }Q^{\nu }}{Q^{2}}\right) +W_{2}U^{\mu }U^{\nu }  \notag \\
&&+W_{3}X^{\mu }X^{\nu }+W_{4}\left( U^{\mu }X^{\nu }+X^{\mu }U^{\nu
}\right) ,  \label{eq-had51}
\end{eqnarray}%
namely, with \emph{four} contributions involving invariant functions $W_{m}$%
, $m=1,2,3,4$ (here we have shifted from using invariant functions $A_{m}$
to more familiar notation, including the minus sign in the $W_{1}$ case,
which is conventional and useful as will become apparent when inclusive
scattering is discussed later). Since the tensors defined above are all
dimensionless, the invariant functions here all have the same dimensions.

\subsubsection{Second-Rank Tensors: Anti-symmetric, Unpolarized\label%
{subsubsec-antiunpol}}

We have only one anti-symmetric contribution that uses $Q^{\mu }$, $U^{\mu }$
and $X^{\nu }$ as a basis, namely

\begin{equation}
W_{1,a}^{\mu \nu }\equiv i(U^{\mu }X^{\nu }-X^{\mu }U^{\nu }),
\label{eq-had31}
\end{equation}%
where here and below the factor $i$ has been included following Eq. (\ref%
{eq-hadd6c}); as above this tensor is dimensionless. Note that again no
contributions such as $Q^{\mu }U^{\nu }-U^{\mu }Q^{\nu }$ or $Q^{\mu }X^{\nu
}-X^{\mu }Q^{\nu }$ are included as these yield zero upon contraction with
the lepton tensor, and that contributions of this form using $D^{\mu }$ are
invalid since they do not behave as vector/vector, and that contributions
such as $\epsilon ^{\mu \nu \alpha \beta }Q_{\alpha }U_{\beta }$, $\epsilon
^{\mu \nu \alpha \beta }Q_{\alpha }V_{\beta }$ and $\epsilon ^{\mu \nu
\alpha \beta }U_{\alpha }V_{\beta }$ are invalid for the same reason. Also
contributions involving the Levi-Civita symbol with $D^{\mu }$ and one of $%
\left( Q^{\mu },U^{\mu },X^{\mu }\right) $ can be shown to be redundant; in
fact one has the following identities:%
\begin{eqnarray}
i\epsilon ^{\mu \nu \alpha \beta }Q_{\alpha }D_{\beta } &=&\frac{1}{M}%
Q^{2}W_{1,a}^{\mu \nu }  \label{eq-had32a} \\
i\epsilon ^{\mu \nu \alpha \beta }U_{\alpha }D_{\beta } &=&-\frac{i}{M}%
U^{2}\left( Q^{\mu }X^{\nu }-X^{\mu }Q^{\nu }\right)  \label{eq-had32b} \\
i\epsilon ^{\mu \nu \alpha \beta }X_{\alpha }D_{\beta } &=&\frac{i}{M}%
X^{2}\left( Q^{\mu }U^{\nu }-U^{\mu }Q^{\nu }\right) .  \label{eq-had32c}
\end{eqnarray}%
Contracting the valid anti-symmetric tensor with $Q_{\mu }$ yields zero and
we find that the anti-symmetric, unpolarized tensor is constructed from the 
\emph{single} basis tensor of the correct type with an invariant functions
here called $W_{5}$: 
\begin{equation}
\left( W_{a}^{\mu \nu }\right) _{unpol}=iW_{5}\left( U^{\mu }X^{\nu }-X^{\mu
}U^{\nu }\right) ,  \label{eq-had55}
\end{equation}%
namely the so-called 5th response (see \cite{twdprogpartnuc} and references therein). We note in passing that in that same reference the general problem of reactions of the type $A(\overrightarrow{e},e'x_1 x_2 \ldots)$ having the target unpolarized, but having any number of particles $x_1$, $x_2$, {\it etc.,} detected in coincidence with the scattered electron was developed.

\subsubsection{Second-Rank Tensors: Symmetric, Polarized\label%
{subsubsec-sympol}}

Let us begin the symmetric polarized developments by starting with a set of
symmetric second-rank tensors that starts with the set of four symmetric
tensors obtained in the unpolarized case, $W_{m,s}^{\mu \nu }$, with $%
m=1\cdots 4$ as in Eqs. (\ref{eq-had14a}--\ref{eq-had19}), multiplied by $%
I_{0}$, namely%
\begin{eqnarray}
W_{1,s}^{\prime \mu \nu } &\equiv &\left( g^{\mu \nu }-\frac{Q^{\mu }Q^{\nu }%
}{Q^{2}}\right) I_{0}  \notag \\
W_{2,s}^{\prime \mu \nu } &\equiv &\left( U^{\mu }U^{\nu }\right) I_{0}
\label{eq-had12cc} \\
W_{3,s}^{\prime \mu \nu } &\equiv &\left( X^{\mu }X^{\nu }\right) I_{0} 
\notag \\
W_{4,s}^{\prime \mu \nu } &\equiv &\left( U^{\mu }X^{\nu }+X^{\mu }U^{\nu
}\right) I_{0}.  \notag
\end{eqnarray}%
Here and below the prime is included to denote the fact that the target spin
is involved. These all have the desired properties, namely, they behave as
vector/vector and are linear in the spin; they are all dimensionless.
Contractions with $Q^{\mu }$ yield zero as above. To these we can add
another set built from $\bar{U}^{\mu }$ and $\overline{X}^{\mu }$ together
with the 4-vectors $Q^{\mu }$, $U^{\mu }$ and $X^{\mu }$.

For the remaining building blocks constructed from tensors containing the
spin we use%
\begin{align}
W_{5,s}^{\prime \mu \nu }& \equiv U^{\mu }\bar{U}^{\nu }+U^{\nu }\bar{U}%
^{\mu }  \label{eq-hhad5} \\
W_{6,s}^{\prime \mu \nu }& \equiv U^{\mu }\overline{X}^{\nu }+U^{\nu }%
\overline{X}^{\mu }  \label{eq-hhad6} \\
W_{7,s}^{\prime \mu \nu }& \equiv X^{\mu }\bar{U}^{\nu }+X^{\nu }\bar{U}%
^{\mu }  \label{eq-hhad8} \\
W_{8,s}^{\prime \mu \nu }& \equiv X^{\mu }\overline{X}^{\nu }+X^{\nu }%
\overline{X}^{\mu },  \label{eq-hhad9}
\end{align}%
again with no contributions that are proportional to $Q^{\mu }$ or $Q^{\nu }$
as these would yield zero when contracted with the electron tensor. Again
these behave as vector/vector and are linear in the spin and all yield zero
when contracted with $Q^{\mu }$. Accordingly, if we expand the symmetric
polarized tensor in this set of basis tensors,%
\begin{equation}
\left( W_{s}^{\mu \nu }\right) _{pol}=\sum_{m=1}^{8}A_{m}^{\prime
}W_{m,s}^{\prime \mu \nu }  \label{eq-hhadz1}
\end{equation}%
with general invariant response functions $A_{m}^{\prime }$, and impose the
continuity equation constraint $Q_{\mu }\left( W_{s}^{\mu \nu }\right)
_{pol}=0$ we obtain the following: 
\begin{eqnarray}
\left( W_{s}^{\mu \nu }\right) _{pol} &=&\left[ -W_{1}^{\prime }\left(
g^{\mu \nu }-\frac{Q^{\mu }Q^{\nu }}{Q^{2}}\right) +W_{2}^{\prime }U^{\mu
}U^{\nu }\right.  \notag \\
&&+\left. W_{3}^{\prime }X^{\mu }X^{\nu }+W_{4}^{\prime }\left( U^{\mu
}X^{\nu }+X^{\mu }U^{\nu }\right) \right] I_{0}  \notag \\
&&+W_{5}^{\prime }\left( U^{\mu }\bar{U}^{\nu }+U^{\nu }\bar{U}^{\mu
}\right) +W_{6}^{\prime }\left( U^{\mu }\overline{X}^{\nu }+U^{\nu }%
\overline{X}^{\mu }\right)  \notag \\
&&+W_{7}^{\prime }\left( X^{\mu }\bar{U}^{\nu }+X^{\nu }\bar{U}^{\mu
}\right) +W_{8}^{\prime }\left( X^{\mu }\overline{X}^{\nu }+X^{\nu }%
\overline{X}^{\mu }\right) ,  \label{eq-hhadz1x}
\end{eqnarray}%
again shifting from generic invariant functions $A_{m}^{\prime }$ to the
more conventional notation involving invariant $W_{m}^{\prime }$. Thus, for
the symmetric, polarized case we are left with \emph{eight} contributions.
All tensors here are dimensionless and consequently all invariant functions
have the same dimensions.

\subsubsection{Second-Rank Tensors: Anti-symmetric, polarized\label%
{subsubsec-antipol}}

In this sector we begin with a basis tensor that involves the Levi-Civita
symbol and is linear in spin:%
\begin{equation}
W_{1,a}^{\prime \mu \nu }\equiv \frac{i}{M}\epsilon ^{\mu \nu \alpha
\beta }\Sigma _{\alpha }Q_{\beta }.  \label{eq-had47x0}
\end{equation}%
Note that one has the following identities,%
\begin{eqnarray}
Q^{2}\epsilon ^{\mu \nu \alpha \beta }\Sigma _{\alpha }X_{\beta }
&=&M\left( Q^{\mu }\bar{U}^{\nu }-Q^{\nu }\bar{U}^{\mu }\right)
\label{eq-had47x1} \\
Q^{2}\epsilon ^{\mu \nu \alpha \beta }\Sigma _{\alpha }U_{\beta }
&=&M\left( Q^{\mu }\overline{X}^{\nu }-Q^{\nu }\overline{X}^{\mu }\right)
\label{eq-had46x2}
\end{eqnarray}%
and hence no terms having the Levi-Civita symbol as here are needed, since
they also yield zero upon contraction with the electron tensor. Since we
want tensors that are linear in spin and of vector/vector form we can also
have the following dimensionless tensors: 
\begin{align}
W_{2,a}^{\prime \mu \nu }& \equiv i(U^{\mu }\bar{U}^{\nu }-U^{\nu }\bar{U}%
^{\mu })  \label{eq-had42} \\
W_{3,a}^{\prime \mu \nu }& \equiv i(U^{\mu }\overline{X}^{\nu }-U^{\nu }%
\overline{X}^{\mu })  \label{eq-had44} \\
W_{4,a}^{\prime \mu \nu }& \equiv i(X^{\mu }\bar{U}^{\nu }-X^{\nu }\bar{U}%
^{\mu })  \label{eq-had45} \\
W_{5,a}^{\prime \mu \nu }& \equiv i(X^{\mu }\overline{X}^{\nu }-X^{\nu }%
\overline{X}^{\mu })  \label{eq-had46}
\end{align}%
with no terms of the form $Q^{\mu }\bar{U}^{\nu }-Q^{\nu }\bar{U}^{\mu }$ or 
$Q^{\mu }\overline{X}^{\nu }-Q^{\nu }\overline{X}^{\mu }$, since, as above,
these yield zero when contracted with the lepton tensor. Finally, as in the
symmetric case we can use the dimensionless, anti-symmetric contribution
above (Eq. (\ref{eq-had31})) multiplied by the invariant $I_{0}$:%
\begin{equation}
W_{6,a}^{\prime \mu \nu }\equiv i(U^{\mu }X^{\nu }-X^{\mu }U^{\nu })I_{0}.
\label{eq-hadyy3}
\end{equation}%
however, one can prove the following identity%
\begin{eqnarray}
-I_{0}\left( U^{\mu }X^{\nu }-U^{\nu }X^{\mu }\right) &=&\frac{1}{M}%
U^{2}X^{2}\epsilon ^{\mu \nu \alpha \beta }\Sigma _{\alpha }Q_{\beta
}+X^{2}\left( U^{\mu }\overline{X}^{\nu }-U^{\nu }\overline{X}^{\mu }\right)
\notag \\
&&+U^{2}\left( X^{\mu }\overline{U}^{\nu }-X^{\nu }\overline{U}^{\mu }\right)
\label{eq-hadyy4}
\end{eqnarray}%
and hence the tensor $W_{6,a}^{\prime \mu \nu }$ is redundant. The remaining
five tensors all yield zero when contracted with $Q^{\mu }$. Accordingly we
have the following \emph{five} independent contributions:%
\begin{eqnarray}
\left( W_{a}^{\mu \nu }\right) _{pol} &=&i\left[ \frac{1}{M}W_{9}^{\prime
}\epsilon ^{\mu \nu \alpha \beta }\Sigma _{\alpha }Q_{\beta }\right. 
\notag \\
&&+W_{10}^{\prime }(U^{\mu }\bar{U}^{\nu }-U^{\nu }\bar{U}^{\mu
})+W_{11}^{\prime }(U^{\mu }\overline{X}^{\nu }-U^{\nu }\overline{X}^{\mu })
\notag \\
&&+\left. W_{12}^{\prime }(X^{\mu }\bar{U}^{\nu }-X^{\nu }\bar{U}^{\mu
})+W_{13}^{\prime }(X^{\mu }\overline{X}^{\nu }-X^{\nu }\overline{X}^{\mu })
\right] \label{eq-had46x7} 
\end{eqnarray}%
As above, we have shifted notation to make this sector coherent with the
previous ones; all tensors are dimensionless, implying that
the invariant functions all have the same dimensions. As an alternative it is also possible to expand the contraction of the leptonic and hadronic tensors in terms of Lorentz scalars rather than employing the 4-vectors as we have here. The resulting form is documented in  \ref{sec-Lscalars}.

Let us end this section with a brief discussion of how the use of time-reversal invariance allows one to separate the four types of contributions into two classes. The basic requirement for the time-reversal operator is to relate a given matrix element to one that describes the process running in the opposite direction, that is to a matrix element where the incoming state now contains all of the particles from the original final state and the final state contains the particles contains the particles from the original initial state. If the original matrix element has a final state with two or more interacting particles this requires that the boundary condition for this state be changed from the incoming boundary condition to the outgoing boundary condition.

The effects of time-reversal on the hadronic tensor have been studied in great detail in the context of multipole expansions for arbitrary target spin (see, for instance, \cite{Raskin:1988kc} and references therein). The result is that the matrix elements must fall into two classes: one where the transition multipole moment is real and another where it is imaginary. These two classes result in response functions that are either even or odd under time-reversal, TRE or TRO, respectively. Note that time-reversal invariance is assumed throughout this work; being TRE or TRO does not imply violation of this symmetry.

For the case of a spin-1/2 particle in the initial or final state the effects of time-reversal can be greatly simplified by the simultaneous application of both time-reversal and parity \cite{Picklesimer:1986wj}. This is particularly useful in the case where the hadronic tensor is written as a linear combination of invariant functions of inner products of the available 4-momenta and second-rank tensors constructed from these four-momenta and the spin vector, such as we have done above.
For the purpose of this discussion let
\begin{align}
	W^{\mu \nu }(Q,P,P_{x},P_{m},S,(-))=&\left<P,S\right|{J^\mu}^\dag(Q)\left|P_{x},P_{m},S,(-)\right> \nonumber\\& \quad \times\left<P_{x},P_{m},S,(-)\right|J^\nu(Q)(-)\left|P,S\right> ,
\end{align}
where $(-)$ denotes the incoming boundary conditions for the final scattering state. 
This trivially implies that
\begin{equation}
	W^{*\mu \nu }(Q,P,P_{x},P_{m},S,(-))=	W^{\nu \mu }(Q,P,P_{x},P_{m},S,(-)) .
\end{equation}
Equations (\ref{eq-had51},\ref{eq-had55},\ref{eq-hhadz1x},\ref{eq-had46x7}) are constructed such that $W_i,\ i=1,\dots,5$ and  $W'_i,\ i=1,\dots,13$ are real.

The components of the hadronic tensor in Eqs. (\ref{eq-had51},\ref{eq-had55},\ref{eq-hhadz1x},\ref{eq-had46x7}) are parameterized in terms of Lorentz 4-vectors.  The result of combining time-reversal and parity causes no change to the momentum 4-vectors while causing the spin 4-vector to change sign.  Most importantly, time-reversal causes a change in the boundary condition of the scattering state from incoming ($(-)$) to outgoing ($(+)$). This gives
\begin{align}
	W^{\mu \nu }(Q,P,P_{x},P_{m},S,(-))\xrightarrow{\mathcal{TP}}&W^{\nu \mu }(Q,P,P_{x},P_{m},-S,(+))\nonumber\\
	=& -W^{*\mu \nu }(Q,P,P_{x},P_{m},S,(+))\,.
\end{align}

Since $Q^\mu$, $U^\mu$ and $X^\mu$ depend only on the momentum 4-vectors one has
\begin{align}
	&Q^\mu\xrightarrow{\mathcal{TP}}Q^\mu\nonumber\\
	&U^\mu\xrightarrow{\mathcal{TP}}U^\mu\nonumber\\
	&X^\mu\xrightarrow{\mathcal{TP}}X^\mu\,.
\end{align}
The vectors $\Sigma^\mu $,  $\overline{X}^{\mu }$ and $\overline{U}^{\mu }$ are linear in $S^\mu$ and thus
\begin{align}
	&\Sigma^{\mu }\xrightarrow{\mathcal{TP}}-\Sigma^{\mu }\nonumber\\
	&\overline{X}^{\mu }\xrightarrow{\mathcal{TP}}-\overline{X}^{\mu }\nonumber\\
	&\overline{U}^{\mu }\xrightarrow{\mathcal{TP}}-\overline{U}^{\mu } .
\end{align}
The scalar $I_0$ is also linear in $S^\mu$ and accordingly
\begin{equation}
	I_0\xrightarrow{\mathcal{TP}}-I_0\,.
\end{equation}
The invariant functions $W_i$ and $W'_i$ are real and the complex conjugation changes the sign of all factors of $i$.

Applying these rules to Eqs. (\ref{eq-had51},\ref{eq-had55},\ref{eq-hhadz1x},\ref{eq-had46x7}) yields
\begin{align}
	&W_i(-)\xrightarrow{\mathcal{TP}} W_i(+), \qquad i=1,\dots,4\label{eq:TP_1}\\
	&W_5(-)\xrightarrow{\mathcal{TP}} -W_5(+)\label{eq:TP_2}\\
	&W'_i(-)\xrightarrow{\mathcal{TP}} -W'_i(+), \qquad i=1,\dots,8\label{eq:TP_3}\\
	&W'_i(-)\xrightarrow{\mathcal{TP}} W'_i(+), \qquad i=9,\dots,13\,.\label{eq:TP_4}
\end{align}

Under conditions where the boundary condition has no effect, such as the plane-wave impulse approximation,
factorization approximations or where the final state is obtained through a single resonance at the energy where only the real part contributes, the invariant functions in Eqs. (\ref{eq:TP_2}) and (\ref{eq:TP_3}) must be zero. In such a special case this reduces the number of invariant functions from 18 to 9 with a similar reduction in the number of reponse functions. Generally speaking, however, all 18 play a role. This is the same as would be obtained by applying the multipole analysis with time-reversal only (see \cite{Raskin:1988kc} and references therein).

\begin{table}
	\begin{center}
\begin{tabular}{|c|c|c|c|}
	\hline
	&  & Number & Time-Reversal  \\ \hline
	Unpolarized & Symmetric & 4 & Even \\ \hline
	& Anti-symmetric & 1 & Odd \\ \hline
	Polarized & Symmetric & 8 & Odd  \\ \hline
	& Anti-symmetric & 5 & Even \\ \hline
\end{tabular}\end{center}
\caption{This table shows the number of invariant functions falling into the four sectors according to polarization and symmetry indicating the time-reversal properties of each sector. }\label{tab:1}\end{table}
\vspace{0.3in}

In summary we have 18 invariant response functions falling into the 
four sectors categorized in Table \ref{tab:1}, with the symmetric contributions entering when the incident
electrons are unpolarized and the anti-symmetric contributions when they are
polarized, in fact, longitudinally polarized when in the ERL$_{e}$. The
sectors are otherwise specified by whether or not the spin-1/2 target is
unpolarized or polarized.

\subsection{Specific Components of the General Hadronic Tensors\label%
{subsec-comp}}

We next proceed to write explicit forms for the hadronic tensors defined
above. We begin with the \textbf{symmetric, unpolarized} case given in Eq. (%
\ref{eq-had51}) which immediately yields the following for the minimal set
of components:%
\begin{eqnarray}
\left( W_{s}^{00}\right) _{unpol} &=&-\frac{1}{\rho }W_{1}+\left(
U^{0}\right) ^{2}W_{2}+\left( X^{0}\right) ^{2}W_{3}+\left(
2U^{0}X^{0}\right) W_{4}  \label{eq-cont9} \\
\left( W_{s}^{01}\right) _{unpol} &=&\left( U^{0}U^{1}\right) W_{2}+\left(
X^{0}X^{1}\right) W_{3}+\left( U^{0}X^{1}+X^{0}U^{1}\right) W_{4}
\label{eq-cont10} \\
\left( W_{s}^{11}\right) _{unpol} &=&W_{1}+\left( U^{1}\right)
^{2}W_{2}+\left( X^{1}\right) ^{2}W_{3}+\left( 2U^{1}X^{1}\right) W_{4}
\label{eq-cont11} \\
\left( W_{s}^{22}\right) _{unpol} &=&W_{1}+\left( U^{2}\right)
^{2}W_{2}+\left( X^{2}\right) ^{2}W_{3}+\left( 2U^{2}X^{2}\right) W_{4}
\label{eq-cont12} \\
\left( W_{s}^{02}\right) _{unpol} &=&\left( U^{0}U^{2}\right) W_{2}+\left(
X^{0}X^{2}\right) W_{3}+\left( U^{0}X^{2}+X^{0}U^{2}\right) W_{4}
\label{eq-cont13} \\
\left( W_{s}^{12}\right) _{unpol} &=&\left( U^{1}U^{2}\right) W_{2}+\left(
X^{1}X^{2}\right) W_{3}+\left( U^{1}X^{2}+X^{1}U^{2}\right) W_{4}.
\label{eq-cont14}
\end{eqnarray}%
Note that, since the symmetric leptonic tensor in Eqs. (\ref{eq-lept38}-\ref{eq-lept41}) has
no $\mu \nu =02$ or 12 components, the last two hadronic contributions (Eqs.
(\ref{eq-cont13}-\ref{eq-cont14})) do not enter when the tensors are
contracted, leaving a total of four terms, as expected for the situation
where only the incident electrons may be polarized and the ERL$_{e}$ is
invoked \cite{Donnelly:1985ry}. Following the nomenclature in \cite{Donnelly:1985ry} we have%
\begin{eqnarray}
W_{unpol}^{L} &\equiv &\left( W_{s}^{00}\right) _{unpol}=-\frac{1}{\rho }%
W_{1}+\left( U^{0}\right) ^{2}W_{2}+\left( X^{0}\right) ^{2}W_{3}  \notag \\
&&+2U^{0}X^{0}W_{4}  \label{eq-cont15} \\
W_{unpol}^{T} &\equiv &\left( W_{s}^{22+11}\right) _{unpol}=2W_{1}+\left[
\left( U^{1}\right) ^{2}+\left( U^{2}\right) ^{2}\right] W_{2}  \notag \\
&&+\left[ \left( X^{1}\right) ^{2}+\left( X^{2}\right) ^{2}\right] W_{3}+2%
\left[ U^{1}X^{1}+U^{2}X^{2}\right] W_{4}  \label{eq-cont16} \\
W_{unpol}^{TT} &\equiv &\left( W_{s}^{22-11}\right) _{unpol}=\left[ -\left(
U^{1}\right) ^{2}+\left( U^{2}\right) ^{2}\right] W_{2}  \notag \\
&&+\left[ -\left( X^{1}\right) ^{2}+\left( X^{2}\right) ^{2}\right] W_{3}+2%
\left[ -U^{1}X^{1}+U^{2}X^{2}\right] W_{4}  \label{eq-cont17} \\
W_{unpol}^{TL} &\equiv &2\sqrt{2}\left( W_{s}^{01}\right) _{unpol}=2\sqrt{2}%
\left[ U^{0}U^{1}W_{2}+X^{0}X^{1}W_{3}\right.  \notag \\
&&\left. +\left( U^{0}X^{1}+X^{0}U^{1}\right) W_{4}\right] .
\label{eq-cont18a}
\end{eqnarray}

Next, for the \textbf{anti-symmetric, unpolarized} case we have the
following from Eq. (\ref{eq-had55}):%
\begin{eqnarray}
\left( W_{a}^{02}\right) _{unpol} &=&iW_{5}\left(
U^{0}X^{2}-X^{0}U^{2}\right)  \label{eq-cont29a} \\
\left( W_{a}^{12}\right) _{unpol} &=&iW_{5}\left(
U^{1}X^{2}-X^{1}U^{2}\right) ,  \label{eq-cont30}
\end{eqnarray}%
yielding%
\begin{eqnarray}
W_{unpol}^{TL^{\prime }} &\equiv &2\sqrt{2}\left( iW_{a}^{02}\right)
_{unpol}=-2\sqrt{2}W_{5}\left( U^{0}X^{2}-X^{0}U^{2}\right)
\label{eq-cont31a} \\
W_{unpol}^{T^{\prime }} &\equiv &2\left( iW_{a}^{12}\right)
_{unpol}=-2W_{5}\left( U^{1}X^{2}-X^{1}U^{2}\right) .  \label{eq-cont32}
\end{eqnarray}%
These can all contribute in a situation where the incident electron is
polarized. However, note the following: if mass terms in the electron tensor
are retained (even in the PWBA) then one finds that the $TL^{\prime }$ and $%
T^{\prime }$ contributions are of leading order whereas the \underline{$TL$}$%
^{\prime }$ contributions go as $1/\gamma _{e}$ or $1/\gamma _{e}^{\prime }$
where $\gamma _{e}=\epsilon /m_{e}$ and $\gamma _{e}^{\prime }=\epsilon
^{\prime }/m_{e}$ and hence may usually be neglected at high energies,
leaving only the $TL^{\prime }$ and $T^{\prime }$ contributions. 

Next we consider the contributions that arise from contractions of the
symmetric leptonic tensor with the symmetric hadronic tensor for the case
where the target is polarized -- the \textbf{symmetric, polarized} case.
From the developments in the last section we find that the following
contributions enter in this sector:%
\begin{eqnarray}
W_{pol}^{L} &\equiv &\left( W_{s}^{00}\right) _{pol}  \notag \\
&=&\left\{ -W_{1}^{\prime }/\rho +\left( U^{0}\right) ^{2}W_{2}^{\prime
}+\left( X^{0}\right) ^{2}W_{3}^{\prime }+2U^{0}X^{0}W_{4}^{\prime }\right\}
I_{0}  \notag \\
&&+2\left\{ U^{0}\bar{U}^{0}W_{5}^{\prime }+U^{0}\overline{X}%
^{0}W_{6}^{\prime }+X^{0}\bar{U}^{0}W_{7}^{\prime }+X^{0}\overline{X}%
^{0}W_{8}^{\prime }\right\}  \label{eq-cont25aa}
\end{eqnarray}%
\begin{eqnarray}
W_{pol}^{T} &\equiv &\left( W_{s}^{22}+W_{s}^{11}\right) _{pol}=\left\{
2W_{1}^{\prime }+\left( \left( U^{2}\right) ^{2}+\left( U^{1}\right)
^{2}\right) W_{2}^{\prime }\right.  \notag \\
&&\left. +\left( \left( X^{2}\right) ^{2}+\left( X^{1}\right) ^{2}\right)
W_{3}^{\prime }+2\left( U^{2}X^{2}+U^{1}X^{1}\right) W_{4}^{\prime }\right\}
I_{0}  \notag \\
&&+2\left\{ \left( U^{2}\bar{U}^{2}+U^{1}\bar{U}^{1}\right) W_{5}^{\prime
}+\left( U^{2}\overline{X}^{2}+U^{1}\overline{X}^{1}\right) W_{6}^{\prime
}\right.  \label{eq-cont25c} \\
&&\left. +\left( X^{2}\bar{U}^{2}+X^{1}\bar{U}^{1}\right) W_{7}^{\prime
}+\left( X^{2}\overline{X}^{2}+X^{1}\overline{X}^{1}\right) W_{8}^{\prime
}\right\}  \notag
\end{eqnarray}%
\begin{eqnarray}
W_{pol}^{TT} &\equiv &\left( W_{s}^{22}-W_{s}^{11}\right) _{pol}=\left\{
\left( \left( U^{2}\right) ^{2}-\left( U^{1}\right) ^{2}\right)
W_{2}^{\prime }\right.  \notag \\
&&\left. +\left( \left( X^{2}\right) ^{2}-\left( X^{1}\right) ^{2}\right)
W_{3}^{\prime }+2\left( U^{2}X^{2}-U^{1}X^{1}\right) W_{4}^{\prime }\right\}
I_{0}  \label{eq-cont25d} \\
&&+2\left\{ \left( U^{2}\bar{U}^{2}-U^{1}\bar{U}^{1}\right) W_{5}^{\prime
}+\left( U^{2}\overline{X}^{2}-U^{1}\overline{X}^{1}\right) W_{6}^{\prime
}\right.  \notag \\
&&\left. +\left( X^{2}\bar{U}^{2}-X^{1}\bar{U}^{1}\right) W_{7}^{\prime
}+\left( X^{2}\overline{X}^{2}-X^{1}\overline{X}^{1}\right) W_{8}^{\prime
}\right\}  \notag
\end{eqnarray}%
\begin{eqnarray}
W_{pol}^{TL} &\equiv &2\sqrt{2}\left( W_{s}^{01}\right) _{pol}  \notag \\
&=&2\sqrt{2}\left[ \left\{ U^{0}U^{1}W_{2}^{\prime }+X^{0}X^{1}W_{3}^{\prime
}+\left( U^{0}X^{1}+U^{1}X^{0}\right) W_{4}^{\prime }\right\} I_{0}\right. 
\notag \\
&&+\left( U^{0}\bar{U}^{1}+U^{1}\bar{U}^{0}\right) W_{5}^{\prime }+\left(
U^{0}\overline{X}^{1}+U^{1}\overline{X}^{0}\right) W_{6}^{\prime }
\label{eq-cont25b} \\
&&+\left. \left( X^{0}\bar{U}^{1}+X^{1}\bar{U}^{0}\right) W_{7}^{\prime
}+\left( X^{0}\overline{X}^{1}+X^{1}\overline{X}^{0}\right) W_{8}^{\prime }%
\right]  \notag
\end{eqnarray}%
following conventional notation.

Finally, we need to develop the \textbf{anti-symmetric, polarized} case.
From Eq. (\ref{eq-had46x7}) we have that%
\begin{eqnarray}
\left( W_{a}^{02}\right) _{pol} &=&i\left[ \frac{1}{M}W_{9}^{\prime
}\epsilon ^{02\alpha \beta }\Sigma _{\alpha }Q_{\beta }\right.  \notag \\
&&+W_{10}^{\prime }(U^{0}\bar{U}^{2}-U^{2}\bar{U}^{0})+W_{11}^{\prime }(U^{0}%
\overline{X}^{2}-U^{2}\overline{X}^{0})  \notag \\
&&+\left. W_{12}^{\prime }(X^{0}\bar{U}^{2}-X^{2}\bar{U}^{0})+W_{13}^{\prime
}(X^{0}\overline{X}^{2}-X^{2}\overline{X}^{0})\right]  \label{eq-cont34b} \\
\left( W_{a}^{12}\right) _{pol} &=&i\left[ \frac{1}{M}W_{9}^{\prime
}\epsilon ^{12\alpha \beta }\Sigma _{\alpha }Q_{\beta }\right.  \notag \\
&&+W_{10}^{\prime }(U^{1}\bar{U}^{2}-U^{2}\bar{U}^{1})+W_{11}^{\prime }(U^{1}%
\overline{X}^{2}-U^{2}\overline{X}^{1})  \notag \\
&&+\left. W_{12}^{\prime }(X^{1}\bar{U}^{2}-X^{2}\bar{U}^{1})+W_{13}^{\prime
}(X^{1}\overline{X}^{2}-X^{2}\overline{X}^{1})\right] ,  \label{eq-cont34c}
\end{eqnarray}%
where no cases with components $\mu \nu =03$, 13 or 23 are needed, since
they can be eliminated using the continuity equation. These yield three
possible responses, of which we only further develop the two below, as ${\underline{TL'}}$
is typically suppressed; see the discussion after Eq. (\ref{eq-cont32}).%
\begin{eqnarray}
W_{pol}^{T^{\prime }} &\equiv &2\left( iW_{a}^{12}\right) _{pol}
\label{eq-cont35c} \\
&=&-2\left[ \frac{1}{M}W_{9}^{\prime }\epsilon ^{12\alpha \beta }\Sigma
_{\alpha }Q_{\beta }\right.  \notag \\
&&+W_{10}^{\prime }(U^{1}\bar{U}^{2}-U^{2}\bar{U}^{1})+W_{11}^{\prime }(U^{1}%
\overline{X}^{2}-U^{2}\overline{X}^{1})  \notag \\
&&+\left. W_{12}^{\prime }(X^{1}\bar{U}^{2}-X^{2}\bar{U}^{1})+W_{13}^{\prime
}(X^{1}\overline{X}^{2}-X^{2}\overline{X}^{1})\right]  \label{eq-cont35c1} \\
W_{pol}^{TL^{\prime }} &\equiv &2\sqrt{2}\left( iW_{a}^{02}\right) _{pol}
\label{eq-contd} \\
&=&-2\sqrt{2}\left[ \frac{1}{M}W_{9}^{\prime }\epsilon ^{02\alpha \beta
}\Sigma _{\alpha }Q_{\beta }\right.  \notag \\
&&+W_{10}^{\prime }(U^{0}\bar{U}^{2}-U^{2}\bar{U}^{0})+W_{11}^{\prime }(U^{0}%
\overline{X}^{2}-U^{2}\overline{X}^{0})  \notag \\
&&+\left. W_{12}^{\prime }(X^{0}\bar{U}^{2}-X^{2}\bar{U}^{0})+W_{13}^{\prime
}(X^{0}\overline{X}^{2}-X^{2}\overline{X}^{0})\right] .  \label{eq-cont35d1}
\end{eqnarray}

It can be shown that, upon knowing the responses $W^J_{unpol,pol}$ with $J=L,T,TT,TL,T',TL'$ and making use of the fact that the target polarization can be arranged to point in various directions, one can invert to obtain the invariant response functions $W_i$ for $i=1,5$ and $W^{\prime}_i$ for $i=1,13$; see  \ref{sec-invert}.

This completes the general structure of both the leptonic and hadronic
tensors in a general frame where the spin-1/2 target is polarized and moving
with some general 4-momentum $P^{\mu }$. 

\section{Semi-inclusive Cross Section for Electron Scattering from a
Polarized Spin-1/2 Target\label{sec-Semi}}

The full semi-inclusive electron scattering cross section in a general frame
of reference may be written in terms of the Mott cross section, some
kinematic factors that arise from using the Feynman rules \cite{Bjorken:1965sts},
together with a general response function $\mathcal{F}^{semi}$. We begin the
discussion in this section by introducing useful notation for the kinematic
variables involved in semi-inclusive scattering.

\subsection{Kinematics for Semi-inclusive Scattering\label%
{subsec-kinematics}}

As discussed above, we are assuming that the initial state has two particles
of masses $m_e$ and $M$ with 4-momenta $K^{\mu }=\left( \epsilon ,\mathbf{k%
}\right) $ and $P^{\mu }=\left( E_p ,\mathbf{p}\right) $, where $\epsilon =%
\sqrt{k^{2}+m_e^{2}}$ and $E_p =\sqrt{p^{2}+M^{2}},$ respectively, which collide,
leaving a particle of mass $m_e$ with 4-momentum $K^{\prime \mu
}=\left( \epsilon^{\prime} ,\mathbf{k}^{\prime }\right) $ where $\epsilon^{\prime}=\sqrt{{k^{\prime}}^{2}+m_e^{2}}$ and producing a final
state with 4-momentum $P^{\prime \mu }=\left( E_{p \prime },\mathbf{p}%
^{\prime }\right) $ and hence invariant mass $W=\sqrt{E^2 _{ p\prime}-p^{\prime
2}}$. In turn, the final state is assumed to be divided into two pieces, one
the specific particle ``x" that is assumed to be detected, having 4-momentum $%
P_{x}^{\mu }=\left( E_{x},\mathbf{p}_{x}\right) $, where $E_{x}=\sqrt{%
p_{x}^{2}+M_{x}^{2}}$, together with the undetected (``missing") parts of the
final state having 4-momentum $P_{m}^{\mu }=\left( E_{m}^{tot},\mathbf{p}%
_{m}\right) $ with missing energy $E_{m}^{tot}$, missing momentum $%
\mathbf{p}_{m}$, and invariant mass $W_{m}=\sqrt{\left(
E_{m}^{tot}\right) ^{2}-p_{m}^{2}}$. Note: for the \emph{total} missing
energy we use $E_{m}^{tot}$, since we reserve the notation $E_{m}$ to denote
a different, but related quantity (see below). See Fig.~\ref{fig:figure3} where conservation of 4-momentum requires that 
\begin{equation*}
Q^{\mu }+P^{\mu }=P^{\prime \mu }=P_{x}^{\mu }+P_{m}^{\mu },
\end{equation*}%
and thus%
\begin{eqnarray}
E_{m}^{tot} &=&E_{p \prime }-E_{x}  \label{eq-semi5x} \\
\mathbf{p}_{m} &=&\mathbf{p}^{\prime }-\mathbf{p}_{x}.
\label{eq-semi6x}
\end{eqnarray}

From above we have that%
\begin{equation}
P_{m}^{\mu }=Q^{\mu }+P^{\mu }-P_{x}^{\mu }  \label{eq-semi8x}
\end{equation}%
and therefore that%
\begin{eqnarray}
E_{m}^{tot} &=&\omega +E_p -E_{x}  \label{eq-semi9x} \\
\mathbf{p}_{m} &=&\mathbf{p}^{\prime }-\mathbf{p}_{x}.
\label{eq-semi10x}
\end{eqnarray}%
Following the procedures adopted in studies of scaling \cite{Day:1990mf} let
us employ as independent kinematic variables the missing momentum $%
\mathbf{p}_{m}$ and, rather than the missing energy $E_{m}$, the
following energy%
\begin{eqnarray}
\mathcal{E}_m(p_{m}) &\equiv &E_{m}^{tot}-\left( E_{m}^{tot}\right) _{T}\geq 0
\label{eq-semi11x} \\
&=&\sqrt{W_{m}^{2}+p_{m}^{2}}-\sqrt{\left( W_{m}^{T}\right) ^{2}+p_{m}^{2}},
\label{eq-semi12x}
\end{eqnarray}%
where the threshold value of the invariant mass of the missing momentum is
denoted $W_{m}^{T}$; examples of this are given later. This quantity has the
merit of taking on the value $\mathcal{E}_m=0$ at threshold. When used in the
context of nuclear physics the missing 3-momentum is typically much smaller
than the invariant masses of either the daughter threshold value (often the
daughter ground-state mass) or any higher-energy daughter state and thus Eq.
(\ref{eq-semi12x}) may be written%
\begin{eqnarray}
\mathcal{E}_m(p_{m}) &=&W_{m}\sqrt{1+\left( \frac{p_{m}}{W_{m}}\right) ^{2}}%
-W_{m}^{T}\sqrt{1+\left( \frac{p_{m}}{W_{m}^{T}}\right) ^{2}}
\label{eq-semi12x1} \\
&=&W_{m}\left( 1+\frac{p_{m}^{2}}{2W_{m}^{2}}+\cdots \right)
-W_{m}^{T}\left( 1+\frac{p_{m}^{2}}{2\left( W_{m}^{T}\right) ^{2}}+\cdots
\right)   \label{eq-semi12x2} \\
&=&\left( W_{m}-W_{m}^{T}\right) \left[ 1-\delta _{m}+\cdots \right] 
\label{eq-semi12x3}
\end{eqnarray}%
where%
\begin{equation}
\delta _{m}\equiv \frac{p_{m}^{2}}{2W_{m}W_{m}^{T}}\ll 1  \label{eq-sem12x4}
\end{equation}%
typically. Often setting $\delta _{m}$ to zero is an excellent
approximation; this correction involves only the difference between the
kinetic energy of recoil when the daughter system is at threshold and when
it is in some excited state. However, it is not necessary ever to make these
approximations and the exact expressions can alway be employed.

In studies of nuclear physics it is common to define a different quantity
(confusingly also called the missing energy) where kinetic energies are
employed, $E_{m}$. Defining the kinetic energies%
\begin{eqnarray}
T &\equiv &E_p -M  \label{eq-semi12x5a} \\
T_{x} &\equiv &E_{x}-M_{x}  \label{eq-semi12x5} \\
T_{m} &\equiv &E_{m}^{tot}-W_{m},  \label{eq-semi12x6}
\end{eqnarray}%
one has%
\begin{eqnarray}
E_{m} &\equiv &\omega -\left( T_{x}+T_{m}\right)   \label{eq-semi12x7} \\
&=&\left( W_{m}-W_{m}^{T}\right) +E_{s}-T  \label{eq-semi12x8} \\
&\simeq &\mathcal{E}_m(p_{m})+E_{s}-T,  \label{eq-semi12x9}
\end{eqnarray}%
where the so-called separation energy%
\begin{equation}
E_{s}\equiv M_{x}+W_{m}^{T}-M\geq 0  \label{eq-semi12x10}
\end{equation}%
has been introduced and the approximation in the third equation above
corresponds to neglecting the correction involving $\delta _{m}$ discussed
above.

\begin{figure}
	\centering
	\includegraphics[height=5cm]{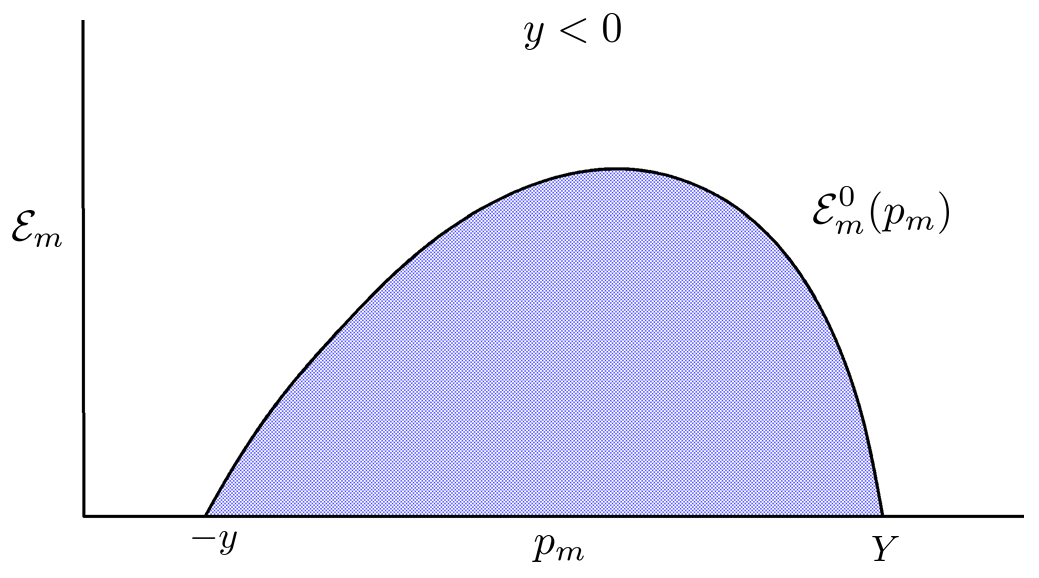} 				
	\caption{Physically allowed region for the situation where $y<0$. The variables employed here are discussed in the text.}
	\label{fig:figure4}
\end{figure}

\begin{figure}
	\centering
	\includegraphics[height=5cm]{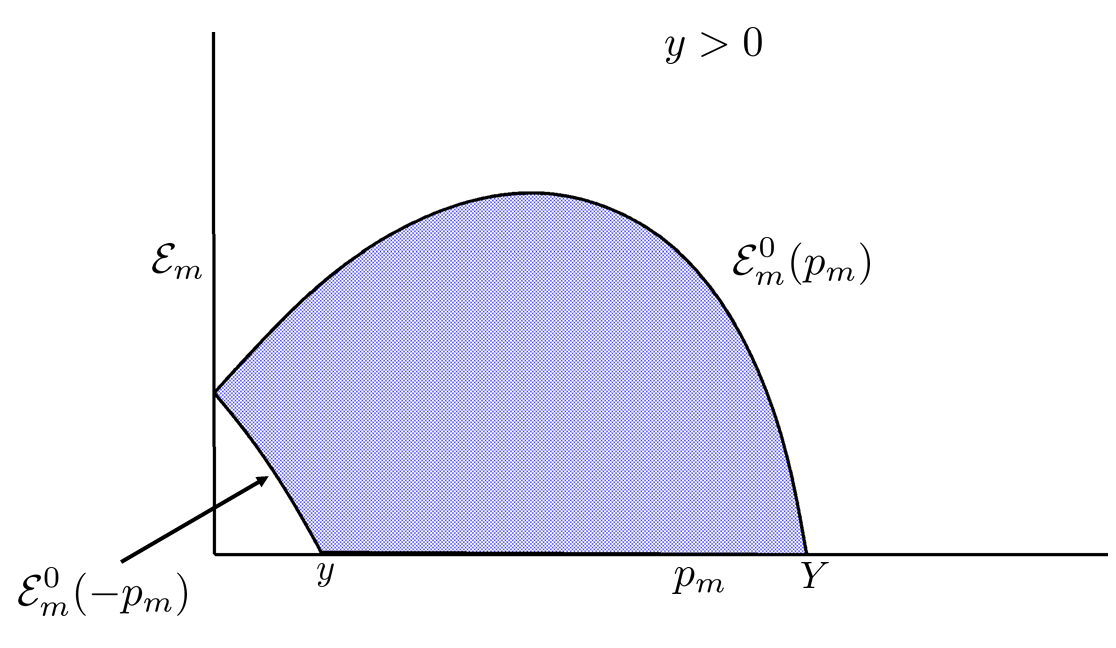} 				
	\caption{Physically allowed region for the situation where $y>0$. The variables employed here are discussed in the text.}
	\label{fig:figure5}
\end{figure}

Using the energy conservation condition in Eq. (\ref{eq-semi9x}) we have%
\begin{equation}
\mathcal{E}_m(p_{m})=\left( E_p +\omega \right) -\left( E_{m}^{tot}\right) _{T}-%
\sqrt{M_{x}^{2}+p^{\prime 2}+p_{m}^{2}-2p_{m}p^{\prime }\cos \theta _{m}},
\label{eq-semi13x}
\end{equation}%
where $\theta _{m}$ is the angle between $\mathbf{p}^{\prime }$ and $%
\mathbf{p}_{m}$ and $p_{m}=|\mathbf{p}_{m}|$. By setting $\mathcal{E}_m
$ to zero and solving the above equation for $p_{m}$ under the limiting
conditions where $\cos \theta _{m}=\pm 1$ it is straightforward to show that
the above equation at $\mathcal{E}_m=0$ has two solutions%
\begin{eqnarray}
p_{m}^{+} &\equiv &Y=\frac{1}{W^{2}}\left[ \left( E_p +\omega \right) \sqrt{%
\Lambda ^{2}-W^{2}\left( W_{m}^{T}\right) ^{2}}+p^{\prime }\Lambda \right] 
\label{eq-semi14x} \\
-p_{m}^{-} &\equiv &y=\frac{1}{W^{2}}\left[ \left( E_p +\omega \right) \sqrt{%
\Lambda ^{2}-W^{2}\left( W_{m}^{T}\right) ^{2}}-p^{\prime }\Lambda \right] ,
\label{eq-semi15x}
\end{eqnarray}%
where, following the notation of \cite{Day:1990mf}  we have introduced the
quantity%
\begin{equation}
\Lambda \equiv \frac{1}{2}\left[ W^{2}+\left( W_{m}^{T}\right) ^{2}-M_{x}^{2}%
\right] .  \label{eq-semi16x}
\end{equation}%
Note that the quantity in the square root may be written%
\begin{equation}
\Lambda ^{2}-W^{2}\left( W_{m}^{T}\right) ^{2}=\frac{1}{4}\left[
W^{2}-\left( W_{m}^{T}+M_{x}\right) ^{2}\right] \left[ W^{2}-\left(
W_{m}^{T}-M_{x}\right) ^{2}\right]   \label{eq-semi17x}
\end{equation}%
and, since the argument of the square root must be non-negative, that%
\begin{equation}
W\geq W^{T}=W_{m}^{T}+M_{x}.  \label{eq-semi18x}
\end{equation}%
Upon setting $y=0$ one finds that%
\begin{equation}
\omega =\omega _{0}\equiv \sqrt{M_{x}^{2}+q^{2}}+W_{m}^{T}-M.
\label{eq-semi19x}
\end{equation}%
Given these relationships it is then straightforward to determine the
physically allowed regions in the $\mathcal{E}_m$-$p_{m}$ plane: for $y\geq 0$
corresponding to $\omega \geq \omega _{0}$ one has 
\begin{equation}
\begin{array}{ll}
\mathcal{E}_m^{0}(-p_{m})\leq \mathcal{E}(p_{m})\leq \mathcal{E}_m^{0}(p_{m}) & 
\mathrm{for\;}0\leq p_{m}\leq y \\ 
0\leq \mathcal{E}(p_{m})\leq \mathcal{E}_m^{0}(p_{m}) & \mathrm{for\;}y\leq
p_{m}\leq Y,%
\end{array}
\label{eq-semi20x}
\end{equation}%
while for $y\leq 0$ corresponding to $\omega \leq \omega _{0}$ one has 
\begin{equation}
\begin{array}{ll}
0\leq \mathcal{E}(p_{m})\leq \mathcal{E}_m^{0}(p_{m}) & \mathrm{for\;-}y\leq
p_{m}\leq Y,%
\end{array}
\label{eq-semi21x}
\end{equation}%
where%
\begin{equation}
\mathcal{E}_m^{0}(p_{m})\equiv \left( E_p +\omega \right) -\left(
E_{m}^{tot}\right) _{T}-\sqrt{M_{x}^{2}+\left( p^{\prime }-p_{m}\right) ^{2}}%
,  \label{eq-semi-22x}
\end{equation}%
namely, the value of $\mathcal{E}_m(p_{m})$ when $\cos \theta_m =+1$. These
regions are shown in Figs.~\ref{fig:figure4} and \ref{fig:figure5}. The region in Fig.~\ref{fig:figure5} is seen to be
bounded from below by the curve $\mathcal{E}_m^{0}(-p_{m})$ which occurs when $%
\theta _{m}=\pi $ and above by the curve $\mathcal{E}_m^{0}(p_{m})$ which
occurs when $\theta _{m}=0$ for $0\leq p_{m}\leq y$, while the other regions
are all bounded by zero from below and by the curve $\mathcal{E}_m^{0}(p_{m})$
from above. When $\mathcal{E}_m(p_{m})=0$ one has from Eq. (\ref{eq-semi13x})
that
\begin{equation}
\cos \theta _{m}=\frac{1}{2p_{m}p^{\prime }}\left\{ M_{x}^{2}+p^{\prime
2}+p_{m}^{2}-\left[ \left( E_p +\omega \right) -\left( E_{m}^{tot}\right) _{T}%
\right] ^{2}\right\} ,  \label{eq-semi23x}
\end{equation}%
which determines $\theta_{m} $ for this boundary.

Thus we have the allowed regions of kinematics in the $\mathcal{E}_m$-$p_{m}$
plane for given values of $q$ and $\omega $ or, equivalently, of $Q^{2}$ and 
$\omega = \nu$ or $q$ and $y$,
where $y=y\left( q,\omega \right) $ given above is often used to replace $%
\omega $ in scaling analyses \cite{Day:1990mf}. In turn these impose limits on the allowed
values of the energy, 3-momentum and polar angle for the detected particle $x
$: first, taking the scalar and cross product of $\mathbf{p}^{\prime }$
with $\mathbf{p}_{x}=\mathbf{p}^{\prime }-\mathbf{p}_{m}$ yields%
\begin{eqnarray}
p_{x}\cos \theta _{x} &=&p^{\prime }-p_{m}\cos \theta _{m}
\label{eq-semi24x} \\
p_{x}\sin \theta _{x} &=&p_{m}\sin \theta _{m}  \label{eq-semi25x}
\end{eqnarray}%
and thus 
\begin{eqnarray}
E_{x} &=&\left( E_p +\omega \right) -\left( \left( E_{m}^{tot}\right) _{T}+%
\mathcal{E}_m(p_{m})\right)   \label{eq-semi27x} \\
p_{x} &=&\sqrt{p^{\prime 2}+p_{m}^{2}-2p_{m}p^{\prime }\cos \theta _{m}}
\label{eq-semi28x} \\
\tan \theta _{x} &=&\frac{p_{m}\sin \theta _{m}}{p^{\prime }-p_{m}\cos
\theta _{m}}.  \label{eq-semi29x}
\end{eqnarray}%
By evaluating these expressions on the above boundaries one can then
determine the physically allowed regions for $P_{x}^{\mu }$. Let us denote
the allowed region for the variables $p_{x}$ (and hence $E_{x}$) and the
polar angle $\theta _{x}$ by $\Gamma _{x}$. The above equations define the
kinematic boundaries within which all values of $(p_{x},\theta _{x})$ are
allowed and outside of which no physically allowed values exist. Later we
discuss the roles played by the azimuthal angle $\phi _{x}$ where all values 
$(0,2\pi )$ are allowed.

These results may be specialized from the general frame to the rest frame
where $p=0$ and thus $T=0$ by making the following replacements: the energy $%
E_p$ and the 3-momentum $\mathbf{p}^{\prime }$ are replaced by $M$ and $%
\mathbf{q}$, respectively, and $\theta _{m}$ becomes the angle between $%
\mathbf{q}$ and $\mathbf{p}_{m}$; $W$ and $\Lambda $ are Lorentz
invariants and so do not change. The results one then obtains are the ones
that are familiar from analyses of scaling \cite{Day:1990mf}.

That said, it should be noted that all of these developments are also valid for studies of particle physics at high energies.

\subsection{Semi-inclusive Cross Section\label{subsec-semicross}}

Having established the allowed regions for the kinematics in semi-inclusive
reactions we may now proceed to a discussion of the cross section. The Feynman rules followed in this work are those of \cite{Bjorken:1965sts}: we provide details in  \ref{sec-6fold} of how the general expression for the six-fold semi-inclusive cross section is obtained. That general answer may be re-written in the following form to connect with the above development of the leptonic and hadronic tensors
\begin{equation}
\left[ \frac{d^{6}\sigma }{d\Omega dk^{\prime }dp_{x}d\cos \theta _{x}d\phi
_{x}}\right] _{x}=\frac{1}{2\pi }\sigma _{\mathrm{Mott}}f\frac{M}{E_{p}}%
\frac{p_{x}^{2}}{E_{x}}\left[ \mathcal{F}^{semi}\right] _{x}  \label{eq-si1}
\end{equation}%
where 
\begin{equation}
\frac{\alpha ^{2}v_{0}k^{\prime }}{Q^{4}k}=\sigma _{\mathrm{Mott}}=\left( 
\frac{\alpha \cos \theta _{e}/2}{2\epsilon \sin ^{2}\theta _{e}/2}\right)
^{2}  \label{eq-semi11}
\end{equation}%
is the Mott cross section and $\left[ \mathcal{F}^{semi}\right] _{x}$ is the invariant called $%
\mathcal{C}=\chi _{\mu \nu }W^{\mu \nu }$ divided by the factor $v_{0}$,
namely%
\begin{eqnarray}
\left[ \mathcal{F}^{semi}\right] _{x} &=&\chi _{\mu \nu }W_{x}^{\mu \nu
}/v_{0}  \label{eq-si2} \\
&=&v_{L}\left[ W_{x}^{L}\right] ^{semi}+v_{T}\left[ W_{x}^{T}\right]
^{semi}+\cdots   \label{eq-si3}
\end{eqnarray}%
as discussed below and where the subscript \textquotedblleft
x\textquotedblright\ has been added to remind us that this forms the
semi-inclusive cross section where particle x is assumed to be detected. The
factor $M/E_{p}$ arises from applying the Feynman rules in a general frame
where the target is moving; this factor becomes unity in the target rest
frame. Furthermore, the factor \cite{moller,goldbergerwatson}\
\begin{equation}
f=\left[ \left( \bm{\beta }_{e}-\bm{\beta }_{p}\right) ^{2}-\left( 
\bm{\beta }_{e}\times \bm{\beta }_{p}\right) ^{2}\right] ^{-1/2},
\label{eq-si4}
\end{equation}%
with $\bm{\beta }_{e}=\bm{k}/\epsilon $ and $\bm{\beta }_{p}=%
\mathbf{p}/E_{p}$ as usual, accounts for the flux of the (in general
colliding) beams. In the rest frame one has $\beta _{p}=0$ and thus $%
f^{R}=1/\beta _{e}$ which equals unity in the ERL$_{e}$. 

In Eq. (\ref{eq-si1}) a specific choice has been made for the normalization.
In particular, while any constants or Lorentz invariants could be absorbed
into the definitions of the invariant functions we choose to fix the
conventions so that upon integrating the semi-inclusive cross section over
the detected particle's 3-momentum and summing over all open channels, 
\textit{i.e.,} all particles x while taking care not to double-count, one
should recover the inclusive cross section with its conventional
normalization. That is, to obtain the contribution of the channel having
particle x to the inclusive cross section one should perform the integral
over $p_{x}$, $\cos \theta _{x}$ and $\phi _{x}$ over the allowed physical
region for the semi-inclusive reaction $(e,e^{\prime }x)$ (see above for
detailed discussion concerning the allowed region) 
\begin{eqnarray}
\left[ \frac{d^{2}\sigma }{d\Omega dk^{\prime }}\right] _{x} &=&\left\{ \int
dp_{x}\int d\cos \theta _{x}\int_{0}^{2\pi }d\phi _{x}\left[ \frac{%
d^{6}\sigma }{d\Omega dk^{\prime }dp_{x}d\cos \theta _{x}d\phi _{x}}\right]
_{x}\right\} _{\mathrm{allowed}}  \label{eq-si5} \\
&=&\frac{1}{2\pi }\sigma _{\mathrm{Mott}}f\frac{M}{E_{p}}\left\{ \int dp_{x}%
\frac{p_{x}^{2}}{E_{x}}\int d\cos \theta _{x}\left[ \mathcal{G}^{semi}\right]
_{x}\right\} _{\mathrm{allowed}},  \label{eq-si6}
\end{eqnarray}%
where%
\begin{equation}
\left[ \mathcal{G}^{semi}\right] _{x}\equiv \int_{0}^{2\pi }d\phi _{x}\left[ 
\mathcal{F}^{semi}\right] _{x}.  \label{eq-si7}
\end{equation}%
Then the full inclusive cross section is obtained by summing over all open
channels, taking care not to double-count (see below for examples):%
\begin{equation}
\frac{d^{2}\sigma }{d\Omega dk^{\prime }}=\widehat{\sum_{x}}\left[ \frac{%
d^{2}\sigma }{d\Omega dk^{\prime }}\right] _{x},  \label{eq-si8}
\end{equation}%
where the requirement not to double-count is indicated by the hat over the
summation. In the next section the full inclusive cross section is also
written in the form%
\begin{equation}
\frac{d^{2}\sigma }{d\Omega dk^{\prime }}=\sigma _{\mathrm{Mott}}f\frac{M}{%
E_{p}}\mathcal{R}^{incl},  \label{eq-si9}
\end{equation}%
where%
\begin{equation}
\mathcal{R}^{incl}=R_{1}^{incl}+\cdots   \label{eq-si10}
\end{equation}%
and%
\begin{equation}
R_{1}^{incl}=\left[ v_{L}R_{unpol}^{L}\right] ^{incl}+\cdots 
\label{eq-si11}
\end{equation}%
Clearly the integral over $\phi _{x}$ for contributions that have no
explicit $\phi _{x}$-dependence simply accounts for the factor $2\pi $ put
in the denominator above.

One may now change variables in the following ways. Since from Eq.~(\ref{eq-semi10x}) $\mathbf{p}_{m}=\mathbf{p}^\prime-\mathbf{p}_{x}$ and we are
keeping $\mathbf{q}$ and $\mathbf{p}$ constant and hence also $\mathbf{p}^\prime = \mathbf{p}+\mathbf{q}$ constant, one has%
\begin{equation}
p_{x}^{2}dp_{x}d\cos \theta _{x}=p_{m}^{2}dp_{m}d\cos \theta _{m}
\label{eq-si12}
\end{equation}%
and thus the semi-inclusive cross section may be written as differential in
the missing-momentum plus changing $p_{x}^{2}$ to $p_{m}^{2}$.\textbf{\ }%
Since we also have from Eq.~(\ref{eq-semi13x}) that%
\begin{equation}
\mathcal{E}_m(p_{m})=\left( E_p+\omega \right) -\left( E_{m}^{tot}\right) _{T}-%
\sqrt{M_{x}^{2}+{p^\prime}^{2}+p_{m}^{2}-2p_{m}p^\prime\cos \theta _{m}} ,  \label{eq-si13}
\end{equation}%
we can change variables from $\cos \theta _{m}$
to $\mathcal{E}$:%
\begin{equation}
\left[ \frac{\partial \mathcal{E}_m}{\partial \cos \theta _{m}}\right]
_{p_{m}}=\frac{p_{m}p^\prime}{E_{x}}  \label{eq-si14}
\end{equation}%
and so%
\begin{equation}
\left[ \frac{d^{6}\sigma }{d\Omega dk^{\prime }dp_{m}d\mathcal{E}_md\phi _{x}}%
\right] _{x}=\frac{1}{2\pi p^\prime}\sigma _{\mathrm{Mott}}f\frac{M}{E_{p}}p_{m}%
\left[ \mathcal{F}^{semi}\right] _{x}.  \label{eq-si15}
\end{equation}%
To form the inclusive cross section one may then proceed to integrate over $%
p_{m}$, $\mathcal{E}_m$ and $\phi _{x}$ (which is unchanged from the previous
treatment), where now the physical region defining the boundaries in the $%
\left( p_{m},\mathcal{E}_m\right) $-plane is that discussed above.

The above has been developed in a general frame; if one wishes to have the results in the target rest frame all that is necessary is to set $p$ to zero, in which case $\mathbf{p}^\prime \rightarrow \mathbf{q}$, $\theta _{m}$ becomes the angle between $\mathbf{q}$ and $\mathbf{p}_m$ and $E_p \rightarrow M$.

As discussed in detail above where the invariant response functions have
been developed, the overall response can be decomposed into the four sectors
that are classified by the types of polarization they involve%
\begin{equation}
\mathcal{F}^{semi}=\mathcal{F}_{1}^{semi}+h\mathcal{F}_{2}^{semi}+h^{\ast }%
\mathcal{F}_{3}^{semi}+hh^{\ast }\mathcal{F}_{4}^{semi}.  \label{eq-semi15}
\end{equation}%
In the semi-inclusive case, as we have seen earlier, the responses here
depend on four scalar invariants, $Q^{2}$, $I_{1,2,3}$, together with the kinematic variables that enter through the lepton tensor. Clearly again the four sectors can be
separated by flipping the electron helicity $h$ and the direction of the
target spin via the factor $h^{\ast }$. Explicitly we have%
\begin{eqnarray}
\mathcal{F}_{1}^{semi} &=&v_{L}\left[ W_{unpol}^{L}\right] ^{semi}+v_{T}%
\left[ W_{unpol}^{T}\right] ^{semi}  \notag \\
&&+v_{TT}\left[ W_{unpol}^{TT}\right] ^{semi}+v_{TL}\left[ W_{unpol}^{TL}%
\right] ^{semi}  \label{eq-semi16} \\
h\mathcal{F}_{2}^{semi} &=&v_{T^{\prime }}\left[ W_{unpol}^{T^{\prime }}%
\right] ^{semi}+v_{TL^{\prime }}\left[ W_{unpol}^{TL^{\prime }}\right]
^{semi}  \label{eq-semi17} \\
h^{\ast }\mathcal{F}_{3}^{semi} &=&v_{L}\left[ W_{pol}^{L}\right]
^{semi}+v_{T}\left[ W_{pol}^{T}\right] ^{semi}  \notag \\
&&+v_{TT}\left[ W_{pol}^{TT}\right] ^{semi}+v_{TL}\left[ W_{pol}^{TL}\right]
^{semi}  \label{eq-semi18} \\
hh^{\ast }\mathcal{F}_{4}^{semi} &=&v_{T^{\prime }}\left[ W_{pol}^{T^{\prime
}}\right] ^{semi}+v_{TL^{\prime }}\left[ W_{pol}^{TL^{\prime }}\right]
^{semi}.  \label{eq-semi19}
\end{eqnarray}%
Here the responses $\left[ W_{unpol}^{K}\right] ^{semi}$ and $\left[
W_{pol}^{K}\right] ^{semi}$ with $K=L$, $T$, $TL$, $TT$, $T^{\prime }$ and $%
TL^{\prime }$ are the semi-inclusive quantities developed earlier, now with
the label $semi$ appended to distinguish them from the inclusive responses
discussed above. As we found earlier, $\mathcal{F}_{1,4}^{semi}$ are TRE
while $\mathcal{F}_{2,3}^{semi}$ are TRO. In turn, the individual responses
are built from the 18 invariant response functions $W_{m}$, $m=1,\ldots ,5$
and $W_{m}^{\prime }$, $m=1,\ldots ,13$. Note: the invariant responses here
are for \emph{semi-inclusive} scattering and depend on the four chosen
scalar invariants; these quantities should not be confused with the \emph{%
inclusive} invariant response functions discussed below.

\subsection{Two Coordinate Systems for the Target Spin\label%
{subsec-primed}}

\begin{figure}
	\centering
	\includegraphics[height=5cm]{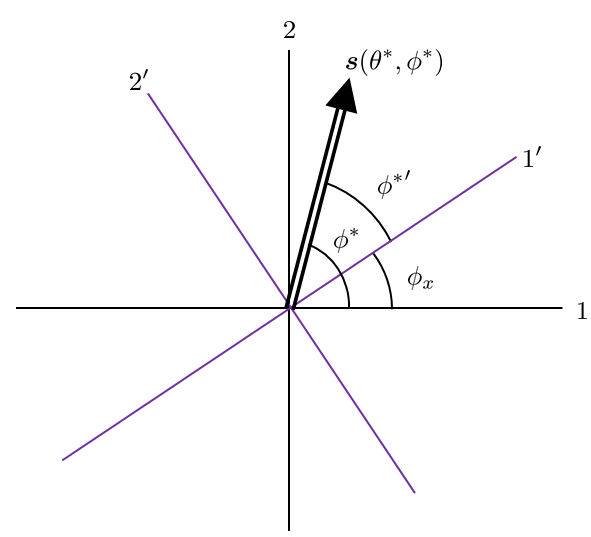} 				
	\caption{Two coordinate systems for the target spin. The original coordinate system is shown in Fig.~\ref{fig:figure1} and here one can see how the primed system is related via a rotation around the 3-direction (the direction of the 3-momentum transfer $\mathbf{q}$) by the azimuthal angle $\phi_x$. Hence in the 123-system the azimuthal angle of the target spin is $\phi^\ast$, while in the $1'2'3'$-system it is ${\phi^\ast}^{\prime} = \phi^\ast - \phi_x$.}
	\label{fig:figure7}
\end{figure}

We will have occasion to use two different coordinate system to specify the
axis of quantization for the target spin. In the discussions above we chose
the lepton-plane oriented coordinate system where $\mathbf{q}$ is along
the 3-axis and the 2-axis is normal to the electron scattering plane (see
Fig.~\ref{fig:figure1}). It proves to be convenient to introduce a rotated (around the
3-direction) coordinate system which we denote with primes, namely one with $%
3^{\prime }$-axis along $\mathbf{q}$ and $2^{\prime }$-axis normal to
the plane formed by $\mathbf{q}$ and $\mathbf{p}_{x}$ (see Fig.~\ref{fig:figure7}). The reason
for this choice of rotated system will become apparent in due course. The
unit vectors in these two systems are related by%
\begin{eqnarray}
\mathbf{u}_{1^{\prime }} &=&\cos \phi _{x}\mathbf{u}_{1}+\sin \phi
_{x}\mathbf{u}_{2}  \label{eq-ss1} \\
\mathbf{u}_{2^{\prime }} &=&-\sin \phi _{x}\mathbf{u}_{1}+\cos \phi
_{x}\mathbf{u}_{2}  \label{eq-ss2} \\
\mathbf{u}_{3^{\prime }} &=&\mathbf{u}_{3}  \label{eq-ss3}
\end{eqnarray}%
and the inverse%
\begin{eqnarray}
\mathbf{u}_{1} &=&\cos \phi _{x}\mathbf{u}_{1^{\prime }}-\sin \phi
_{x}\mathbf{u}_{2^{\prime }}  \label{eq-ss4} \\
\mathbf{u}_{2} &=&\sin \phi _{x}\mathbf{u}_{1^{\prime }}+\cos \phi
_{x}\mathbf{u}_{2^{\prime }}  \label{eq-ss5} \\
\mathbf{u}_{3} &=&\mathbf{u}_{3^{\prime }}.  \label{eq-ss6}
\end{eqnarray}%
One has that%
\begin{equation}
\mathbf{q}=q\mathbf{u}_{3}=q\mathbf{u}_{3^{\prime }}
\label{eq-ss6a}
\end{equation}%
while%
\begin{equation}
\mathbf{p}_{x}=p_{x}\left[ \sin \theta _{x}\mathbf{u}_{1^{\prime
}}+\cos \theta _{x}\mathbf{u}_{3^{\prime }}\right]   \label{eq-ss6b}
\end{equation}%
with no $2^{\prime }$ component, by construction. A simple result (which we
use below) is accordingly%
\begin{eqnarray}
\mathbf{q}\times \mathbf{p}_{x} &=&qp_{x}\sin \theta _{x}\left(
-\sin \phi _{x}\mathbf{u}_{1}+\cos \phi _{x}\mathbf{u}_{2}\right) 
\label{eq-ss6c} \\
&=&qp_{x}\sin \theta _{x}\mathbf{u}_{2^{\prime }},  \label{eq-ss6c1}
\end{eqnarray}%
namely having only a $2^{\prime }$ component. The spin 4-vector may then be
written in either the 123 system or the $1^{\prime }2^{\prime }3^{\prime }$
system. One may define projections of the spin 3-vector in the two systems
in the following way: the L, S and N directions are obtained by setting $%
\theta ^{\ast }=0$ (for L), $\theta ^{\ast }=\pi /2$ with $\phi ^{\ast }=0$
(for S) and $\phi ^{\ast }=$ $\pi /2$ (for N), namely, making projections
along the 123 system unit vectors%
\begin{eqnarray}
\mathcal{P}_{L} &\equiv &\mathbf{u}_{3}\cdot \mathbf{s}=h^{\ast
}s\cos \theta ^{\ast }  \label{eq-ss8} \\
\mathcal{P}_{S} &\equiv &\mathbf{u}_{1}\cdot \mathbf{s}=h^{\ast
}s\sin \theta ^{\ast }\cos \phi ^{\ast }  \label{eq-ss9} \\
\mathcal{P}_{N} &\equiv &\mathbf{u}_{2}\cdot \mathbf{s}=h^{\ast
}s\sin \theta ^{\ast }\sin \phi ^{\ast }  \label{eq-ss10}
\end{eqnarray}%
or doing the same, but for the unit vectors in the $1^{\prime }2^{\prime
}3^{\prime }$ system 
\begin{eqnarray}
\mathcal{P}_{L^{\prime }} &\equiv &\mathbf{u}_{3^{\prime }}\cdot 
\mathbf{s}=h^{\ast }s\cos \theta ^{\ast }  \label{eq-ss11} \\
\mathcal{P}_{S^{\prime }} &\equiv &\mathbf{u}_{1^{\prime }}\cdot 
\mathbf{s}=h^{\ast }s\sin \theta ^{\ast }\cos \phi^{\ast ^{\prime }}
\label{eq-ss12} \\
\mathcal{P}_{N^{\prime }} &\equiv &\mathbf{u}_{2^{\prime }}\cdot 
\mathbf{s}=h^{\ast }s\sin \theta ^{\ast }\sin \phi ^{\ast ^{\prime }}.
\label{eq-ss13}
\end{eqnarray}%
The magnitude of the spin 3-vector is given in Eq. (\ref{eq-ff7a1}). Using
the relationships amongst the unit vectors above one has that%
\begin{eqnarray}
\mathcal{P}_{L} &=&\mathcal{P}_{L^{\prime }}  \label{eq-sss1} \\
\mathcal{P}_{S} &=&\cos \phi _{x}\mathcal{P}_{S^{\prime }}-\sin \phi _{x}%
\mathcal{P}_{N^{\prime }}  \label{eq-sss2} \\
\mathcal{P}_{N} &=&\sin \phi _{x}\mathcal{P}_{S^{\prime }}+\cos \phi _{x}%
\mathcal{P}_{N^{\prime }}  \label{eq-sss3} \\
\mathcal{P}_{S^{\prime }} &=&\cos \phi _{x}\mathcal{P}_{S}+\sin \phi _{x}%
\mathcal{P}_{N}  \label{eq-sss4} \\
\mathcal{P}_{N^{\prime }} &=&-\sin \phi _{x}\mathcal{P}_{S}+\cos \phi _{x}%
\mathcal{P}_{N}.  \label{eq-sss5}
\end{eqnarray}%
Note that $\mathcal{P}_{L^{\prime }}=\mathcal{P}_{L}$ contains no dependence
on $\phi _{x}$. 

\section{Inclusive Cross Section\label{sec-CrossSec}}

For inclusive scattering one simply needs to eliminate all contributions
that contain the 4-vectors $V^{\mu }$ or $X^{\mu }$, as well as the
invariant $I_{0}$ as they involve the 4-vector $P_{x}^{\mu }$ which does not
enter in the inclusive case.\ All invariant response functions depend only
on two scalar quantities, for example, $Q^{2}$ and $Q\cdot P = Q^2 I_1$. Accordingly one obtains the following:%
\begin{eqnarray}
\left( W_{s}^{\mu \nu }\right) _{unpol}^{incl} &=&-\left( W_{1}\right)
^{incl}\left( g^{\mu \nu }-\frac{Q^{\mu }Q^{\nu }}{Q^{2}}\right) +\left(
W_{2}\right) ^{incl}U^{\mu }U^{\nu }  \label{eq-had62} \\
\left( W_{a}^{\mu \nu }\right) _{unpol}^{incl} &=&0  \label{eq-had62a} \\
\left( W_{s}^{\mu \nu }\right) _{pol}^{incl} &=&\left( W_{6}^{\prime
}\right) ^{incl}\left( U^{\mu }\overline{X}^{\nu }+U^{\nu }\overline{X}^{\mu
}\right)   \label{eq-had63a} \\
-i\left( W_{a}^{\mu \nu }\right) _{pol}^{incl} &=&\frac{1}{M}\left(
W_{9}^{\prime }\right) ^{incl}\epsilon ^{\mu \nu \alpha \beta }\Sigma
_{\alpha }Q_{\beta }  \notag \\
&&\;\;+\left( W_{11}^{\prime }\right) ^{incl}(U^{\mu }\overline{X}^{\nu
}-U^{\nu }\overline{X}^{\mu })  \label{eq-had63b}
\end{eqnarray}%
with 5 inclusive invariant functions $\left( W_{m}\right) ^{incl}$, $m=1,2$
and $\left( W_{m}^{\prime }\right) ^{incl}$, $m=6,9,11$. Using our previous
results for semi-inclusive scattering but now dropping all contributions
containing $V^{\mu }$ or $X^{\mu }$ we obtain the following: for the \textbf{%
symmetric, unpolarized }cases (now not continuing to develop the \underline{$%
TL$} and \underline{$TT$} cases)\textbf{\ }%
\begin{eqnarray}
\left[ W_{unpol}^{L}\right] ^{incl} &=&-\frac{1}{\rho }\left( W_{1}\right)
^{incl}+\left( U^{0}\right) ^{2}\left( W_{2}\right) ^{incl}  \label{eq-inc1}
\\
\left[ W_{unpol}^{T}\right] ^{incl} &=&2\left( W_{1}\right) ^{incl}+\left[
\left( U^{1}\right) ^{2}+\left( U^{2}\right) ^{2}\right] \left( W_{2}\right)
^{incl}  \label{eq-inc2} \\
\left[ W_{unpol}^{TT}\right] ^{incl} &=&\left[ -\left( U^{1}\right)
^{2}+\left( U^{2}\right) ^{2}\right] \left( W_{2}\right) ^{incl}
\label{eq-inc3a} \\
\left[ W_{unpol}^{TL}\right] ^{incl} &=&2\sqrt{2}U^{0}U^{1}\left(
W_{2}\right) ^{incl},  \label{eq-inc4}
\end{eqnarray}%
no results for the \textbf{anti-symmetric, unpolarized} case%
\begin{equation}
\left( W_{a}^{\mu \nu }\right) _{unpol}^{incl}=0,  \label{eq-inc4a}
\end{equation}%
as can be seen above in discussing the semi-inclusive responses. All
contributions there contained explicit factors involving $V^{\mu }$ or $%
X^{\mu };$ in fact, potential contributions of this type are
parity-violating when electrons are polarized longitudinal or sideways. For the \textbf{symmetric, polarized} cases (now not
continuing to develop the \underline{$TL$} and \underline{$TT$} cases) we
have%
\begin{eqnarray}
\left[ W_{pol}^{L}\right] ^{incl} &=&U^{0}\overline{X}^{0}W_{6}^{\prime }
\label{eq-inc8a} \\
\left[ W_{pol}^{T}\right] ^{incl} &=&\left( U^{2}\overline{X}^{2}+U^{1}%
\overline{X}^{1}\right) W_{6}^{\prime }  \label{eq-had8aa} \\
\left[ W_{pol}^{TT}\right] ^{incl} &=&\left( U^{2}\overline{X}^{2}-U^{1}%
\overline{X}^{1}\right) W_{6}^{\prime }  \label{eq-inc8c} \\
\left[ W_{pol}^{TL}\right] ^{incl} &=&2\sqrt{2}\left( U^{0}\overline{X}%
^{1}+U^{1}\overline{X}^{0}\right) W_{6}^{\prime },  \label{eq-inc8d}
\end{eqnarray}%
all of which are proportional to the same invariant response function $%
W_{6}^{\prime }$. And, finally, for the \textbf{anti-symmetric, polarized}
situation (now not continuing to develop the \underline{$TL$}$^{\prime }$
case, although it is very similar to the $TL^{\prime }$ case below, simply
having 2 replaced by 1; as noted earlier, this term can occur when only the
incident electron is polarized but when the scattered electron's
polarization is not measured although the leptonic factor goes as $1/\gamma $
and hence this contribution may be safely neglected at high energies -- we
do so in the following) we have%
\begin{eqnarray}
\left[ W_{pol}^{T^{\prime }}\right] ^{incl} &=&-2\left[ \frac{1}{M}\left(
W_{9}^{\prime }\right) ^{incl}\epsilon ^{12\alpha \beta }\Sigma _{\alpha
}Q_{\beta }\right.   \notag \\
&&\left. +\left( W_{11}^{\prime }\right) ^{incl}(U^{1}\overline{X}^{2}-%
\overline{X}^{1}U^{2})\right]   \label{eq-inc9} \\
\left[ W_{pol}^{TL^{\prime }}\right] ^{incl} &=&-2\sqrt{2}\left[ \frac{1}{M}%
\left( W_{9}^{\prime }\right) ^{incl}\epsilon ^{02\alpha \beta }\Sigma
_{\alpha }Q_{\beta }\right.   \notag \\
&&\left. +\left( W_{11}^{\prime }\right) ^{incl}(U^{0}\overline{X}^{2}-%
\overline{X}^{0}U^{2})\right] .  \label{eq-inc10}
\end{eqnarray}%
In total we find that 5\ invariant response functions enter, $W_{1,2}$ and $%
W_{9,11}^{\prime }$ in contributions that are TRE, plus the contributions
that involve the invariant response function $W_{6}^{\prime }$ and are TRO.

The general inclusive cross section may then be written in the following
form:%
\begin{equation}
\frac{d^{2}\sigma }{d\Omega _{e}dk^{\prime }}\equiv \sigma _{Mott}f\frac{M}{%
E_{p}}\mathcal{R}^{incl}  \label{eq-inc10a}
\end{equation}%
where $\sigma _{Mott}$ is the Mott cross section given in Eq. (\ref%
{eq-semi11}) and the full inclusive response is given by%
\begin{equation}
\mathcal{R}^{incl}=\mathcal{R}_{1}^{incl}+h\mathcal{R}_{2}^{incl}+h^{\ast }%
\mathcal{R}_{3}^{incl}+hh^{\ast }\mathcal{R}_{4}^{incl},  \label{eq-incx3}
\end{equation}%
in which the four contributions correspond to completely unpolarized,
electron polarization only, target polarization only, and double
polarization, respectively. As above all responses here depend on two scalar
invariants such as $Q^{2}$ and $Q\cdot P$ together with the electron
scattering angle $\theta _{e}$ which enters via the leptonic factors.
Clearly the four sectors can be separated by flipping the electron helicity $%
h$ and the direction of the target spin via the factor $h^{\ast }$.
Explicitly we have%
\begin{eqnarray}
\mathcal{R}_{1}^{incl} &=&v_{L}\left[ W_{unpol}^{L}\right] ^{incl}+v_{T}%
\left[ W_{unpol}^{T}\right] ^{incl}  \notag \\
&&+v_{TL}\left[ W_{unpol}^{TL}\right] ^{incl}+v_{TT}\left[ W_{unpol}^{TT}%
\right] ^{incl}  \label{eq-incx4} \\
h\mathcal{R}_{2}^{incl} &=&0  \label{eq-incx5} \\
h^{\ast }\mathcal{R}_{3}^{incl} &=&v_{L}\left[ W_{pol}^{L}\right]
^{incl}+v_{T}\left[ W_{pol}^{T}\right] ^{incl}  \notag \\
&&+v_{TL}\left[ W_{pol}^{TL}\right] ^{incl}+v_{TT}\left[ W_{pol}^{TT}\right]
^{incl}  \label{eq-incx6} \\
hh^{\ast }\mathcal{R}_{4}^{incl} &=&v_{TL^{\prime }}\left[
W_{pol}^{TL^{\prime }}\right] ^{incl}+v_{T^{\prime }}\left[
W_{pol}^{T^{\prime }}\right] ^{incl},  \label{eq-incx7}
\end{eqnarray}%
where, as above, we have dropped the small \underline{$TL$}$^{\prime }$
contribution. The leptonic factors are given in Sect. \ref{sec-lept} while
the inclusive hadronic response functions are given above.

\subsection{The Transition from Semi-Inclusive to Inclusive Scattering 
\label{subsec-semitoincl}}

While the above developments yield the structure of the general inclusive
cross section directly, it is also instructive to follow a different
strategy and proceed from the semi-inclusive cross section for a given
channel (\textit{i.e.,} for a specific particle x detected in coincidence
with the scattered electron), integrating over the allowed kinematics of the
4-momentum that goes with that particle, and then summing over all open
channels, of course, paying close attention to issues of double-counting.

We start with the general forms for the semi-inclusive cross section for the
specific channel where particle x is assumed to be detected given above in Secs. \ref{subsec-semicross} and \ref{subsec-primed}. The dependence on the azimuthal angle $\phi _{x}$
occurs in the explicit factors $\cos \phi _{x}$, $\cos 2\phi _{x}$ and $\sin
\phi _{x}$ in Eqs. (\ref{eq-r20},\ref{eq-r21},\ref{eq-r22aa}) for the cases
where the target is unpolarized. Clearly, upon performing the integrals over 
$\phi _{x}$ over the range $(0,2\pi )$ yields zero for the $TT$, $TL$ and $%
TL^{\prime }$ cases, verifying the above inclusive structure (see Eqs. (\ref%
{eq-inc14}), for example). The $L$ and $T$ cases in Eqs. (\ref{eq-r19aa},\ref%
{eq-r19}) simply pick up a factor $2\pi $ when the azimuthal integral is
performed. In summary, for the target unpolarized situation one finds that
each channel yields only $L$ and $T$ responses, as we have already seen
above (see Eqs. (\ref{eq-inc11},\ref{eq-inc12})).

The situation where the target is polarized is a little more complicated.
There one finds that as well as explicit factors $\cos \phi _{x}$, $\cos
2\phi _{x}$, $\sin \phi _{x}$ and $\sin 2\phi _{x}$ in Eqs. (\ref{eq-r24x6},%
\ref{eq-r24x8},\ref{eq-r24y4}) one has implicit dependence on $\phi _{x}$
via the factors $\mathcal{P}_{S^{\prime }}$ and $\mathcal{P}_{N^{\prime }}$
in those equations together with Eqs. (\ref{eq-r24x2},\ref{eq-r24x4},\ref%
{eq-r24y2}). In this scenario it, of course, makes no sense to use the
primed spin-projection variables, since the plane in which the momentum of
particle x lies is being integrated over and accordingly we must go back to
the original unprimed spin projections which are referred to the electron
scattering frame. Two of the symmetric, polarized cases are simple: the $L$
and $T$ results in Eqs. (\ref{eq-r24x2}) and (\ref{eq-r24x4}), respectively,
depend on the azimuthal angle solely through the factor $\mathcal{P}%
_{N^{\prime }}$, which, by Eq. (\ref{eq-sss5}) only has dependences $\sin
\phi _{x}$ and $\cos \phi _{x}$ and accordingly upon integrations over $\phi
_{x}$ yield zero, in accordance with Eqs. (\ref{eq-inc18}). The remaining
cases require somewhat more work. The symmetric $TL$ response in Eq. (\ref%
{eq-r24x8}) has three contributions%
\begin{eqnarray}
x_{1} &\sim &\cos \phi _{x}\mathcal{P}_{N^{\prime }}=\frac{1}{2}\left[ -\sin
2\phi _{x}\mathcal{P}_{S}+\left( 1+\cos 2\phi _{x}\right) \mathcal{P}_{N}%
\right]   \label{eq-incff3} \\
x_{2} &\sim &\sin \phi _{x}\mathcal{P}_{L^{\prime }}=\sin \phi _{x}\mathcal{P%
}_{L}  \label{eq-incff4} \\
x_{3} &\sim &\sin \phi _{x}\mathcal{P}_{S^{\prime }}=\frac{1}{2}\left[ \sin
2\phi _{x}\mathcal{P}_{S}+\left( 1-\cos 2\phi _{x}\right) \mathcal{P}_{N}%
\right] .  \label{eq-incff5}
\end{eqnarray}%
Upon integrating over $\phi _{x}$ one then finds that the $x_{1}$ and $x_{3}$
cases yield $\pi \mathcal{P}_{N}$, while the $x_{2}$ case yields zero, in
accord with Eq. (\ref{eq-inc16a}), namely, a nonzero result that goes as $%
\mathcal{P}_{N}$. Similarly, the symmetric $TT$ response in Eq. (\ref%
{eq-r24x6}) also has three contributions%
\begin{eqnarray}
y_{1} &\sim &\cos 2\phi _{x}\mathcal{P}_{N^{\prime }}=\frac{1}{2}\left[
-\left( \sin 3\phi _{x}-\sin \phi _{x}\right) \mathcal{P}_{S}\right.   \notag
\\
&&\left. +\left( \cos 3\phi _{x}+\cos \phi _{x}\right) \mathcal{P}_{N}\right]
\label{eq-incff6} \\
y_{2} &\sim &\sin 2\phi _{x}\mathcal{P}_{L^{\prime }}=\sin 2\phi _{x}%
\mathcal{P}_{L}  \label{eq-incff7} \\
y_{3} &\sim &\sin 2\phi _{x}\mathcal{P}_{S^{\prime }}=\frac{1}{2}\left[
\left( \sin 3\phi _{x}+\sin \phi _{x}\right) \mathcal{P}_{S}\right.   \notag
\\
&&\left. +\left( -\cos 3\phi _{x}+\cos \phi _{x}\right) \mathcal{P}_{N}%
\right] ,  \label{eq-incff8}
\end{eqnarray}%
all of which integrate to zero and yield no contribution for the $TT$ term,
in accord with Eq. (\ref{eq-inc18}). Next, the anti-symmetric polarized
cases are handled similarly: for the $T^{\prime }$ response in Eq. (\ref%
{eq-r24y2}) the contribution that involves $\mathcal{P}_{S^{\prime }}$
yields zero upon integration over $\phi _{x}$ while the contribution that
involves $\mathcal{P}_{L^{\prime }}$ and hence no dependence on $\phi _{x}$
yields a nonzero result arising from the factor $2\pi $ coming from the
integral. Thus the $T^{\prime }$ response yields a nonzero result that is
proportional to $\mathcal{P}_{L}$, as in Eq. (\ref{eq-inc24}). Finally, the $%
TL^{\prime }$ response in Eq. (\ref{eq-r24y4}) involves three contributions%
\begin{eqnarray*}
z_{1} &\sim &\cos \phi _{x}\mathcal{P}_{L^{\prime }}=\cos \phi _{x}\mathcal{P%
}_{L} \\
z_{2} &\sim &\cos \phi _{x}\mathcal{P}_{S^{\prime }}=\frac{1}{2}\left[
\left( 1+\cos 2\phi _{x}\right) \mathcal{P}_{S}+\sin 2\phi _{x}\mathcal{P}%
_{N}\right]  \\
z_{3} &\sim &\sin \phi _{x}\mathcal{P}_{N^{\prime }}=\frac{1}{2}\left[
-\left( 1-\cos 2\phi _{x}\right) \mathcal{P}_{S}+\sin 2\phi _{x}\mathcal{P}%
_{N}\right] .
\end{eqnarray*}%
As above, the term involving $z_{1}$ integrates to zero, while the $z_{2}$
and $z_{3}$ terms yields factors of $\pi $ and $-\pi $, respectively, and
involve the spin projection $\mathcal{P}_{S}$, in agreement with Eq. (\ref%
{eq-inc23}). Thus exactly the structure found above when proceeding to the
inclusive cross section directly is found by integrating the semi-inclusive
responses over $\phi _{x}$.

\section{Rest System Variables\label{sec-restSystem}}

One can now proceed to use the general expressions given above in any
coordinate system, since everything is written in covariant form. In
particular, one major goal in making the developments above is to have the
semi-inclusive cross section both in the collider frame and also in the
target rest frame. Typically one will develop some model for the cross
section in the target rest frame and thereby identify the invariant
functions this entails. This then immediately yields the cross section in
the general collider frame, since these response functions are, by
construction, invariant, and all of the kinematic factors discussed above
are covariant.

The most straightforward way to obtain the required target rest frame
variables is to use the original 123-system expressions obtained above, but
to assume first that $\theta =0$ so that $\mathbf{p}$ and $\mathbf{q}
$ are collinear and second that $p$ is set to zero. One then has%
\begin{eqnarray}
Q_{R}^{\mu } &=&\left( \omega _{R},0,0,q_{R}\right) =q_{R}\left( \nu^{\prime}
_{R},0,0,1\right)  \label{eq-r1} \\
P_{R}^{\mu } &=&M(1,0,0,0)  \label{eq-r2} \\
U_{R}^{\mu } &=&\frac{1}{\rho _{R}}\left( 1,0,0,\nu^{\prime} _{R}\right)
\label{eq-r3}
\end{eqnarray}%
with%
\begin{eqnarray}
\nu^{\prime} _{R} &=&\frac{\omega _{R}}{q_{R}}  \label{eq-r4} \\
\rho _{R} &=&-\frac{Q^{2}}{q_{R}^{2}}=1-{\nu^{\prime} _{R}}^{2}.  \label{eq-r5}
\end{eqnarray}%
Clearly one has $Q_{R}\cdot U_{R}=0$ as required by construction. Next, one
has%
\begin{equation}
P_{x,R}^{\mu }=\left( E_{x,R},\mathbf{p}_{x,R}\right)  \label{eq-r6}
\end{equation}%
with%
\begin{eqnarray}
\mathbf{p}_{x,R} &=&p_{x,R}\left[ \sin \theta _{x,R}\left( \cos \phi
_{x,R}\mathbf{u}_{1}+\sin \phi _{x,R}\mathbf{u}_{2}\right) +\cos
\theta _{x,R}\mathbf{u}_{3}\right]  \label{eq-r7} \\
E_{x,R} &=&\sqrt{p_{x,R}^{2}+M_{x}^{2}},  \label{eq-r8}
\end{eqnarray}%
which yields%
\begin{equation}
V_{R}^{\mu }=\left( \frac{1}{M\rho _{R}}\mathcal{E}_{x,R},V_{R}^{1}%
\mathbf{,}V_{R}^{2},\frac{\nu^{\prime} _{R}}{M\rho _{R}}\mathcal{E}_{x,R}\right)
\label{eq-r9}
\end{equation}%
with%
\begin{eqnarray}
\mathcal{E}_{x,R} &\equiv &E_{x,R}-\nu^{\prime} _{R}p_{x,R}\cos \theta _{x,R}
\label{eq-r10} \\
V_{R}^{1} &=&\frac{P_{x,R}^{1}}{M}=\frac{p_{x,R}}{M}\sin \theta _{x,R}\cos
\phi _{x,R}=\eta _{x,R}\cos \phi _{x,R}  \label{eq-r10aa} \\
V_{R}^{2} &=&\frac{P_{x,R}^{2}}{M}=\frac{p_{x,R}}{M}\sin \theta _{x,R}\sin
\phi _{x,R}=\eta _{x,R}\sin \phi _{x,R},  \label{eq-r10bb}
\end{eqnarray}%
where%
\begin{equation}
\eta _{x,R}\equiv \frac{p_{x,R}}{M}\sin \theta _{x,R},  \label{eq-r10b1}
\end{equation}%
and again, as required by construction, one has $Q_{R}\cdot V_{R}=0$. Upon
finding that%
\begin{eqnarray}
U_{R}^{2} &=&\frac{1}{\rho _{R}}  \label{eq-r10a} \\
U_{R}\cdot V_{R} &=&\frac{\mathcal{E}_{x,R}}{M\rho _{R}}  \label{eq-r10b}
\end{eqnarray}%
and using Eq. (\ref{eq-had5x6}) one has that%
\begin{eqnarray}
X_{R}^{\mu } &=&V_{R}^{\mu }-\left( \frac{\mathcal{E}_{x,R}}{M}\right)
U_{R}^{\mu }  \label{eq-r10c} \\
&=&\left( 0,V_{R}^{1},V_{R}^{2},0\right) =\frac{1}{M}\left(
0,P_{x,R}^{1},P_{x,R}^{2},0\right)  \label{eq-r10d} \\
&=&\eta _{x,R}\left( 0,\cos \phi _{x,R},\sin \phi _{x,R},0\right) .
\label{eq-r10dd}
\end{eqnarray}

First, we have from above that%
\begin{equation}
S_{R}^{0}=0  \label{eq-ss7}
\end{equation}%
and that%
\begin{eqnarray}
\mathbf{s}_{R} &=&h^{\ast }\left[ \sin \theta _{R}^{\ast }\left( \cos
\phi _{R}^{\ast }\mathbf{u}_{1}+\sin \phi _{R}^{\ast }\mathbf{u}%
_{2}\right) +\cos \theta _{R}^{\ast }\mathbf{u}_{3}\right] 
\label{eq-r12} \\
&=&h^{\ast }\left[ \sin \theta _{R}^{\ast }\left( \cos \phi _{R}^{\ast
^{\prime }}\mathbf{u}_{1^{\prime }}+\sin \phi _{R}^{\ast ^{\prime }}%
\mathbf{u}_{2^{\prime }}\right) +\cos \theta _{R}^{\ast }\mathbf{u}%
_{3^{\prime }}\right] .  \label{eq-r12aa}
\end{eqnarray}%
Clearly from Fig.~\ref{fig:figure6} one has that%
\begin{equation}
\phi _{R}^{\ast }=\phi _{x,R}+\phi _{R}^{\ast ^{\prime }}.  \label{eq-r12bb}
\end{equation}%
One may then employ Eqs. (\ref{eq-ss1}-\ref{eq-sss5}) in the rest system
(indicated by adding the label R). In particular, the target spin 4-vector
becomes%
\begin{equation}
\Sigma _{R}^{\mu }=h^{\ast }\left( -\frac{\nu^{\prime} _{R}}{\rho _{R}}\cos \theta
_{R}^{\ast },\sin \theta _{R}^{\ast }\cos \phi _{R}^{\ast },\sin \theta
_{R}^{\ast }\sin \phi _{R}^{\ast },-\frac{{\nu^{\prime} _{R}}^{2}}{\rho _{R}}\cos
\theta _{R}^{\ast }\right) ,  \label{eq-r12b}
\end{equation}%
using the fact that%
\begin{equation}
Q_{R}\cdot S_{R}=-q_{R}h^{\ast }\cos \theta _{R}^{\ast }.  \label{eq-r12a}
\end{equation}%
One has $Q_{R}\cdot \Sigma _{R}=0$ as required by construction and also%
\begin{equation}
U_{R}\cdot \Sigma _{R}=-h^{\ast }\frac{\nu^{\prime} _{R}}{\rho _{R}}\cos \theta
_{R}^{\ast }  \label{eq-r12e}
\end{equation}%
as well as%
\begin{eqnarray}
\left[ I_{0}\right] _{R} &=&\frac{1}{M^{2}}\left( \mathbf{q}_{R}\times 
\mathbf{p}_{x,R}\right) \cdot \mathbf{s}_{R}  \label{eq-r12ee} \\
&=&\frac{q_{R}p_{x,R}}{M^{2}}\sin \theta _{x,R}\mathcal{P}_{N^{\prime }}^{R}
\label{eq-r12ee1} \\
&=&\frac{q_{R}}{M}\eta _{x,R}\mathcal{P}_{N^{\prime }}^{R}  \label{eq-r12ee2}
\end{eqnarray}%
namely, in the $1^{\prime }2^{\prime }3^{\prime }$ system this invariant is
especially simple in that it involves only the N$^{\prime }$ projection of
the spin, motivating the rotation from the 123 system to the $1^{\prime
}2^{\prime }3^{\prime }$ system introduced above.

Then we can find $\overline{X}_{R}^{\mu }$ in terms of these 4-vectors. We
have from Eq. (\ref{eq-f42a}) that%
\begin{equation}
\overline{X}^{\mu }=\frac{1}{M^{2}}\epsilon ^{\mu \alpha \beta \gamma
}S_{\alpha }Q_{\beta }P_{\gamma },  \label{eq-r13}
\end{equation}%
where, using the above expressions for $Q_{\beta }$ and $P_{\gamma }$, one
must have $\gamma =0$ and hence $\beta =3$; accordingly only the cases with $%
\mu \alpha =12$ and 21 occur. Using Eqs. (\ref{eq-ff29}) and (\ref{eq-ff30}%
), in the target rest system we then have specifically that%
\begin{eqnarray}
\overline{X}_{R}^{0} &=&\overline{X}_{R}^{3}=0  \label{eq-ff30a} \\
\overline{X}_{R}^{1} &=&-\frac{1}{M}\left( \mathbf{q}_{R}\times 
\mathbf{s}_{R}\right) ^{1}=\frac{q_{R}}{M}\mathcal{P}_{N}^{R}=h^{\ast }%
\frac{q_{R}}{M}\sin \theta _{R}^{\ast }\sin \phi _{R}^{\ast }
\label{eq-ff30b} \\
\overline{X}_{R}^{2} &=&-\frac{1}{M}\left( \mathbf{q}_{R}\times 
\mathbf{s}_{R}\right) ^{2}=-\frac{q_{R}}{M}\mathcal{P}_{S}^{R}=-h^{\ast }%
\frac{q_{R}}{M}\sin \theta _{R}^{\ast }\cos \phi _{R}^{\ast },
\label{eq-ff30c}
\end{eqnarray}%
namely%
\begin{eqnarray}
\overline{X}_{R}^{\mu } &=&h^{\ast }\frac{q_{R}}{M}\sin \theta _{R}^{\ast
}\left( 0,\sin \phi _{R}^{\ast },-\cos \phi _{R}^{\ast },0\right) 
\label{eq-r14} \\
&=&\frac{q_{R}}{M}\left( 0,\mathcal{P}_{N},-\mathcal{P}_{S},0\right) .
\label{eq-r14y}
\end{eqnarray}%
From these results in the 123 system the corresponding results in the $%
1^{\prime }2^{\prime }3^{\prime }$ system are immediate:%
\begin{eqnarray}
\overline{X}_{R}^{1^{\prime }} &=&h^{\ast }\frac{q_{R}}{M}\sin \theta
_{R}^{\ast }\sin \phi _{R}^{\ast \prime }  \label{eq-r14x1} \\
\overline{X}_{R}^{2^{\prime }} &=&-h^{\ast }\frac{q_{R}}{M}\sin \theta
_{R}^{\ast }\cos \phi _{R}^{\ast \prime },  \label{eq-r14x2}
\end{eqnarray}%
where here we use primes on the transverse Lorentz components to indicate
that they are in the rotated coordinate system. Finally, we have the last
remaining 4-vector $\overline{U}_{R}^{\mu }$ which is given in terms of $%
\overline{T}_{R}^{\mu }$ and $\overline{V}_{R}^{\mu }$:%
\begin{equation}
\overline{U}_{R}^{\mu }=\overline{T}_{R}^{\mu }-\left( \frac{\mathcal{E}%
_{x,R}}{M}\right) \overline{X}_{R}^{\mu },  \label{eq-r14x2a}
\end{equation}%
where%
\begin{eqnarray}
\overline{T}_{R}^{0} &=&-h^{\ast }\frac{q_{R}p_{x,R}}{M^{2}}\sin \theta
_{x,R}\sin \theta _{R}^{\ast }\sin \left( \phi _{x,R}-\phi _{R}^{\ast
}\right)   \label{eq-r14a1} \\
\overline{T}_{R}^{3} &=&\nu^{\prime} _{R}\overline{T}_{R}^{0}  \label{eq-r14a2} \\
\overline{T}_{R}^{1} &=&h^{\ast }\frac{q_{R}}{M^{2}}\left[ \mathcal{E}%
_{x,R}\sin \theta _{R}^{\ast }\sin \phi _{R}^{\ast }\right.   \notag \\
&&\left. +\nu^{\prime} _{R}p_{x,R}\sin \theta _{x,R}\sin \phi _{x,R}\cos
\theta _{R}^{\ast }\right]   \label{eq-r14a3} \\
\overline{T}_{R}^{2} &=&-h^{\ast }\frac{q_{R}}{M^{2}}\left[ \mathcal{E}%
_{x,R}\sin \theta _{R}^{\ast }\cos \phi _{R}^{\ast }\right.   \notag \\
&&\left. +\nu^{\prime} _{R}p_{x,R}\sin \theta _{x,R}\cos \phi _{x,R}\cos
\theta _{R}^{\ast }\right] .  \label{eq-r14a4}
\end{eqnarray}%
Then, assembling all of the developments in the above section, using in
particular Eq. (\ref{eq-ss6c}), in the target rest system we find that%
\begin{eqnarray}
\overline{U}_{R}^{0} &=&\frac{1}{\nu^{\prime} _{R}}\overline{U}_{R}^{3}=\frac{1}{M^{2}%
}\left( \mathbf{q}_{R}\times \mathbf{p}_{x,R}\right) \cdot 
\mathbf{s}_{R}  \label{eq-fg15a} \\
&=&\frac{q_{R}}{M}\eta _{x,R}\mathcal{P}_{N^{\prime }}^{R}  \label{eq-fg15a1}
\\
\overline{U}_{R}^{1} &=&-\frac{\omega _{R}}{M^{2}q_{R}^{2}}\left( 
\mathbf{q}_{R}\times \mathbf{p}_{x,R}\right) ^{1}\left( \mathbf{q%
}_{R}\cdot \mathbf{s}_{R}\right)   \label{eq-fg15b} \\
&=&\frac{\omega _{R}}{M}\eta _{x,R}\sin \phi _{x,R}\mathcal{P}_{L^{\prime
}}^{R}  \label{eq-fg15b1} \\
\overline{U}_{R}^{2} &=&-\frac{\omega _{R}}{M^{2}q_{R}^{2}}\left( 
\mathbf{q}_{R}\times \mathbf{p}_{x,R}\right) ^{2}\left( \mathbf{q%
}_{R}\cdot \mathbf{s}_{R}\right)   \label{eq-fg15c} \\
&=&-\frac{\omega _{R}}{M}\eta _{x,R}\cos \phi _{x,R}\mathcal{P}_{L^{\prime
}}^{R},  \label{eq-fg15c1}
\end{eqnarray}%
that is,%
\begin{equation}
\overline{U}_{R}^{\mu }=\frac{q_{R}}{M}\eta _{x,R}\left( \mathcal{P}%
_{N^{\prime }}^{R},\nu^{\prime} _{R}\sin \phi _{x,R}\mathcal{P}_{L^{\prime
}}^{R},-\nu^{\prime} _{R}\cos \phi _{x,R}\mathcal{P}_{L^{\prime }}^{R},\nu^{\prime} _{R}%
\mathcal{P}_{N^{\prime }}^{R}\right) .  \label{eq-fg15c2}
\end{equation}%
Note that%
\begin{equation}
\overline{U}_{R}^{0}=\left[ I_{0}\right] _{R}.  \label{eq-fg15d}
\end{equation}

We are now in a position to write explicit expressions for the hadronic
tensors in the rest system. 

\subsection{Semi-inclusive Tensors in the Rest System\label%
{subsec-semiRest}}

For the \textbf{symmetric, unpolarized} case we immediately have%
\begin{eqnarray}
\left[ W_{unpol}^{L}\right] _{R} &=&\frac{1}{\rho _{R}^{2}}\left( -\rho
_{R}W_{1}+W_{2}\right)   \label{eq-r19aa} \\
\left[ W_{unpol}^{T}\right] _{R} &=&2W_{1}+\eta _{x,R}^{2}W_{3}
\label{eq-r19} \\
\left[ W_{unpol}^{TT}\right] _{R} &=&-\eta _{x,R}^{2}\cos 2\phi _{x,R}W_{3}
\label{eq-r20} \\
\left[ W_{unpol}^{TL}\right] _{R} &=&2\sqrt{2}\frac{1}{\rho _{R}}\eta
_{x,R}\cos \phi _{x,R}W_{4},  \label{eq-r21}
\end{eqnarray}%
all of which are TRE. For the \textbf{anti-symmetric, unpolarized} case one
has%
\begin{equation}
\left[ W_{unpol}^{TL^{\prime }}\right] _{R}=-2\sqrt{2}\frac{1}{\rho _{R}}%
\eta _{x,R}\sin \phi _{x,R}W_{5}  \label{eq-r22aa}
\end{equation}%
namely, the usual $TL^{\prime }$ so-called TRO 5th response function \cite{twdprogpartnuc}
which goes as $\sin \phi _{x,R}$, as expected; there is no $\mu \nu =12$, $%
T^{\prime }$ response in the rest frame. Each of these has explicit
dependence on the azimuthal angle $\phi _{x,R}$ and consequently when one
wishes to relate any specific model for the semi-inclusive reaction to the
invariant response functions the procedures here are clear. For the
anti-symmetric, unpolarized case only a single invariant response enters and
thus $W_{5}$ may immediately be extracted. In the symmetric, unpolarized
case the $\phi _{x,R}$ dependences allow the $W_{3}$ and $W_{4}$ invariant
responses to be isolated using Eqs. (\ref{eq-r20}) and (\ref{eq-r21}),
respectively. Then knowing $W_{3}$ one can deduce $W_{1}$ from Eq. (\ref%
{eq-r19}) which contains a linear combination of $W_{1}$ and $W_{3}$.
Finally, using Eq. (\ref{eq-r19aa}) which involves a linear combination of $%
W_{1}$ and $W_{2}$, the latter can also be extracted. This procedure can, in
principle, be used with experimental data; however, frequently one or more of
the responses may be so small that the extractions become very difficult. In
contrast, when the goal is to relate some model to the invariant responses,
these procedures can always be followed.

For the cases where the target is polarized it is helpful now to label the
response functions by the polarizations. Each of the six types of response ($%
L$, $T$, $TT$, $TL$, $T^{\prime }$, $TL^{\prime }$), in addition to
depending on the kinematic variables in the problem ($Q^{2}$, $x$, {\it etc.;} see
the previous discussions), also depends on the polar and azimuthal angles
that specify the direction in which the polarization axis of quantization
points, {\it i.e.,} the angles $\theta _{R}^{\ast }$ and $\phi _{R}^{\ast }$. Or,
equivalently, one may use the three particular directions given in Eqs. (\ref%
{eq-ss1}--\ref{eq-ss13}); here we do the latter and write the responses in
the form $\left[ W_{pol}^{K}\right] _{R}^{\Lambda ^{\prime }}$, where $K=$ $%
L $, $T$, $TT$, $TL$, $T^{\prime }$ or $TL^{\prime }$ and $\Lambda ^{\prime
}=L^{\prime }$, $S^{\prime }$ or $N^{\prime }$.

For the \textbf{symmetric, polarized} case one has%
\begin{eqnarray}
\left[ W_{pol}^{L}\right] _{R}^{\Lambda ^{\prime }} &=&\frac{1}{\rho _{R}^{2}%
}\left( -\rho _{R}W_{1}^{\prime }+W_{2}^{\prime }\right) \left[ I_{0}\right]
_{R}+\frac{2}{\rho _{R}}\overline{U}_{R}^{0}W_{5}^{\prime }  \label{eq-r24x1}
\\
&=&\frac{1}{\rho _{R}^{2}}\frac{q_{R}}{M}\eta _{x,R}\left( \rho _{R}\left(
2W_{5}^{\prime }-W_{1}^{\prime }\right) +W_{2}^{\prime }\right) \mathcal{P}%
_{N^{\prime }}^{R}  \label{eq-r24x2} \\
\left[ W_{pol}^{T}\right] _{R}^{\Lambda ^{\prime }} &=&\left( 2W_{1}^{\prime
}+\eta _{x,R}^{2}W_{3}^{\prime }\right) \left[ I_{0}\right] _{R}  \label{eq-r24x3}\\
&&+2\left\{ \left( X_{R}^{2}\overline{U}_{R}^{2}+X_{R}^{1}\overline{U}%
_{R}^{1}\right) W_{7}^{\prime }+\left( X_{R}^{2}\overline{X}%
_{R}^{2}+X_{R}^{1}\overline{X}_{R}^{1}\right) W_{8}^{\prime }\right\} 
\notag  \\
&=&\frac{q_{R}}{M}\eta _{x,R}\left( 2W_{1}^{\prime }+2W_{8}^{\prime }+\eta
_{x,R}^{2}W_{3}^{\prime }\right) \mathcal{P}_{N^{\prime }}^{R}
\label{eq-r24x4} \\
\left[ W_{pol}^{TT}\right] _{R}^{\Lambda ^{\prime }} &=&\left( -\eta
_{x,R}^{2}\cos 2\phi _{x,R}W_{3}^{\prime }\right) \left[ I_{0}\right] _{R} 
\label{eq-r24x5}\\
&&+2\left\{ \left( X_{R}^{2}\overline{U}_{R}^{2}-X_{R}^{1}\overline{U}%
_{R}^{1}\right) W_{7}^{\prime }+\left( X_{R}^{2}\overline{X}%
_{R}^{2}-X_{R}^{1}\overline{X}_{R}^{1}\right) W_{8}^{\prime }\right\} 
\notag  \\
&=&-\frac{q_{R}}{M}\eta _{x,R}\left[ \left( 2W_{8}^{\prime }+\eta
_{x,R}^{2}W_{3}^{\prime }\right) \cos 2\phi _{x,R}\mathcal{P}_{N^{\prime
}}^{R}\right.   \notag \\
&&\left. +2\sin 2\phi _{x,R}\left( \nu^{\prime} _{R}\eta _{x,R}\mathcal{P}_{L^{\prime
}}W_{7}^{\prime }+\mathcal{P}_{S^{\prime }}W_{8}^{\prime }\right) \right] 
\label{eq-r24x6} \\
\left[ W_{pol}^{TL}\right] _{R}^{\Lambda ^{\prime }} &=&2\sqrt{2}\left[
\left( \frac{1}{\rho _{R}}\eta _{x,R}\cos \phi _{x,R}W_{4}^{\prime }\right) %
\left[ I_{0}\right] _{R}\right.   \notag \\
&&+\frac{1}{\rho _{R}}\left( \overline{U}_{R}^{1}W_{5}^{\prime }+\overline{X}%
_{R}^{1}W_{6}^{\prime }\right)   \notag \\
&&\left. +\eta _{x,R}\cos \phi _{x,R}\left( \overline{U}_{R}^{0}W_{7}^{%
\prime }\right) \right]   \label{eq-r24x7} \\
&=&2\sqrt{2}\frac{1}{\rho _{R}}\frac{q_{R}}{M}\left[ \cos \phi _{x,R}%
\mathcal{P}_{N^{\prime }}^{R}\left( W_{6}^{\prime }+\eta _{x,R}^{2}\left(
W_{4}^{\prime }+\rho _{R}W_{7}^{\prime }\right) \right) \right.   \notag \\
&&\left. +\nu^{\prime} _{R}\eta _{x,R}\sin \phi _{x,R}\mathcal{P}_{L^{\prime
}}^{R}W_{5}^{\prime }+\sin \phi _{x,R}\mathcal{P}_{S^{\prime
}}^{R}W_{6}^{\prime }\right] ,  \label{eq-r24x8}
\end{eqnarray}%
all of which are TRO. Clearly $W_{5,6,7,8}^{\prime }$ may immediately be
isolated by choosing $\Lambda ^{\prime }=L^{\prime }$ and $S^{\prime }$ in
Eqs. (\ref{eq-r24x6}) and (\ref{eq-r24x8}). Then, choosing $\Lambda ^{\prime
}=N^{\prime }$ in Eq. (\ref{eq-r24x6}), $W_{3}^{\prime }$ can be determined.
Following this, $W_{1}^{\prime }$ may be deduced from Eq. (\ref{eq-r24x4})
and $W_{1}^{\prime }$ may be deduced from Eq. (\ref{eq-r24x2}), thereby
yielding the full set of symmetric, polarized invariant response functions.

Finally, for the \textbf{anti-symmetric, polarized} case the required
results are the following:%
\begin{eqnarray}
\left[ W_{pol}^{T^{\prime }}\right] _{R}^{\Lambda ^{\prime }} &=&-2\left[ 
\frac{1}{M}W_{9}^{\prime }\epsilon ^{12\alpha \beta }\Sigma _{\alpha
}Q_{\beta }\right.   \notag \\
&&\left. +W_{12}^{\prime }(X_{R}^{1}\bar{U}_{R}^{2}-X_{R}^{2}\bar{U}%
_{R}^{1})+W_{13}^{\prime }(X_{R}^{1}\overline{X}_{R}^{2}-X_{R}^{2}\overline{X%
}_{R}^{1})\right]   \label{eq-r24y1} \\
&=&2\frac{q_{R}}{M}\left[ \nu^{\prime} _{R}\left( W_{9}^{\prime }+\eta
_{x,R}^{2}W_{12}^{\prime }\right) \mathcal{P}_{L^{\prime }}^{R}+\eta
_{x,R}W_{13}^{\prime }\mathcal{P}_{S^{\prime }}^{R}\right]   \label{eq-r24y2}
\\
\left[ W_{pol}^{TL^{\prime }}\right] _{R}^{\Lambda ^{\prime }} &=&-2\sqrt{2}%
\left[ \frac{1}{M}W_{9}^{\prime }\epsilon ^{02\alpha \beta }\Sigma
_{\alpha }Q_{\beta }\right.   \notag \\
&&+U^{0}\left( \bar{U}_{R}^{2}W_{10}^{\prime }+\overline{X}%
_{R}^{2}W_{11}^{\prime }\right) -\left. X_{R}^{2}\bar{U}_{R}^{0}W_{12}^{%
\prime }\right]   \label{eq-r24y3} \\
&=&2\sqrt{2}\frac{q_{R}}{M}\left[ \left( \frac{1}{\rho _{R}}\nu^{\prime} _{R}\eta
_{x,R}W_{10}^{\prime }\right) \cos \phi _{x,R}\mathcal{P}_{L^{\prime
}}^{R}\right.   \notag \\
&&-\left( W_{9}^{\prime }-\frac{1}{\rho _{R}}W_{11}^{\prime }\right) \cos
\phi _{x,R}\mathcal{P}_{S^{\prime }}^{R}  \notag \\
&&\left. +\left( W_{9}^{\prime }-\frac{1}{\rho _{R}}W_{11}^{\prime }+\eta
_{x,R}^{2}W_{12}^{\prime }\right) \sin \phi _{x,R}\mathcal{P}_{N^{\prime
}}^{R}\right] ,  \label{eq-r24y4}
\end{eqnarray}%
all of which are TRE. By choosing $\Lambda ^{\prime }=S^{\prime }$ in Eq. (%
\ref{eq-r24y2}) and $\Lambda ^{\prime }=L^{\prime }$ in Eq. (\ref{eq-r24y4}) 
$W_{10,13}^{\prime }$ may both be isolated. Then by choosing $\Lambda
^{\prime }=S^{\prime }$ in Eq. (\ref{eq-r24y4}) the combination $%
W_{9}^{\prime }-\frac{1}{\rho _{R}}W_{11}^{\prime }$ may be extracted, and
choosing $\Lambda ^{\prime }=N^{\prime }$ in Eq. (\ref{eq-r24y4}) the
response $W_{12}^{\prime }$ determined. Finally, using Eq. (\ref{eq-r24y2})
and knowing $W_{12}^{\prime }$ the response $W_{9}^{\prime }$ may be
determined from the $\Lambda ^{\prime }=L^{\prime }$ there, and hence the $%
W_{11}^{\prime }$ response, since the combination $W_{9}^{\prime }-\frac{1}{%
\rho _{R}}W_{11}^{\prime }$ has been determined above. Thus, when the goal
is to provide relationships between any model responses for the
semi-inclusive cross section and the full set of invariant response
functions, these procedures provide the proof that such can be accomplished.

Note the behavior of the 18 types of contributions when the
results are written in terms of the $1^{\prime }2^{\prime }3^{\prime }$
system with polarizations $\mathcal{P}_{L^{\prime }}$, $\mathcal{P}%
_{S^{\prime }}$ and $\mathcal{P}_{N^{\prime }}$ times explicit dependence on
the angle $\phi _{x,R}$ are summarized in Table \ref{tab:2}. 

\begin{table}
	\begin{center}
\begin{tabular}{|c|c|c|c|c|}
\hline
& $unpol$ & $\mathcal{P}_{L^{\prime }}^{R}$ & $\mathcal{P}_{S^{\prime }}^{R}$
& $\mathcal{P}_{N^{\prime }}^{R}$ \\ \hline
$L$ & 1 & - & - & 1 \\ \hline
$T$ & 1 & - & - & 1 \\ \hline
$TL$ & $\cos \phi _{x,R}$ & $\sin \phi _{x,R}$ & $\sin \phi _{x,R}$ & $\cos
\phi _{x,R}$ \\ \hline
$TT$ & $\cos 2\phi _{x,R}$ & $\sin 2\phi _{x,R}$ & $\sin 2\phi _{x,R}$ & $%
\cos 2\phi _{x,R}$ \\ \hline
$T^{\prime }$ & - & 1 & 1 & - \\ \hline
$TL^{\prime }$ & $\sin \phi _{x,R}$ & $\cos \phi _{x,R}$ & $\cos \phi _{x,R}$
& $\sin \phi _{x,R}$ \\ \hline
\end{tabular}%
\end{center}
\caption{This table summarizes the dependence of the response functions on the angle $\phi _{x,R}$}. \label{tab:2}
\end{table}

For the $L$, $T$, $TT$ and $TL$ cases, the unpolarized responses are TRE and
the polarized responses are TRO, while for $T^{\prime }$ and $TL^{\prime }$
cases the reverse is true with the unpolarized case being TRO and the
polarized cases being TRE.

Using a very different procedure where the goal was to develop
semi-inclusive electron scattering with polarizations for situations where
the target could have any spin in \cite{Raskin:1988kc}, the case of pion
electroproduction was developed as an example of applying that approach.
Clearly this is an alternative to the present approach for the case where
the target spin is 1/2. The results in that cited work were vetted against
much earlier studies specifically of pion electroproduction (see the
references in \cite{Raskin:1988kc}). The behavior summarized in the above table was exactly
what was found in the earlier studies.

Again, the strategy in the present work is the following: given some model
for the polarized semi-inclusive cross section in the rest system one can
deduce what are the invariant response functions for that model. With these
the expressions in a general system immediately yield results for any choice
of kinematics. The key feature is having everything written in terms of
kinematic factors and invariant responses, since the latter are independent
of the choice of frame. So, for example, while the earlier studies referred
to above are completely general, they must be re-cast in terms of invariant
response functions if one wishes to relate the results in different frames
of reference.

\subsection{Inclusive Tensors and Cross Section in the Rest System\label{subsec-inclRest}%
}

In the rest frame the results are relatively simple: there one obtains the
following for the \textbf{symmetric cases without and with spin}:%
\begin{eqnarray}
\left[ W_{unpol}^{L}\right] _{R}^{incl} &=&\frac{1}{\rho _{R}^{2}}\left(
-\rho _{R}\left( W_{1}\right) ^{incl}+\left( W_{2}\right) ^{incl}\right) 
\label{eq-inc11} \\
\left[ W_{unpol}^{T}\right] _{R}^{incl} &=&2\left( W_{1}\right) ^{incl}
\label{eq-inc12} \\
\left[ W_{unpol}^{TT}\right] _{R}^{incl} &=&\left[ W_{unpol}^{TL}\right]
_{R}^{incl}=0.  \label{eq-inc14}
\end{eqnarray}%
\begin{eqnarray}
\left[ W_{pol}^{TL}\right] _{R}^{incl} &=&2\sqrt{2}\frac{q_{R}h^{\ast }}{%
M\rho _{R}}\left( W_{6}^{\prime }\right) ^{incl}\sin \theta _{R}^{\ast }\sin
\phi _{R}^{\ast }  \label{eq-inc16a} \\
\left[ W_{pol}^{L}\right] _{R}^{incl} &=&\left[ W_{pol}^{T}\right]
_{R}^{incl}=\left[ W_{pol}^{TT}\right] _{R}^{incl}=0;  \label{eq-inc18}
\end{eqnarray}%
here the only non-zero contribution goes as $\mathcal{P}_{N}=h^{\ast }\sin
\theta _{R}^{\ast }\sin \phi _{R}^{\ast }$. Note that one cannot in general
assume that $\left[ W_{pol}^{TL}\right] _{R}^{incl}$ is zero. Indeed, the
final states reached via inelastic scattering in general contain interfering
channels with complex amplitudes. An example of how this can occur is, for
instance, in the region where the $\Delta $ is important and one might model
the final states as containing a resonant $\Delta $ and non-resonant pion
production with different phase shifts. Or, at high energies one might go
beyond the lowest-order approximation for the inelastic processes involved
and incorporate higher-order loop diagrams, which are in general complex. As
discussed in \cite{Raskin:1988kc} and references therein, in such cases the TRO response $\left[
W_{pol}^{TL}\right] _{R}^{incl}$ is found to be non-zero.

As above no \textbf{anti-symmetric unpolarized} case survives and finally
for the \textbf{anti-symmetric, polarized} situation the required results
are the following:%
\begin{eqnarray}
\left[ W_{pol}^{T^{\prime }}\right] _{R}^{incl} &=&2h^{\ast }\frac{\omega
_{R}}{M}\left( W_{9}^{\prime }\right) ^{incl}\cos \theta _{R}^{\ast }  
\label{eq-inc24} \\
\left[ W_{pol}^{TL^{\prime }}\right] _{R}^{incl} &=&-2\sqrt{2}h^{\ast }\frac{%
q_{R}}{M}\left[ \left( W_{9}^{\prime }\right) ^{incl}-\frac{1}{\rho _{R}}%
\left( W_{11}^{\prime }\right) ^{incl}\right] \sin \theta _{R}^{\ast }\cos
\phi _{R}^{\ast };  \notag\\
&& \label{eq-inc23}
\end{eqnarray}%
in this sector the $TL^{\prime }$ contribution goes as $\mathcal{P}%
_{S}=h^{\ast }\sin \theta _{R}^{\ast }\cos \phi _{R}^{\ast }$ while the $%
T^{\prime }$ contribution goes as $\mathcal{P}_{L}=h^{\ast }\cos \theta
_{R}^{\ast }$.

Let us assemble these results into the cross section for inclusive
scattering in the target rest frame. First note that for the completely
unpolarized contributions we have%
\begin{equation}
\mathcal{R}_{1,R}^{incl}\equiv v_{L}^{R}\left[ W_{unpol}^{L}\right]
_{R}^{incl}+v_{T}^{R}\left[ W_{unpol}^{T}\right] _{R}^{incl}=\left(
W_{2}\right) ^{incl}+2\left( W_{1}\right) ^{incl}\tan ^{2}\theta _{e}^{R}/2
\label{eq-iinc1}
\end{equation}%
which, upon implementing the Feynman rules in the standard way, yields the
following familiar form for the unpolarized inclusive cross section:%
\begin{equation}
\left[ \frac{d^{2}\sigma }{d\Omega _{e}dk^{\prime }}\right]
_{R}^{unpol}=\sigma _{Mott}^{R}\left[ \left( W_{2}\right) ^{incl}+2\left(
W_{1}\right) ^{incl}\tan ^{2}\theta _{e}^{R}/2\right] =\sigma _{Mott}^{R}%
\mathcal{R}_{1,R}^{incl},  \label{eq-iinc2}
\end{equation}%
where the Mott cross section in the rest frame is given by%
\begin{equation}
\sigma _{Mott}^{R}=\left( \frac{\alpha \cos \theta _{e}^{R}/2}{2\epsilon
_{R}\sin ^{2}\theta _{e}^{R}/2}\right) ^{2}.  \label{eq-had99}
\end{equation}%
Here the invariant response functions $\left( W_{1,2}\right) ^{incl}$ have
dimensions of GeV$^{-1}$. In \ref{sec-inclusive} we develop the inclusive cross section in more detail as this may help the reader by making contact with more familiar expressions.

\section{\protect\bigskip Summary\label{sec-summary}}

The present study has focused on the scattering of polarized electrons from polarized spin-1/2 targets in situations where the scattered electron and some (unpolarized) particle $x$ are detected in coincidence, {\it viz.,} semi-inclusive scattering. Together with the well-known leptonic tensor that arises from products of the electron EM current matrix elements the EM hadronic tensor has been constructed using specific general basis sets of 4-vectors. When the target is unpolarized, following standard procedures these are taken to be the mutually orthogonal set $Q^\mu$, $U^\mu$ and $X^\mu$ given in Eqs. (\ref{eq-lept16}), (\ref{eq-had2}) and (\ref{eq-had5x4}), respectively. When the target is polarized and the target spin 4-vector $S^\mu$ is involved (see Eqs. (\ref{eq-ff8}) and (\ref{eq-f14})) it proves to be convenient to employ the set ${\bar X}^\mu$, ${\bar U}^\mu$, given in Eqs. (\ref{eq-had9}) and (\ref{eq-had7}), respectively, together with $U^\mu$ and $X^\mu$ along with an invariant $I_0$ (Eq. (\ref{eq-had11b})) and a special tensor obtained using the Levi-Civita symbol (Eq. (\ref{eq-had46x7})). In total one finds that there are 18 basis tensors, four symmetric ones when both the electron and target are unpolarized, a single anti-symmetric one when the electron is longitudinally polarized while the target is unpolarized, eight symmetric ones when the electron is unpolarized but the target is polarized, and five anti-symmetric ones when the electron and the target are both polarized.

The contraction of the leptonic and hadronic tensors that enters when applying the Feynman rules, which is a Lorentz invariant, is then formed as a linear combination involving these 18 hadronic tensors weighted with 18 invariant response functions, $W_i$, $i=1,5$ when the target is unpolarized and $W_i^{\prime}$, $i=1,13$ when the target is polarized. Each of these invariant responses is a function of four Lorentz scalars $(Q^2, I_{1,2,3})$ (see Eqs. (\ref{eq-f9}--\ref{eq-f14a})). Thus one has the kinematics of the reaction and the target spin dependence expressed in terms of the basis 4-vectors while the dynamics are contained in the 18 invariant response functions. Clearly the former are frame-dependent while the latter are not.

Given the Lorentz invariant contraction of the leptonic and hadronic tensors one can proceed using the Feynman rules to obtain the semi-inclusive cross section in a general frame where both the incident electron and the target are assumed to be moving, the latter with momentum $\mathbf{p}$. All of the kinematic factors summarized above must then be evaluated in this specific frame. One may obtain the corresponding results in a different frame where the target has a different value for its momentum simply by choosing a different value for $\mathbf{p}$; all other kinematic variables are then to be evaluated in that different frame. Specifically, one can express the semi-inclusive cross section in the target rest frame by setting $p=0$ and the results of doing so are detailed in the paper. Importantly, the dynamical content in the problem, which is encapsulated in the invariant response functions summarized above does not change when changing frames. Also, the 18 invariant response functions are functions only of the four Lorentz scalars listed above; these are also invariant.

The semi-inclusive cross section separates into four sectors according to the electron and target polarizations, namely, (I) both unpolarized, (II) electron polarized, target unpolarized, (III) target polarized, electron unpolarized, and (IV) both polarized. Having control of these polarizations then immediately allows the four sectors to be isolated. Furthermore, the cross section has explicit dependence on several kinematic variables that may be evaluated in principle to obtain enough linear equations in the 18 unknowns --- the 18 invariant response functions --- to invert and thereby determine those response functions. Specifically, the dependences on the electron scattering angle $\theta_e$, on the azimuthal angle for the 3-momentum of the detected particle, $\phi_x$, and on the angles $\theta^*$ and $\phi^*$ that specify the axis of quantization of the target spin can be used to isolate the required linear equations (an appendix is provided with the details).

Hence several strategies are available. In one approach where measurements are made in two different types of experiments the experimental results could be used in principle to isolate the 18 invariant response functions for the kinematical situation involved in the two experiments. Specifically, one could envision one experiment being performed in the target rest frame (fixed-target experiments) and from those measurements the 18 invariant response functions or some subset thereof being determined. One might then have a different experiment where the electron and target are both in motion (collider experiments): nevertheless, the same strategy could be followed and the 18 invariant response functions determined, albeit, perhaps for non-overlapping kinematics. The two sets of invariant responses could then be analyzed in a universal way.

A similar strategy occurs when using theory to make predictions of the semi-inclusive cross section. For instance, one may be forced to work in the target rest frame when modeling the dynamics using ingredients that are not ``boostable’’, which is almost always the case in nuclear physics for nuclei other than the deuteron. However, one could deduce the corresponding invariant response functions working in the target rest frame and then employ them in, say, the collider frame. Specific modeling of this sort will be undertaken by the authors in the future.

To make contact with other approaches, in the process of developing the semi-inclusive cross section we have chosen to express the results in terms of specific Lorentz components of the general hadronic tensor which are governed by the helicity projections of the exchanged virtual photon. We have included an appendix where this step is skipped and the contraction of leptonic and hadronic tensors is expressed directly in terms of invariant quantities. The two approaches are completely equivalent, but each may have advantages in particular applications.

Finally, we have shown how the inclusive scattering of polarized electrons from polarized spin-1/2 targets is related to integrations of the semi-inclusive cross sections plus sums over all open channels. We have included another appendix containing a few more details on inclusive scattering to help the reader find more familiar ground to aid in navigating the much more intricate problem of semi-inclusive scattering.

{\bf Acknowledgements}

This work was supported in part by the Office of Nuclear Physics of the U.S. Department of Energy under Grant Contract DE-FG02-94ER40818 (T. W. D.), in part by funds provided by the National Science Foundation under grant No. PHY-1913261 (S. J.), and by Jefferson Science Associates, LLC under U.S. DOE Contract DE-AC05-06OR23177 and U.S. DOE Grant DE-FG02-97ER41028 (J. W. V. O.).

\appendix

\section{Conventions\label{sec-conventions}}

In this work we employ the following conventions: 4-vectors are written $%
A^{\mu }=(A^{0},A^{1},A^{2},A^{3})=(A^{0},\mathbf{a})$ with capital letters
for the 4-vectors and lower-case letters for 3-vectors. The magnitude of a
3-vector is written as $a=|\mathbf{a}|$. One also has $A_{\mu }=g_{\mu \nu
}A^{\mu }=(A^{0},-A^{1},-A^{2},-A^{3})$ with%
\begin{equation}
g_{\mu \nu }=g^{\mu \nu }=\left( 
\begin{array}{llll}
1 & 0 & 0 & 0 \\ 
0 & -1 & 0 & 0 \\ 
0 & 0 & -1 & 0 \\ 
0 & 0 & 0 & -1%
\end{array}%
\right) .  \label{eq-app-conv-1}
\end{equation}%
The scalar product of two 4-vectors is given by $A\cdot B=A_{\mu }B^{\mu
}=(A^{0})^{2}-a^{2}$, following the conventions of \cite{Bjorken:1965sts}. For instance,
for the 4-momentum of an on-shell particle of mass $M$, energy $E_p$ and
3-momentum $p$ we have $P^{\mu }=(E_p ,\mathbf{p})$ and hence $P^{2}=P_{\mu
}P^{\mu }=E_p^{2}-p^{2}=M^{2}$. One problem occurs with these conventions, 
\textit{viz.} for the momentum transfer 4-vector we have $%
Q^{2}=(Q^{0})^{2}-q^{2}$ which, for electron scattering is spacelike, and
accordingly $Q^{2}<0$. One should be careful not to confuse our sign convention for this quantity with the so-called SLAC convention which has the opposite sign. The totally anti-symmetric Levi-Civita symbol follows the conventions
of \cite{Bjorken:1965sts} where%
\begin{equation}
\epsilon _{0123}=-\epsilon ^{0123}=+1.  \label{eq-app-conv-2}
\end{equation}%
When applying the Feynman rules we also employ the conventions of \cite{Bjorken:1965sts}.

\section{Contracted Tensors}\label{sec-Lscalars}

The contraction of the electron and hadron tensors can be written as

\begin{equation}
\eta_{\mu\nu}\chi^{\mu\nu}=\sum_{i=1}^5 C_i W_i+\sum_{i=1}^{13}C'_iW'_i .
\end{equation}
Since this is a Lorentz scalar, as are the $W_i$ and $W'_i$, the coefficients $C_i$ and $C'_i$ are also Lorentz invariants. From Eqs.~(\ref{eq-lept10},\ref{eq-lept13},\ref{eq-had51},\ref{eq-had55},\ref{eq-hhadz1x},\ref{eq-had46x7}) these coefficients can be written in terms of inner products of Lorentz 4-vectors as:
\begin{equation}
C_1=Q^2
\end{equation}
\begin{equation}
C_2=\frac{-4 K\cdot P P\cdot Q+4
	K\cdot P^2+M^2 Q^2}{2 M^2}
\end{equation}
\begin{align}
C_3=&\frac{1}{2 M^2
	\left(P\cdot Q^2-M^2 Q^2\right)^2}(-4 K\cdot P_x \left(M^2
	Q^2-P\cdot Q^2\right)\nonumber\\
	&\times \left(P_x\cdot Q
	\left(M^2 Q^2-2 K\cdot P
	P\cdot Q\right)+P\cdot P_x Q^2 (2
	K\cdot P-P\cdot Q)\right)\nonumber\\
	&+2 P\cdot P_x
	P_x\cdot Q Q^2 \left(2 K\cdot P \left(M^2
	Q^2+P\cdot Q^2\right)-4 K\cdot P^2
	P\cdot Q-P\cdot Q^3\right)\nonumber\\
	&+P\cdot P_x^2
	Q^4 \left(-4 K\cdot P P\cdot Q+4
	K\cdot P^2-M^2 Q^2+2
	P\cdot Q^2\right)\nonumber\\
	&+P\cdot Q P_x\cdot Q^2
	\left(P\cdot Q \left(4 K\cdot P^2+M^2
	Q^2\right)-4 K\cdot P M^2 Q^2\right)\nonumber\\
	&+4
	K\cdot P_x^2 \left(P\cdot Q^2-M^2
	Q^2\right)^2+M_x^2 Q^2
	\left(P\cdot Q^2-M^2 Q^2\right)^2)
\end{align}
\begin{align}
C_4=&\frac{(2 K\cdot P-P\cdot Q)}{M^4 Q^2-M^2
	P\cdot Q^2}
	\left[K\cdot P \left(2 P\cdot Q
	P_x\cdot Q-2 P\cdot P_x Q^2\right)\right.\nonumber\\
	&\left. +2
	K\cdot P_x \left(M^2
	Q^2-P\cdot Q^2\right)+Q^2
	\left(P\cdot P_x P\cdot Q-M^2
	P_x\cdot Q\right)\right]
\end{align}\begin{equation}
C_5=\frac{2  h\epsilon_{\alpha\beta\gamma\delta}K^\alpha P^\beta P_x^\gamma Q^\delta}{M^2}
\end{equation}
\begin{equation}
C'_1=-\frac{ h^* Q^2\epsilon_{\alpha\beta\gamma\delta}P^\alpha P_x^\beta Q^\gamma S^\delta}{M^3}
\end{equation}
\begin{equation}
C'_2=\frac{h^*\epsilon_{\alpha\beta\gamma\delta}P^\alpha P_x^\beta Q^\gamma S^\delta  \left(-4
	K\cdot P P\cdot Q+4 K\cdot P^2+M^2
	Q^2\right)}{2 M^5}
\end{equation}
\begin{align}
C'_3=&\frac{h^*\epsilon_{\alpha\beta\gamma\delta}P^\alpha P_x^\beta Q^\gamma S^\delta}{2 M^5
	\left(P\cdot Q^2-M^2 Q^2\right)^2} \left(-4
	K\cdot P_x \left(M^2
	Q^2-P\cdot Q^2\right)\right.\nonumber\\
	&\left.\times \left(P_x\cdot Q
	\left(M^2 Q^2-2 K\cdot P
	P\cdot Q\right)+P\cdot P_x Q^2 (2
	K\cdot P-P\cdot Q)\right)\right.\nonumber\\
	&\left.+2 P\cdot P_x
	P_x\cdot Q Q^2 \left(2 K\cdot P \left(M^2
	Q^2+P\cdot Q^2\right)-4 K\cdot P^2
	P\cdot Q-P\cdot Q^3\right)\right.\nonumber\\
	&\left.+P\cdot P_x^2
	Q^4 \left(-4 K\cdot P P\cdot Q+4
	K\cdot P^2-M^2 Q^2+2
	P\cdot Q^2\right)\right.\nonumber\\
	&\left.+P\cdot Q P_x\cdot Q^2
	\left(P\cdot Q \left(4 K\cdot P^2+M^2
	Q^2\right)-4 K\cdot P M^2 Q^2\right)\right.\nonumber\\
	&\left.+4
	K\cdot P_x^2 \left(P\cdot Q^2-M^2
	Q^2\right)^2+M_x^2 Q^2
	\left(P\cdot Q^2-M^2
	Q^2\right)^2\right)
\end{align}
\begin{align}
C'_4=& h^*\frac{\epsilon_{\alpha\beta\gamma\delta}P^\alpha P_x^\beta Q^\gamma S^\delta (2
	K\cdot P-P\cdot Q)}{M^7 Q^2-M^5
	P\cdot Q^2} \left[P_x\cdot Q
	\left(2 K\cdot P P\cdot Q-M^2
	Q^2\right)\right.\nonumber\\
	&\left.+P\cdot P_x Q^2 (P\cdot Q-2
	K\cdot P)+2 K\cdot P_x \left(M^2
	Q^2-P\cdot Q^2\right)\right]
\end{align}
\begin{align}
C'_5=&\frac{h^*}{M^5 Q^2-M^3 P\cdot Q^2} \left[2 (2
	K\cdot P-P\cdot Q)
	\left(\epsilon_{\alpha\beta\gamma\delta}K^\alpha P^\beta Q^\gamma S^\delta\right.\right.\nonumber\\
	&\left.\left. \times  \left(P\cdot P_x
	Q^2-P\cdot Q
	P_x\cdot Q\right)+\epsilon_{\alpha\beta\gamma\delta}K^\alpha P_x^\beta Q^\gamma S^\delta
	\left(P\cdot Q^2-M^2
	Q^2\right)\right)\right.\nonumber\\
	&\left.+\epsilon_{\alpha\beta\gamma\delta}P^\alpha P_x^\beta Q^\gamma S^\delta Q^2
	\left(P\cdot Q^2-M^2
	Q^2\right)\right]
\end{align}
\begin{equation}
C'_6=\frac{2 h^* \epsilon_{\alpha\beta\gamma\delta}K^\alpha P^\beta Q^\gamma S^\delta 
	(P\cdot Q-2 K\cdot P)}{M^3}
\end{equation}
\begin{align}
C'_7=&\frac{2 h^*}{M^3
	\left(P\cdot Q^2-M^2 Q^2\right)^2} \left(\epsilon_{\alpha\beta\gamma\delta}K^\alpha P^\beta Q^\gamma S^\delta
	\left(P\cdot Q P_x\cdot Q-P\cdot P_x
	Q^2\right)\right.\nonumber\\
	&\left.+\epsilon_{\alpha\beta\gamma\delta}K^\alpha P_x^\beta Q^\gamma S^\delta  \left(M^2
	Q^2-P\cdot Q^2\right)\right)
	\left(P_x\cdot Q \left(M^2 Q^2-2 K\cdot P
	P\cdot Q\right)\right.\nonumber\\
	&\left.+P\cdot P_x Q^2 (2
	K\cdot P-P\cdot Q)+2 K\cdot P_x
	\left(P\cdot Q^2-M^2
	Q^2\right)\right)
\end{align}
\begin{align}
C'_8=&\frac{h^*}{M^5 Q^2-M^3 P\cdot Q^2} \left[2 \epsilon_{\alpha\beta\gamma\delta}K^\alpha P^\beta Q^\gamma S^\delta
	\left(P_x\cdot Q \left(M^2 Q^2-2 K\cdot P
	P\cdot Q\right)\right.\right.\nonumber\\
	&\left.\left.+P\cdot P_x Q^2 (2
	K\cdot P-P\cdot Q)+2 K\cdot P_x
	\left(P\cdot Q^2-M^2
	Q^2\right)\right)\right.\nonumber\\
	&\left.+\epsilon_{\alpha\beta\gamma\delta}P^\alpha P_x^\beta Q^\gamma S^\delta
	\left(M^2 Q^4-P\cdot Q^2
	Q^2\right)\right]
\end{align}
\begin{equation}
C'_9=\frac{h h^* Q^2 (Q\cdot S-2
	K\cdot S)}{M}
\end{equation}
\begin{align}
C'_{10}=&\frac{h h^* P\cdot Q Q\cdot S}{M^3
	\left(P\cdot Q^2-M^2 Q^2\right)}
	\left[P_x\cdot Q \left(M^2 Q^2-2 K\cdot P
	P\cdot Q\right)\right.\nonumber\\
	&\left.+P\cdot P_x Q^2 (2
	K\cdot P-P\cdot Q)+2 K\cdot P_x
	\left(P\cdot Q^2-M^2
	Q^2\right)\right]
\end{align}
\begin{equation}
C'_{11}=\frac{h h^* \left(Q\cdot S \left(2
	K\cdot P P\cdot Q-M^2 Q^2\right)+2
	K\cdot S \left(M^2
	Q^2-P\cdot Q^2\right)\right)}{M^3}
\end{equation}
\begin{align}
C'_{12}=&\frac{h h^* Q^2}{M^3
	\left(P\cdot Q^2-M^2 Q^2\right)^2} \Bigl\{-Q\cdot S \Bigl[2
	P\cdot P_x \bigl(P_x\cdot Q
	\bigl(K\cdot P \left(M^2
	Q^2+P\cdot Q^2\right)\nonumber\\
	&-P\cdot Q^3\bigr)+
	K\cdot P_x P\cdot Q (M^2
	Q^2-P\cdot Q^2)\bigr)\nonumber\\
	&-P\cdot P_x^2
	Q^2 (2 P\cdot Q
	(K\cdot P-P\cdot Q)+M^2 Q^2)\nonumber\\
	&+M^2
	P\cdot Q P_x\cdot Q^2 (P\cdot Q-2
	K\cdot P)+2 K\cdot P_x M^2 P_x\cdot Q
	\left(P\cdot Q^2-M^2
	Q^2\right)\nonumber\\
	&+M_x^2
	\left(P\cdot Q^2-M^2
	Q^2\right)^2\Bigr]-P_x\cdot S \left(M^2
	Q^2-P\cdot Q^2\right)\nonumber\\
	&\times \bigl(P_x\cdot Q
	\left(2 K\cdot P P\cdot Q-M^2
	Q^2\right)+P\cdot P_x Q^2 (P\cdot Q-2
	K\cdot P)\nonumber\\
	&+2 K\cdot P_x \left(M^2
	Q^2-P\cdot Q^2\right)\bigr)+2 K\cdot S
	\left(M^2 Q^2-P\cdot Q^2\right)\nonumber\\
	&\times
	\left(M_x^2 \left(M^2
	Q^2-P\cdot Q^2\right)-M^2 P_x\cdot Q^2+2
	P\cdot P_x P\cdot Q
	P_x\cdot Q-P\cdot P_x^2
	Q^2\right)\Bigr\}
\end{align}
\begin{align}
C'_{13}=&\frac{h h^* Q^2 (2 K\cdot P-P\cdot Q)}{M^5 Q^2-M^3 P\cdot Q^2}
	\left[Q\cdot S \left(M^2
	P_x\cdot Q-P\cdot P_x
	P\cdot Q\right)\right.\nonumber\\
	&\left.+P_x\cdot S
	\left(P\cdot Q^2-M^2
	Q^2\right)\right]
\end{align}

\section{Invariant Functions}\label{sec-invert}

\subsection{Semi-inclusive}

Using Eqs.~(\ref{eq-ss11}--\ref{eq-ss13}), Eqs.~(\ref{eq-r19aa}--\ref{eq-r24y4}) can be inverted to give the invariant functions in terms of the response functions as
\begin{equation}
W_1=
\frac{1}{2} (\left[W^{TT}_{unpol}\right] \sec 2
\phi_x +\left[W^{T}_{unpol}\right])
\end{equation}
\begin{equation}
W_2=
\frac{1}{2} \rho  (\left[W^{TT}_{unpol}\right]
\sec 2 \phi_x +2 \rho 
\left[W^{L}_{unpol}\right]+\left[W^{T}_{unpol}\right])
\end{equation}
\begin{equation}
W_3=
-\frac{\left[W^{TT}_{unpol}\right] \sec 2 \phi_x
}{\eta_x^2}
\end{equation}
\begin{equation}
W_4= \frac{\rho
	\left[W^{TL}_{unpol}\right] \sec \phi_x }{2
	\sqrt{2}
	\eta_x}
\end{equation}
\begin{equation}
W_5=
-\frac{\rho  \left[W^{TL'}_{unpol}\right] \csc
	\phi_x }{2 \sqrt{2}
	\eta_x}
\end{equation}
\begin{equation}
W'_1= \frac{M
	(\left[W^{TT}_{pol}\right]^{N'} \sec 2 \phi_x
	+\left[W^{TT}_{pol}\right]^{N'})}{2 \eta_x
	q}
\end{equation}
\begin{align}
W'_2=& \frac{M
	\rho} {2 \eta_x \nu^{\prime}
	q} \left(\nu^{\prime}
	\left[W^{TT}_{pol}\right]^{N'} \sec 2 \phi_x +2
	\nu^{\prime} \rho 
	\left[W^{L}_{pol}\right]^{N'}+\nu^{\prime}
	\left[W^{TT}_{pol}\right]^{N'}\right.\nonumber\\
	&\left.-\sqrt{2} \rho 
	\left[W^{TL}_{pol}\right]^{L'} \csc \phi_x
	\right)
\end{align}
\begin{equation}
W'_3= -\frac{M
	\sec (2 \phi_x )
	(\left[W^{TT}_{pol}\right]^{N'}-\left[W^{TT}_{pol}\right]^{S'} \cot
	2 \phi_x )}{\eta_x^3
	q}
\end{equation}
\begin{align}
W'_4=& \frac{M \rho  \sec \phi_x} {2 \sqrt{2}
	\eta_x^2 \nu^{\prime}
	q}
	\left(-\nu^{\prime} \left[W^{TL}_{pol}\right]^S 
	\cot \phi_x +\nu^{\prime}
	\left[W^{TL}_{pol}\right]^{N'}\right.\nonumber\\
	&\left. +\sqrt{2}
	\left[W^{TT}_{pol}\right]^{L'} \cos \phi_x  \csc
	2 \phi_x \right)
\end{align}
\begin{equation}
W'_5= \frac{M
	\rho  \left[W^{TL}_{pol}\right]^{L'} \csc \phi_x
}{2 \sqrt{2} \eta_x
	\nu^{\prime}
	q}
\end{equation}
\begin{equation}
W'_6= -\frac{M
	\rho  \left[W^{TL}_{pol}\right]^{S'} \csc \phi_x
}{2 \sqrt{2}
	q}
\end{equation}
\begin{equation}
W'_7= -\frac{M
	\left[W^{TT}_{pol}\right]^{L'} \csc 2 \phi_x }{2
	\eta_x^2 \nu^{\prime}
	q}
\end{equation}
\begin{equation}
W'_8= -\frac{M
	\left[W^{TT}_{pol}\right]^{S'} \csc 2 \phi_x }{2
	\eta_x
	q}
\end{equation}
\begin{equation}
W'_9= -\frac{M
	\left(\sqrt{2} \nu^{\prime}
	\left[W^{TL'}_{pol}\right]^{N'} \csc \phi_x
	+\sqrt{2} \nu^{\prime}
	\left[W^{TL}_{pol}\right]^{S'} \sec \phi_x -2
	\left[W^{T'}_{pol}\right]^{L'}\right)}{4 \nu^{\prime}
	q}
\end{equation}
\begin{equation}
W'_{10}= \frac{M
	\rho  \left[W^{TL'}_{pol}\right]^{N'} \sec \phi_x
}{2 \sqrt{2} \eta_x
	\nu^{\prime}
	q}
\end{equation}
\begin{equation}
W'_{11}= -\frac{M
	\rho  \left(\sqrt{2} \nu^{\prime}
	\left[W^{TL'}_{pol}\right]^{N'} \csc \phi_x -2
	\left[W^{T'}_{pol}\right]^{L'}\right)}{4 \nu^{\prime}
	q}
\end{equation}
\begin{equation}
W'_{12}= \frac{M
	(\left[W^{TL'}_{pol}\right]^{N'} \csc \phi_x
	+\left[W^{TL}_{pol}\right]^{S'} \sec \phi_x )}{2
	\sqrt{2} \eta_x^2
	q}
\end{equation}
\begin{equation}
W'_{13}= \frac{M
	\left[W^{T'}_{pol}\right]^{S'}}{2 \eta_x
	q}
\end{equation}

Note that although Eqs.~(\ref{eq-r19aa}--\ref{eq-r24y4}) are derived in the rest frame, the expressions a valid in all frames where $\bm{q}$ is defined to be parallel to the z-axis. Therefore, we have dropped the subscript $R$ in the expressions given above.

\subsection{Inclusive}


Inverting Eqs.~(\ref{eq-inc11}--\ref{eq-inc23})) gives the inclusive invariant functions in terms of the response functions
\begin{equation}
\left(W_1\right)^{incl}=
\frac{\left[W^T_{unpol}\right]^{incl}}{2}
\end{equation}
\begin{equation}
\left(W_2\right)^{incl}=
\frac{1}{2} \rho  \left(2 \rho
\left[W^L_{unpol}\right]^{incl}+\left[W^T_{unpol}\right]^{incl}\right)
\end{equation}
\begin{equation}
\left(W'_6\right)^{incl}=
\frac{M \rho  \left[W^{TL}_{pol}\right]^{incl}\csc
	\theta^*\csc \phi^* }{2
	\sqrt{2}
	q}
\end{equation}
\begin{equation}
\left(W'_9\right)^{incl}=
\frac{M \left[W^{T'}_{pol}\right]^{incl}\sec\theta^*}{2
	\nu^{\prime}
	q}
\end{equation}
\begin{equation}
\left(W'_{11}\right)^{incl}=
\frac{M \rho  \left(\nu^{\prime}
	\left[W^{TL'}_{pol}\right]^{incl}\csc\theta^*\sec\phi^*+\sqrt{2}
	\left[W^{T'}_{pol}\right]^{incl}\sec\theta^*\right)}{2
	\sqrt{2} \nu
	q}
\end{equation}
\begin{figure}
	\centering
	\includegraphics[height=5cm]{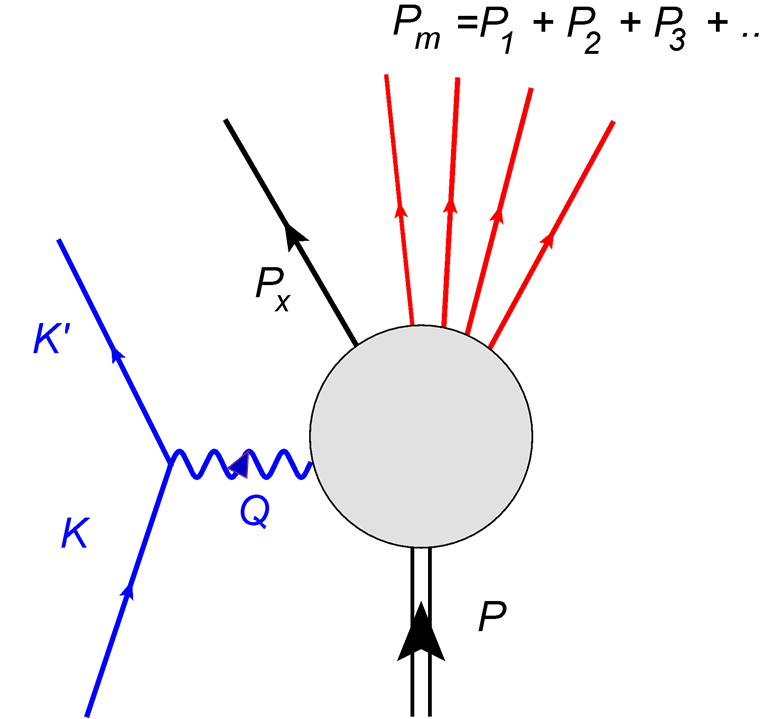} 				
	\caption{Feynman diagram for semi-inclusive electron scattering. The 4-momenta here are discussed in the text. In particular, particle x is assumed to be detected in coincidence with the scattered electron and thus $P_x^\mu$ is assumed to be known. Since the total final-state momentum ${P'}^\mu$ is known (see Fig.~\ref{fig:figure2} for inclusive scattering) this implies that the missing 4-momentum is also known via the relationship $P_m^\mu = {P^{\prime}}^\mu - P_x^\mu$ (see Fig.~\ref{fig:figure3}). Furthermore, for given kinematics the missing momentum is the sum of a set of momenta for the individual particles that constitute that unobserved part of the final state.
	}	
	\label{fig:figure6}
\end{figure}

\section{General Semi-Inclusive Cross Sections}\label{sec-6fold}

Consider the case of electron scattering from a hadronic target with 4-momentum $P^\mu$ producing $\mathcal{N}+1$ hadrons in the final state. For semi-inclusive scattering, the hadron $P_x^\mu$ is detected while the remaining $\mathcal{N}$ hadrons are not detected. This process is represented by the diagram in Fig. \ref{fig:figure6}.

We will use the conventions of \cite{Bjorken:1965sts} giving the differential cross section as
\begin{align}
	d\sigma_\mathcal{N}=&f\frac{m_e}{\epsilon}\frac{M}{E_p}\frac{d^3k'm_e}{(2\pi)^3 \epsilon'}\frac{d^3p_x\zeta_x}{(2\pi)^3\sqrt{p_x^2+M_x^2}}\left(\prod_{i=1}^\mathcal{N}\int \frac{d^3 p_i\zeta_i}{(2\pi)^3\sqrt{p_i^2+m_i^2}}\right)\nonumber\\
	&\times|\mathcal{M}|^2(2\pi)^4\delta^4(K+P-K'-P_x-\sum_{j=1}^\mathcal{N} P_j)\label{eq:gcx_1}
\end{align}
where  $\zeta_{i(x)}=m_{i(x)}$ for Fermions,  $\zeta_{i(x)}=1/2$ for Bosons. The flux factor is given by  \cite{moller,goldbergerwatson}
\begin{equation}
	f=\frac{1}{\sqrt{(\bm{\beta_e}-\bm{\beta_P})^2-(\bm{\beta_e}\times\bm{\beta_P})^2}}\,,
\end{equation}
where
\begin{equation}
	\bm{\beta_e}=\frac{\bm{k}}{\epsilon}
\end{equation}	
and
\begin{equation}
	\bm{\beta_p}=\frac{\bm{p}}{E_p}\,.
\end{equation}
This is the general form of this factor whereas Bjorken and Drell omit the cross product constraining the electron and target velocities to be collinear. Equation (\ref{eq:gcx_1}) is correct in all Lorentz frames \cite{Bjorken:1965sts}.

It is convenient to use
\begin{equation}
	\int\frac{d^3p_i}{(2\pi)^3\sqrt{p_i^2+m_i^2}}=2\int\frac{d^4P_i}{(2\pi)^3}\delta(P_i^2-m_i^2)\theta(P_i^0)
\end{equation}

\begin{align}
	d\sigma_\mathcal{N}=&f\frac{m_e}{\epsilon}\frac{M}{E_p}\frac{d^3k'm_e}{(2\pi)^3 \epsilon'}\frac{d^3p_x\zeta_x}{(2\pi)^3\sqrt{p_x^2+M_x^2}}\left(\prod_{i=1}^\mathcal{N}\frac{2\zeta_i}{(2\pi)^3}\int d^4 P_i \delta(P_i^2-m_i^2)\theta(P^0_i)\right)\nonumber\\
	&\times|\mathcal{M}|^2(2\pi)^4\delta^4(K+P-K'-P_x-\sum_{j=1}^\mathcal{N} P_j)\label{eq:gcx_2}
\end{align}
Now define the missing 4-momentum as
\begin{equation}
	P_m=\sum_n^\mathcal{N} P_n
\end{equation}
\begin{align}
	d\sigma_\mathcal{N}=&f\frac{m_e}{\epsilon}\frac{M}{E_p}\frac{d^3k'm_e}{(2\pi)^3 \epsilon'}\frac{d^3p_x\zeta_x}{(2\pi)^3\sqrt{p_x^2+M_x^2}}\nonumber\\
	&\times\int d^4P_m(2\pi)^4\delta^4(K+P-K'-P_x-P_m)\nonumber\\
	&\times\left(\prod_{i=1}^\mathcal{N}\frac{2\zeta_i}{(2\pi)^3}\int d^4 P_i \delta(P_i^2-m_i^2)\theta(P^0_i)\right)\delta(P_m-\sum_{i=1}^\mathcal{N}P_n)|\mathcal{M}|^2\label{eq:gcx_3}
\end{align}
Writing
\begin{equation}
	P_n=\frac{P_m}{\mathcal{N}}+\mathcal{L}_n\,,
\end{equation}
then
\begin{equation}
	P_m=\sum_{n=1}^\mathcal{N} P_n=\sum_{n=1}^\mathcal{N} \left(\frac{P_m}{\mathcal{N}}+\mathcal{L}_n\right)=P_m+\sum_{n=1}^\mathcal{N} \mathcal{L}_n\,.
\end{equation}
This then implies that
\begin{equation}
	\sum_{n=1}^\mathcal{N} \mathcal{L}_n=0\,.
\end{equation}

The differential cross section then becomes
\begin{align}
	d\sigma_\mathcal{N}=&f\frac{m_e}{\epsilon}\frac{M}{E_p}\frac{d^3k'm_e}{(2\pi)^3 \epsilon'}\frac{d^3p_x\zeta_x}{(2\pi)^3\sqrt{p_x^2+M_x^2}}\nonumber\\
	&\times\int d^4P_m(2\pi)^4\delta^4(K+P-K'-P_x-P_m)\nonumber\\
	&\times\left(\prod_{i=1}^\mathcal{N}\frac{2\zeta_i}{(2\pi)^3}\int d^4 \mathcal{L}_i \delta((\frac{P_m}{\mathcal{N}}+\mathcal{L}_i)^2-m_i^2)\theta(\frac{P^0_m}{\mathcal{N}}+\mathcal{L}_i^0)\right)\nonumber\\
	&\times\delta(P_m-\sum_{n=1}^\mathcal{N}(\frac{P_m}{\mathcal{N}}+\mathcal{L}_n))|\mathcal{M}|^2\nonumber\\
	=&f\frac{m_e}{\epsilon}\frac{M}{E_p}\frac{d^3k'm_e}{(2\pi)^3 \epsilon'}\frac{d^3p_x\zeta_x}{(2\pi)^3\sqrt{p_x^2+M_x^2}}\nonumber\\
	&\times\int d^4P_m(2\pi)^4\delta^4(K+P-K'-P_x-P_m)\nonumber\\
	&\times\left(\prod_{i=1}^\mathcal{N}\frac{2\zeta_i}{(2\pi)^3}\int d^4 \mathcal{L}_i \delta((\frac{P_m}{\mathcal{N}}+\mathcal{L}_i)^2-m_i^2)\theta(\frac{P^0_m}{\mathcal{N}}+\mathcal{L}_i^0)\right)\nonumber\\
	&\times\delta(\sum_{n=1}^\mathcal{N}\mathcal{L}_n)|\mathcal{M}|^2\,.\nonumber\\
	\label{eq:gcx_4}
\end{align}

The minimum value of the invariant mass of the undetected particles is
\begin{equation}
	W^T_m=\sum_{j=1}^\mathcal{N}  m_j >0
\end{equation}
Using
\begin{equation}
	1=\int_{{W^T_m}^2}^\infty dW_m^2\delta(P_m^2-W_m^2)\theta(P_m^0)=2\int_{W^T_m}^\infty dW_m W_m\delta(W_m^2-P_m^2)\theta(P_m^0)
\end{equation}
and
\begin{equation}
	\int d^4P_m\delta(p_m^2-W_m^2)\theta(P_m^0)=\frac{1}{2}\int \frac{d^3p_m}{\sqrt{p_m^2+W_m^2}}\,,
\end{equation}
the differential cross section becomes
\begin{align}
	d\sigma_\mathcal{N}=&f\frac{m_e}{\epsilon}\frac{M}{E_p}\frac{d^3k'm_e}{(2\pi)^3 \epsilon'}\frac{d^3p_x\zeta_x}{(2\pi)^3\sqrt{p_x^2+M_x^2}}\nonumber\\
	&\times\int_{{W_m^T}}^\infty dW_mW_m\int \frac{d^3p_m}{\sqrt{p_m^2+W_m^2}}(2\pi)^4\delta^4(K+P-K'-P_x-P_m)\nonumber\\
	&\times\left(\prod_{i=1}^\mathcal{N}\frac{2\zeta_i}{(2\pi)^3}\int d^4 \mathcal{L}_i \delta((\frac{P_m}{\mathcal{N}}+\mathcal{L}_i)^2-m_i^2)\theta(\frac{P^0_m}{\mathcal{N}}+\mathcal{L}_i^0)\right)\nonumber\\
	&\times\delta(\sum_{n=1}^\mathcal{N}\mathcal{L}_n)|\mathcal{M}|^2\,.\nonumber\\
	\label{eq:gcx_5}
\end{align}

The absolute square of the reduced scattering matrix is given by
\begin{equation}
	m_e^2|\mathcal{M}|^2=\frac{4\pi^2\alpha^2}{Q^4}\chi_{\mu\nu}W^{\mu\nu}_\mathcal{N}
\end{equation}
where the hadronic tensor is
\begin{align}
	W^{\mu\nu}_\mathcal{N}=&\sum_{s_x}\sum_{s_1}\cdots\sum_{s_\mathcal{N}}\left<P,s_R\right|J^\mu(Q)\left|P_x,s_x;P_1,s_1;\dots;P_\mathcal{N},s_\mathcal{N};(-)\right>^*\nonumber\\
	&\times\left<P_x,s_x;P_1,s_1;\dots;P_\mathcal{N},s_\mathcal{N};(-)\right|J^\nu(Q)\left|P,s_R\right>\,,
\end{align}
where $(-)$ indicates that the many-particle final state must be constructed with incoming scatting boundary conditions. The final state must have the complete symmetry associated with the combination of Fermions and Bosons contributing to this state. Note that the current operator $J(Q)$ appearing in the matrix element may consist of a complete set of one-body and many-body contributions appropriate for any particular system.

Now define
\begin{align}
	W^{\mu\nu}=&(2\pi)^3\left(\prod_{i=1}^\mathcal{N}\frac{2\zeta_i}{(2\pi)^3}\int d^4 \mathcal{L}_i \delta((\frac{P_m}{\mathcal{N}}+\mathcal{L}_i)^2-m_i^2)\theta(\frac{P^0_m}{\mathcal{N}}+\mathcal{L}_i^0)\right)\nonumber\\
	&\times\delta(\sum_{n=1}^\mathcal{N}\mathcal{L}_n)W^{\mu\nu}_\mathcal{N}\,.
\end{align}
The differential cross section can then be written as
\begin{align}
	d\sigma_\mathcal{N}=&f\frac{1}{\epsilon}\frac{M}{E_p}\frac{4\pi^2\alpha^2}{Q^4}\frac{d^3k'}{(2\pi)^3 \epsilon'}\frac{d^3p_x\zeta_x}{(2\pi)^3\sqrt{p_x^2+M_x^2}}\int_{{W_m^T}}^\infty dW_mW_m\int \frac{d^3p_m}{(2\pi)^3\sqrt{p_m^2+W_m^2}}\nonumber\\
	&\times (2\pi)^4\delta^4(K+P-K'-P_x-P_m)\chi_{\mu\nu}W^{\mu\nu}\nonumber\\
	=&f\frac{1}{(2\pi)^3}\frac{1}{\epsilon}\frac{M}{E_p}\frac{\alpha^2v_0}{Q^4}\frac{d^3k'}{ \epsilon'}\frac{d^3p_x\zeta_x}{\sqrt{p_x^2+M_x^2}}\int_{{W_m^T}}^\infty dW_mW_m\int \frac{d^3p_m}{\sqrt{p_m^2+W_m^2}}\nonumber\\
	&\times \delta^4(K+P-K'-P_x-P_m)\frac{\chi_{\mu\nu}W^{\mu\nu}}{v_0}		
	\label{eq:gcx_6}
\end{align}

In the extreme relativistic limit let
\begin{align}
	v_0=&4kk'\cos^2\frac{\theta_e}{2}\,,
\end{align}
and 
\begin{align}
	Q^2\cong& (k-k')^2-q^2=k^2-2kk'+{k'}^2-k^2+2kk'\cos\theta_e+k'^2\nonumber\\
	=&2kk'(-1+\cos\theta_e)=-4kk'\sin^2\frac{\theta_e}{2}\,.
\end{align}

Using combination of constants
\begin{equation}
\frac{1}{k}\frac{\alpha^2v_0}{Q^4}\cong\frac{1}{k'}\frac{\alpha^2\cos^2\frac{\theta_e}{2}}{4 k^2\sin^4\frac{\theta_e}{2}}=\frac{1}{k'}\sigma_{Mott}\,,
\end{equation}
the differential cross section becomes
\begin{align}
	d\sigma_\mathcal{N}=&\frac{f}{(2\pi)^3}\sigma_{Mott}\frac{M}{E_p}\frac{d^3k'}{ {k'}^2}\frac{d^3p_x\zeta_x}{\sqrt{p_x^2+M_x^2}}\int_{{W_m^T}}^\infty dW_mW_m\int \frac{d^3p_m}{\sqrt{p_m^2+W_m^2}}\nonumber\\
	&\times \delta^4(K+P-K'-P_x-P_m)\frac{\chi_{\mu\nu}W^{\mu\nu}}{v_0}\,.		
	\label{eq:gcx_7}
\end{align}

The six-fold differential cross section is then
\begin{align}
	\frac{d^6\sigma_\mathcal{N}}{dk'd\Omega'dp_x d\Omega_x}=&\frac{f}{(2\pi)^3}\sigma_{Mott}\frac{M}{E_p}\frac{p_x^2\zeta_x}{\sqrt{p_x^2+M_x^2}}\int_{{W_m^T}}^\infty dW_mW_m\int \frac{d^3p_m}{\sqrt{p_m^2+W_m^2}}\nonumber\\
	&\times \delta^4(K+P-K'-P_x-P_m)\frac{\chi_{\mu\nu}W^{\mu\nu}}{v_0}\,.		
	\label{eq:gcx_8}
\end{align}

For some reactions it is possible that the residual system contains only one particle. A particular example of this is the case of semi-inclusive scattering from nuclei where the residual system may consist of one or more stable states of the daughter nucleus with masses $M_i$. Using Eq.~(\ref{eq:gcx_1}) for $\mathcal{N}=1$ with the unmeasured particle with mass $M_i$ 
\begin{align}
	d\sigma_i=&f\frac{m_e}{\epsilon}\frac{M}{E_p}\frac{d^3k'm_e}{(2\pi)^3 \epsilon'}\frac{d^3p_x\zeta_x}{(2\pi)^3\sqrt{p_x^2+M_x^2}}\int \frac{d^3 p_m\zeta_m}{(2\pi)^3\sqrt{p_m^2+M_i^2}}\nonumber\\
	&\times|\mathcal{M}|^2(2\pi)^4\delta^4(K+P-K'-P_x-P_m)\nonumber\\
	\cong&\frac{f}{(2\pi)^3}\sigma_{Mott}\frac{M}{E_p}\frac{d^3k'}{ {k'}^2}\frac{d^3p_x\zeta_x}{\sqrt{p_x^2+M_x^2}}\int \frac{d^3 p_m\zeta_m}{\sqrt{p_m^2+M_i^2}}\nonumber\\
	&\times\frac{\chi_{\mu\nu}W^{\mu\nu}}{v_0}\delta^4(K+P-K'-P_x-P_m)
	\label{eq:gcx_9}
\end{align}

The six-fold differential cross section is then
\begin{align}
	\frac{d^6\sigma_i}{dk'd\Omega'dp_x d\Omega_x}=&\frac{f}{(2\pi)^3}\sigma_{Mott}\frac{M}{E_p}\frac{p_x^2\zeta_x}{\sqrt{p_x^2+M_x^2}}\int \frac{d^3 p_m\zeta_m}{\sqrt{p_m^2+M_i^2}}\nonumber\\
	&\times\frac{\chi_{\mu\nu}W^{\mu\nu}}{v_0}\delta^4(K+P-K'-P_x-P_m)
	\label{eq:gcx_10}
\end{align}

\section{Kinematic Variables}\label{sec-kinevar}

Here we have collected some useful kinematical variables. From the energy
and momentum transfer variables we can define the following dimensionless
quantities \cite{Alberico:1988bv}:%
\begin{eqnarray}
\lambda &\equiv &\frac{\omega }{2M}  \label{eq-v1} \\
\kappa &\equiv &\frac{q}{2M}  \label{eq-v2} \\
\tau &\equiv &\frac{-Q^{2}}{4M^{2}}  \label{eq-v3}
\end{eqnarray}%
where then%
\begin{equation}
\tau =\kappa ^{2}-\lambda ^{2}.  \label{eq-v4}
\end{equation}%
In the rest system we have%
\begin{eqnarray}
\lambda _{R} &\equiv &\frac{\omega _{R}}{2M}  \label{eq-v5} \\
\kappa _{R} &\equiv &\frac{q_{R}}{2M}  \label{eq-v6} \\
\tau &=&\kappa _{R}^{2}-\lambda _{R}^{2},  \label{eq-v7}
\end{eqnarray}%
where, of course, $\tau $ is an invariant. In the target rest frame the $x$%
-variable is given by (see the following appendix)%
\begin{equation}
x_{R}=\frac{-Q^{2}}{2M\omega _{R}}=\frac{\tau }{\lambda _{R}}.  \label{eq-v8}
\end{equation}%
It is often convenient to use $\tau $ and $x_{R}$ as two independent
variables; Eq. (\ref{eq-v8}) then yields%
\begin{equation}
\lambda _{R}=\frac{\tau }{x_{R}}  \label{eq-v9}
\end{equation}%
and using Eq. (\ref{eq-v7}) one has%
\begin{equation}
\kappa _{R}=\frac{\tau }{x_{R}}\sqrt{1+\frac{x_{R}^{2}}{\tau }}.
\label{eq-v10}
\end{equation}%
This results in the following:%
\begin{eqnarray}
\frac{\lambda _{R}}{\kappa _{R}} &=&\frac{\omega _{R}}{q_{R}}=\frac{\nu _{R}}{q_{R}}=\nu^{\prime}=\frac{1}{\sqrt{%
1+\frac{x_{R}^{2}}{\tau }}}  \label{eq-v11} \\
\rho _{R} &\equiv &\frac{-Q^{2}}{q_{R}^{2}}=1-{\nu^{\prime}_R}^2%
=\frac{x_{R}^{2}}{\tau +x_{R}^{2}}.  \label{eq-v12}
\end{eqnarray}

One has that%
\begin{equation}
0<x_{R}<1,  \label{eq-v13}
\end{equation}%
as discussed in the following appendix. Also one can define the "high-energy
regime (HER)" as being where%
\begin{equation}
\tau \gg 1.  \label{eq-v14}
\end{equation}%
Accordingly, from the above identities, we find that in this regime%
\begin{equation}
\lambda _{R}\simeq \kappa _{R},  \label{eq-v15}
\end{equation}%
implying that%
\begin{equation}
\omega _{R}\simeq q_{R}  \label{eq-v16}
\end{equation}%
and that%
\begin{equation}
\rho _{R}\simeq \frac{x_{R}^{2}}{\tau }\ll 1.  \label{eq-v17}
\end{equation}

\section{Inclusive Scattering}\label{sec-inclusive}
We continue with some developments of the inclusive cross section: following standard practice, the expressions in Sec.~\ref{subsec-inclRest} can be related to
dimensionless invariant functions via%
\begin{eqnarray}
F_{1}^{incl}\left( x_R,Q^{2}\right)  &\equiv &M\left( W_{1}\left(
x_R,Q^{2}\right) \right) ^{incl}  \label{eq-hid1} \\
F_{2}^{incl}\left( x_R,Q^{2}\right)  &\equiv &\omega_R \left( W_{2}\left(
x_R,Q^{2}\right) \right) ^{incl} =\nu_R \left( W_{2}\left(
x_R,Q^{2}\right) \right) ^{incl} .  \label{eq-hid2}
\end{eqnarray}%
Note that these definitions are specific to the rest frame. To make the expressions invariant one should use $x \equiv |Q^2|/2 P\cdot Q$ and instead of $\omega_R=\nu_R$ use $P\cdot Q/M$. At very high momentum transfers one finds
reasonable (Bjorken) scaling:%
\begin{eqnarray}
&&F_{1}^{incl}\left( x_R,Q^{2}\right) \underset{Bj}{\longrightarrow }%
F_{1}^{incl}\left( x_R\right)   \label{eq-hid3} \\
&&F_{2}^{incl}\left( x_R,Q^{2}\right) \underset{Bj}{\longrightarrow }%
F_{2}^{incl}\left( x_R\right) ,  \label{eq-hid4}
\end{eqnarray}%
namely, these two responses become functions only of $x_R$. Moreover, let us define%
\begin{eqnarray}
\mathcal{R}_{L}^{R} &\equiv &v_{L}^R\left[ W_{unpol}^{L}\right] _{R}^{incl}
\label{eq-hid5} \\
\mathcal{R}_{T}^{R} &\equiv &v_{T}^R\left[ W_{unpol}^{T}\right] _{R}^{incl}
\label{eq-hid6}
\end{eqnarray}%
so that%
\begin{eqnarray}
\mathcal{R}_{1,R}^{incl} &=&\mathcal{R}_{L}^{R}+\mathcal{R}_{T}^{R}
\label{eq-hid7} \\
&=&\mathcal{R}_{T}^{R}\left( 1+\delta _{R}\right) ,  \label{eq-hid8}
\end{eqnarray}%
where%
\begin{equation}
\delta _{R}\equiv \frac{\mathcal{R}_{L}^{R}}{\mathcal{R}_{T}^{R}}.
\label{eq-hid9}
\end{equation}%
In principle $\mathcal{R}_{L}^{R}$ and $\mathcal{R}_{T}^{R}$ can be
separated by making a Rosenbluth plot of the unpolarized cross section
versus $\tan ^{2}\theta _{e}^{R}/2$ which occurs in $v_{T}^R$ but not in $v_{L}^R
$. Substituting from above one then finds that%
\begin{eqnarray}
\delta _{R} &=&\left( \frac{q_{R}}{\omega _{R}}\right) ^{2}\left[ \frac{\rho
_{R}F_{1}^{incl}+\frac{1}{2x_{R}}\left(
F_{2}^{incl}-2x_{R}F_{1}^{incl}\right) }{F_{1}^{incl}\left( 1+\frac{2}{\rho
_{R}}\tan ^{2}\theta _{e}^{R}/2\right) }\right]   \label{eq-hid10} \\
&=&\frac{1}{{\nu _{R}^{\prime}}^{2}}\mathcal{E}_{R}\left[ \rho _{R}+\frac{1}{2x_{R}}%
\left( F_{2}^{incl}-2x_{R}F_{1}^{incl}\right) /F_{1}^{incl}\right] ,
\label{eq-hid10a}
\end{eqnarray}%
where the kinematical variables here are discussed in  \ref{sec-kinevar} and $%
\mathcal{E}_{R}$ is the so-called longitudinal photon polarization given in
Eq. (\ref{eq-lept54}). In the very high-energy regime (HER) one finds that%
\begin{equation}
\mathcal{R}_{L}\ll \mathcal{R}_{T},  \label{eq-hid11}
\end{equation}%
namely, given that the usual conditions obtain where $\mathcal{E}_{R}$ is
not especially small, then%
\begin{equation}
\delta _{R}\ll 1.  \label{eq-hid12}
\end{equation}%
In this regime one has from the developments in  \ref{sec-kinevar} that%
\begin{equation}
\frac{q_{R}}{\omega _{R}}\simeq 1  \label{eq-hid13}
\end{equation}%
and that%
\begin{equation}
\rho _{R}\ll 1;  \label{eq-hid14}
\end{equation}%
accordingly one has that%
\begin{equation}
\frac{1}{2x_{R}}\left( F_{2}^{incl}-2x_{R}F_{1}^{incl}\right) \ll 1,
\label{eq-hid15}
\end{equation}%
namely, the Callan-Gross relationship%
\begin{equation}
F_{2}^{incl}\simeq 2x_{R}F_{1}^{incl}.  \label{eq-hid16}
\end{equation}%
However, if extreme conditions obtain where $\mathcal{E}_{R}\ll 1$ then $%
\delta _{R}$ may also be small even when Eq. (\ref{eq-hid16}) is not
satisfied.

To the above unpolarized results we now add the contributions that involve
the target polarization. We can define%
\begin{eqnarray}
\mathcal{R}_{TL}^{R} &\equiv &v_{TL}^{R}\left[ W_{pol}^{TL}\right]
_{R}^{incl}  \label{eq-hid17} \\
\mathcal{R}_{T^{\prime }}^{R} &\equiv &v_{T^{\prime }}^{R}\left[
W_{pol}^{T^{\prime }}\right] _{R}^{incl}  \label{eq-hid18} \\
\mathcal{R}_{TL^{\prime }}^{R} &\equiv &v_{TL^{\prime }}^{R}\left[
W_{pol}^{TL^{\prime }}\right] _{R}^{incl},  \label{eq-hid19}
\end{eqnarray}%
where the first does not involve polarized electrons, whereas the second and
third do and one has%
\begin{eqnarray}
h^{\ast }\left[ \mathcal{R}_{3}^{incl}\right] _{R} &=&\mathcal{R}_{TL}^{R}
\label{eq-hid18a} \\
hh^{\ast }\left[ \mathcal{R}_{4}^{incl}\right] _{R} &=&v_{TL^{\prime }}^{R}%
\left[ W_{pol}^{TL^{\prime }}\right] _{R}^{incl}+v_{T^{\prime }}^{R}\left[
W_{pol}^{T^{\prime }}\right] _{R}^{incl}.  \label{eq-hid18b}
\end{eqnarray}%
From the identities above together with identities involving the leptonic
factors introduced in Sect. \ref{sec-lept} one can show that the above parts
of the response involve the following:%
\begin{eqnarray}
\mathcal{R}_{TL}^{R} &\equiv &-2\left( \frac{\epsilon _{R}+\epsilon
_{R}^{\prime }}{M}\right) \tan \theta _{e}^{R}/2h^{\ast }\sin \theta
_{R}^{\ast }\sin \phi _{R}^{\ast }\left( W_{6}^{\prime }\right) ^{incl}
\label{eq-hid19a} \\
\mathcal{R}_{T^{\prime }}^{R} &\equiv &2\left( \frac{\omega _{R}}{q_{R}}%
\right) \left( \frac{\epsilon _{R}+\epsilon _{R}^{\prime }}{M}\right) \tan
^{2}\theta _{e}^{R}/2h^{\ast }\cos \theta _{R}^{\ast }\left( W_{9}^{\prime
}\right) ^{incl}  \label{eq-hid19b} \\
\mathcal{R}_{TL^{\prime }}^{R} &\equiv &2\left( \frac{q_{R}}{M}\right) \tan
\theta _{e}^{R}/2h^{\ast }\sin \theta _{R}^{\ast }\cos \phi _{R}^{\ast }%
\left[ \rho _{R}\left( W_{9}^{\prime }\right) ^{incl}-\left( W_{11}^{\prime
}\right) ^{incl}\right] .  \notag\\
&& \label{eq-hid19c}
\end{eqnarray}%
As noted above, clearly the three sectors $(\mathcal{R}_{1,R}^{incl}$, $%
\mathcal{R}_{3,R}^{incl}$ and $\mathcal{R}_{4,R}^{incl})$ can in principle
be separated by flipping the electron helicity $h$ and the direction of the
target spin via $h^{\ast }$. Then $\mathcal{R}_{TL^{\prime }}^{R}$ and $%
\mathcal{R}_{T^{\prime }}^{R}$ can be separated by pointing the target spin
in different directions as seen from Eqs. (\ref{eq-hid19b}--\ref{eq-hid19c}%
). Accordingly, all five invariant response functions $\left( W_{1,2}\right)
^{incl}$ and $\left( W_{6,9,11}^{\prime }\right) ^{incl}$ may be determined
separately either experimentally or via specific modeling in the rest
frame.We end this section by rewriting the single and double-polarized
results in a form that is closer to that in Eq. (\ref{eq-iinc1}):%
\begin{eqnarray}
\left[ \mathcal{R}_{3}^{incl}\right] _{R} &=&-2h^{\ast }\frac{1}{{\rho^{\prime}_R}}\left( 
\frac{q_{R}}{M}\right) \tan \theta _{e}^{R}/2\left( W_{6}^{\prime }\right)
^{incl}\sin \theta _{R}^{\ast }\sin \phi _{R}^{\ast }  \label{eq-hid20a} \\
\left[ \mathcal{R}_{4}^{incl}\right] _{R} &=&2hh^{\ast }\left( \frac{q_{R}}{M%
}\right) \tan \theta _{e}^{R}/2\left[ \nu^{\prime} _{R}\eta _{R}\tan \theta
_{e}^{R}\left( W_{9}^{\prime }\right) ^{incl}\cos \theta _{R}^{\ast }\right. 
\notag \\
&&\left. +\left( \rho _{R}\left( W_{9}^{\prime }\right) ^{incl}-\left(
W_{11}^{\prime }\right) ^{incl}\right) \sin \theta _{R}^{\ast }\cos \phi
_{R}^{\ast }\right] ,  \label{eq-hid20}
\end{eqnarray}%
where as earlier we have%
\begin{equation}
\rho^{\prime}_{R}=\frac{q_R}{\epsilon_R +\epsilon^{\prime}_R}.
\label{eq-hid21}
\end{equation}%
In the high-energy regime, as discussed above one has $\rho _{R}\ll 1$ and
accordingly the term above involving $\left( W_{9}^{\prime }\right) ^{incl}$
in that regime becomes negligible if $\left( W_{9}^{\prime }\right) ^{incl}$
and $\left( W_{11}^{\prime }\right) ^{incl}$ are comparable in size.

As for the symmetric case, the anti-symmetric (double-polarized) case may be
written in terms of other conventionally-defined invariant response
functions. From \cite{Filippone_2001} and \cite{Jeschonnek:2005nk}
\begin{eqnarray}
\left[ W_{pol}^{TL^{\prime }}\right] _{R}^{incl} &\equiv &-\frac{2\sqrt{2}}{%
M\nu^{\prime} _{R}}\left( g_{1}+g_{2}\right) \cdot \mathcal{P}_{S}  \label{eq-ip1} \\
\left[ W_{pol}^{T^{\prime }}\right] _{R}^{incl} &\equiv &-\frac{2}{M}\left(
g_{1}-\frac{\rho _{R}}{{\nu^{\prime} _{R}}^{2}}g_{2}\right) \cdot \mathcal{P}_{L}
\label{eq-ip2}
\end{eqnarray}%
and hence, using Eqs. (\ref{eq-inc24},\ref{eq-inc23})%
\begin{eqnarray}
g_{1}+g_{2} &=&\omega _{R}\left[ \left( W_{9}^{\prime }\right) ^{incl}-\frac{%
1}{\rho _{R}}\left( W_{11}^{\prime }\right) ^{incl}\right]   \label{eq-ip3}
\\
g_{1}-\frac{\rho _{R}}{{\nu^{\prime} _{R}}^{2}}g_{2} &=&-\omega _{R}\left(
W_{9}^{\prime }\right) ^{incl}.  \label{eq-ip4}
\end{eqnarray}%
For reference recall that%
\begin{eqnarray}
\nu^{\prime} _{R} &=&\frac{\omega _{R}}{q_{R}}=\frac{\nu_{R}}{q_{R}}  \label{eq-ip5} \\
\rho _{R} &=&\left\vert \frac{Q^{2}}{q_{R}^{2}}\right\vert =1-{\nu^{\prime} _{R}}^{2}.
\label{eq-ip6}
\end{eqnarray}%
This yields the following identities%
\begin{eqnarray}
g_{1} &=&\omega _{R}\left[ \left( 2\rho _{R}-1\right) \left( W_{9}^{\prime
}\right) ^{incl}-\left( W_{11}^{\prime }\right) ^{incl}\right] 
\label{eq-ip7} \\
g_{2} &=&\omega _{R}\frac{1-\rho _{R}}{\rho _{R}}\left[ 2\rho _{R}\left(
W_{9}^{\prime }\right) ^{incl}-\left( W_{11}^{\prime }\right) ^{incl}\right] 
\label{eq-ip8}
\end{eqnarray}%
and their inverses%
\begin{eqnarray}
\left( W_{9}^{\prime }\right) ^{incl} &=&-\frac{1}{\omega _{R}}\left[ g_{1}-%
\frac{\rho _{R}}{1-\rho _{R}}g_{2}\right]   \label{eq-ip9} \\
\left( W_{11}^{\prime }\right) ^{incl} &=&-\frac{1}{\omega _{R}}\rho _{R}%
\left[ 2g_{1}+\frac{1-2\rho _{R}}{1-\rho _{R}}g_{2}\right] .  \label{eq-ip10}
\end{eqnarray}%
Note that if $\left( W_{9}^{\prime }\right) ^{incl}$ and $\left(
W_{11}^{\prime }\right) ^{incl}$ are similar in magnitude and one is in the
HER where $\rho _{R}\ll 1$ then one finds that%
\begin{equation}
\left\vert \frac{g_{1}}{g_{2}}\right\vert \ll 1.  \label{eq-ip11}
\end{equation}%
Conversely, if $g_{1}$ and $g_{2}$ are similar in magnitude and one is in
the HER then one finds that%
\begin{equation}
\left\vert \frac{\left( W_{11}^{\prime }\right) ^{incl}}{\left(
W_{9}^{\prime }\right) ^{incl}}\right\vert \ll 1.  \label{eq-ip12}
\end{equation}
We note that all of the developments in this study are for completely general kinematics, aside from the fact that the ERL$_e$ has been evoked, and even that can easily be extended to inclusion of corrections arising from keeping the electron mass finite (see \cite{Donnelly:1985ry}). Thus, for example if the polarized target is assumed to be a proton and one is studying charged-pion electroproduction, in the resonance region one type of behavior may be observed while at very high energies a different type may pertain.

Finally, we note that these developments are easily inter-related to the treatment of the special case of elastic scattering of polarized electrons from polarized protons given in \cite{Sofiatti:2011yi}.



\end{document}